\documentclass[11pt]{amsart}
\usepackage{amsaddr}
\usepackage{graphicx} 
\usepackage{amsmath}
\usepackage{mathpazo}
\usepackage{hyperref}
\usepackage{etoolbox}
\usepackage{setspace}
\usepackage{amssymb,amscd}

\usepackage{mathrsfs}

\numberwithin{equation}{section}

\newcommand{\bea}{\begin{eqnarray}\displaystyle}
\newcommand{\eea}{\end{eqnarray}}

\newcommand{\mft}{\mathfrak{T}}

\newcommand{\del}{\partial}
\newcommand{\ov}{\overline}

\newcommand{\be}{\begin{equation}}
\newcommand{\ee}{\end{equation}}

\newcommand{\IC}{\mathbb{C}}
\newcommand{\IP}{\mathbb{P}}
\newcommand{\IR}{\mathbb{R}}

\usepackage{amsthm}

\newtheorem{theorem}{Theorem}[section] 
\newtheorem{lemma}[theorem]{Lemma}     

\theoremstyle{definition} 

\newtheorem{remark}[theorem]{Remark}

\newenvironment{myproof}[1][Proof]{\begin{proof}[#1]}{\end{proof}}

\begin{document}

\vspace{-10cm}
\title{On the Algebra of the Infrared with Twisted Masses}

\enlargethispage{2\baselineskip}
 
\author{Ahsan Z. Khan}

\address{Institute for Advanced Study, \\ School of Natural Sciences, \\  1 Einstein Drive,  Princeton, NJ, 08540, USA}
\email{khan@ias.edu}

\author{Gregory W. Moore}
\address{ Rutgers University, \\ NHETC and Department of Physics and Astronomy,  \\ 126 Frelinghuysen Rd., Piscataway, NJ, 08854, USA} 
\email{gwmoore@physics.rutgers.edu}

\begin{abstract}

The Algebra of the Infrared   \cite{Gaiotto:2015aoa} is a framework to construct local observables, interfaces, and categories of supersymmetric boundary conditions of massive $\mathcal{N}=(2,2)$ theories in two dimensions by using information only about the BPS sector. The resulting framework is known as the ``web-based formalism.'' In this paper we initiate the generalization of the web-based formalism to include a much wider class of $\mathcal{N}=(2,2)$ quantum field theories than was discussed in  \cite{Gaiotto:2015aoa}: theories with non-trivial twisted masses. The essential new ingredient is the presence of BPS particles within a fixed vacuum sector. In this paper we work out the web-based formalism for the simplest class of theories that allow for such BPS particles: theories with a single vacuum and a single twisted mass. We show that even in this simple setting there are interesting  new phenomenon including the emergence of Fock spaces of closed solitons and a natural appearance of Koszul dual algebras. Mathematically, studying theories with twisted masses includes studying the Fukaya-Seidel category of A-type boundary conditions for Landau-Ginzburg models defined by a closed holomorphic one-form. This paper sketches a web-based construction for the category of A-type boundary conditions for one-forms with a single Morse zero and a single non-trivial period. We demonstrate our formalism explicitly in a particularly instructive example.   \today
\end{abstract}

\maketitle

\tableofcontents

\section{Introduction}

The observables of a quantum field theory form remarkably rich algebraic structures. 
 For a generic quantum field theory a full non-perturbative construction of the algebra of observables is typically an intractable problem\footnote{While the algebra of observables in a quantum field theory is often a type III von Neumann algebra and these algebras are isomorphic in a structural sense, the practical construction of the detailed physical content encoded within these algebras remains highly challenging.}. More progress can be made in quantum field theories with special properties such as topological, conformal, or supersymmetric field theories. The basic philosophy of the algebra of the infrared is that for the special case of massive $\mathcal{N}=(2,2)$ quantum field theories in two dimensions the algebra of supersymmetric observables should be constructable entirely from data available in the infrared regime \footnote{As we discuss below, the far infrared of a massive two-dimensional $\mathcal{N}=(2,2)$ field theory consists of its BPS sector. That the BPS state counting indices can be used to construct numerical ``wall-crossing invariants" containing non-trivial information about supersymmetric observables is a relatively old idea that is explored in many papers (see \cite{Cecotti:1992rm, Hori:2000ck, Cecotti:2010qn, Iqbal:2012xm, Cordova:2015nma} for a small sample).
The algebra of the infrared can be viewed as a \textit{categorification} of this idea to the statement that one can recover the complete algebra of supersymmetric observables from the BPS data. The result is a sort of ``algebra of BPS particles," compatible with the general idea that the BPS sector of a quantum field theory with extended supersymmetry has interesting algebraic properties \cite{Harvey:1996gc}.}. 

In more detail, suppose our $\mathcal{N}=(2,2)$ quantum field theory has
\begin{itemize}
    \item An unbroken $U(1)_R$-symmetry,
    \item A finite set of vacua $\mathbb{V} = \{i,j, \dots\}$ each of which is massive, 
    \item A central charge $Z_{ij} \in \mathbb{C}$ for each ordered pair $(i,j)$ of vacua such that 
    \begin{equation}
    Z_{ij} + Z_{jk} = Z_{ik}
    \end{equation} for each triple $i,j,k$ of vacua.
\end{itemize} 

Given such a field theory, the particle spectrum has soliton sectors labeled by ordered pairs of vacua, where the energy in the $ij$ sector satisfies the BPS inequality
\begin{equation}
E_{ij} \geq |Z_{ij}|.
\end{equation}
The particles that saturate this bound are known as the BPS particles of type $ij$. The BPS particles are well-known to preserve a fraction (half in this case) of supersymmetry, and will therefore be of central importance in this line of inquiry. Thus in addition to the vacuum set $\mathbb{V}$ and central charges $\{Z_{ij}\}$, such a field theory has a spectrum of BPS particles. Finally, as we will discuss in more detail in later sections, these BPS particles have certain ``interaction vertices"\footnote{We will define what we mean by interaction vertices of BPS particles more precisely in what follows.} which are meant to capture the supersymmetric couplings of these BPS particles. The authors of \cite{Gaiotto:2015aoa} demonstrated how, by using this basic ``infrared data", consisting of the vacuum set, central charges, BPS spectrum and BPS interaction vertices, one can construct an $L_{\infty}$-algebra of supersymmetric bulk observables 
for the underlying $\mathcal{N}=(2,2)$ quantum field theory \footnote{The general theory of factorization algebras as discussed in Section 6.4 of \cite{Costello:2021jvx} predicts that the algebra of bulk observables of a cochain valued two-dimensional topological field theory is actually an $E_2$-algebra, an algebraic notion refining that of an $L_{\infty}$-algebra. We will say more about this refinement in Remark \ref{foot:E2}. We also refer to \cite{Gaiotto:2024gii} for a general discussion of the role of $L_{\infty}$-structures in the context of (perturbative) field theory}. 

In addition, the infrared data also gives a framework for discussing the category of boundary conditions: from it, one constructs an $A_{\infty}$ category of ``thimble" boundary conditions, which then generate (in an appropriate sense) the category of half-supersymmetric boundary conditions. The specific way these algebraic structures are constructed relies on a diagrammatic formalism involving ``webs" - graphs drawn on a plane or half-plane, where the slopes of edges are subject to certain constraints involving the central charges. 

More concretely, one can consider the example of Landau-Ginzburg models, where a massive $\mathcal{N}=(2,2)$ quantum field theory can be defined by specifying a target manifold $X$, assumed to be a K\"{a}hler, and a holomorphic superpotential $W: X \rightarrow \mathbb{C}$ assumed to be Morse\footnote{The Morse condition ensures that the theory is massive.}. For such models, the BPS soliton spectrum can be determined by solving the $\zeta$-soliton equation, and the interaction vertices of these BPS particles can be determined from solving the $\zeta$-instanton equation
with ``fan" boundary conditions. 
\footnote{The $\zeta$-instanton equation is a deformation of the Cauchy-Riemann equation. In some circles it is referred to as ``the Witten equation.'' In references such as \cite{Wang:2022xdn} it is also known as the ``complex gradient flow equation."}
The $A_{\infty}$-category of boundary conditions one builds from the web-based rules in \cite{Gaiotto:2015aoa} is then argued to be $A_{\infty}$-equivalent to the Fukaya-Seidel category of the pair $(X,W)$. The algebra of the infrared therefore in particular gives an independent ``web-based" construction of the category of A-type boundary conditions for massive Landau-Ginzburg models. In addition, it also gives a powerful framework for constructing supersymmetric interfaces, which in particular leads to the categorification of wall-crossing formulas of Picard-Lefschetz and Cecotti-Vafa \cite{Gaiotto:2015aoa, Khan:2020hir}.
\footnote{It also leads to a categorification of Stokes' phenomenon 
\cite{Gaiotto:2015aoa, Kapranov:2020zoa}. }
We will give a more detailed review of some of these developments in Section \ref{review}.

One expects the basic philosophy of the web-based formalism to extend to a much wider class of theories than the ones that fit the framework of \cite{Gaiotto:2015aoa}. A simple way to go beyond the assumptions required for the web formalism is as follows. If our $\mathcal{N}=(2,2)$ quantum field theory has a global $U(1)$-symmetry with a generating charge $q$, so that $q$ commutes with all the generators of the supersymmetry algebra, one can perform a relevant deformation by turning on a ``twisted mass" $m \in \mathbb{C}$ for this symmetry \cite{AlvarezGaume:1983ab,Hanany:1997vm}. 
The resulting theory with the twisted mass deformation has a deformed central charge which has an additive contribution from the generator $q$ of the form $\Delta Z \sim mq.$ The original assumptions of the web formalism imply that the central charge in the $i$th vacuum sector vanishes, $Z_{ii} = 0$. However, with a non-trivial twisted mass one can have an excitation around the $i$th vacuum that is charged under the $U(1)$ global symmetry. Such an excitation will carry a central charge of the form
\begin{equation}
Z = mq.
\end{equation}
In particular, with a non-trivial twisted mass there can be BPS particles in a fixed vacuum sector, leading to new phenomena not captured by the original web formalism.

For Landau-Ginzburg models the twisted mass deformation has a transparent geometric interpretation. Suppose the target space $X$ has a non-vanishing first Betti number $b_1(X) \neq 0,$ so that $X$ has some homologically non-trivial one-cycles. One can then define a $\mathcal{N}=(2,2)$ supersymmetric field theory by specifying a closed holomorphic one-form $\alpha$ on $X$ instead of a holomorphic function $W$. In such a case a ``twisted mass" simply refers to a period integral of the one-form $\alpha$ around a cycle $\gamma \in H_1(X,\mathbb{Z})$, and the global symmetry of the previous paragraph is given simply by the topological/winding symmetry associated to the non-trivial one-cycles. We recover the previous case of a ``Landau-Ginzburg model" associated to $(X,W)$ when $\alpha$ is an exact form being written as $\text{d}W$ for a global holomorphic function $W$ on $X$. Thus a ``Landau-Ginzburg model with twisted masses" just refers to a theory defined by a closed holomorphic  one-form $\alpha.$ The $ii$-BPS particles then come from quantizing $\textit{closed}$ BPS solitons: paths in $X$ which solve BPS soliton equation and that go to the same zero, say $\phi_i$, of $\alpha$ at both ends of $\mathbb{R}$. 
\footnote{An analogous situation is the generalization of Morse theory to Morse-Novikov theory, where instead of considering a Morse function $h$ on a (Riemannian) manofold $M$ one instead considers a closed one-form $\mu$. When $\mu$ has a single non-trivial period, Morse-Novikov theory is also known as ``circle-valued Morse theory". In circle-valued Morse theory the closed gradient trajectories also play a crucial role, allowing us to connect Morse theory with the notion of Reidemeister torsion \cite{hutchings1, hutchings2, hutchings3}.} 

Indeed, one of our motivations for writing the present paper is to give a web-based construction of the ``Fukaya-Seidel" category associated to the K\"{a}hler manifold $X$ and a closed holomorphic one-form $\alpha$. For some initial remarks in this direction see Section 1.7 of \cite{Wang:2022xdn}. See also the discussion of closed one-forms and the associated ``wall-crossing structures" in \cite{Kontsevich:2024esg}.

Many of the Landau-Ginzburg models that are of interest in low dimensional topology  \cite{Witten:2011zz, Haydys:2010dv, Galakhov:2016cji, Gaiotto:2011nm, Aganagic:2021ubp} have non-trivial twisted masses. This includes for instance the Landau-Ginzburg model where the role of the superpotential is played by the complex Chern-Simons functional on a three-manifold $M_3$. In that context, the twisted mass is the same as the ``instanton number" of a field configuration on $M_3 \times \mathbb{R}$ (or $M_3 \times \mathbb{R}_+$ in the presence of boundaries).  In Witten's gauge theoretic formulation of knot homology \cite{Witten:2011zz}, it is precisely this non-trivial twisted mass that is responsible for the extra gradation one is familiar with in Khovanov homology.

The purpose of this (series of) paper(s) is to extend the web-based infrared framework to $\mathcal{N}=(2,2)$ quantum field theories with non-trivial twisted masses. In the present paper we will focus our attention exclusively on theories with a single vacuum, and a single twisted mass, with a more general discussion left to a future paper \cite{WIP}. Even in the simple setup of the present paper we find some novel and interesting phenomena. One of our main results in this setting is a Koszul duality theorem for the web-based algebras associated to the ``left" and ``right" thimble boundary conditions. We demonstrate this result in a simple and illustrative toy example: the Landau-Ginzburg model which is the mirror dual to the theory of $N$ free chiral superfields with a uniform twisted mass. In this example we reproduce the Koszul duality between the exterior and symmetric algebra of an $N$-dimensional complex vector space.  Our web-based discussion of this Koszul duality result elucidates a number of previous observations of a similar nature \cite{Aganagic:2021ubp, Bullimore:2016nji}. We also find an interesting emergence of Fock space combinatorics in the description of closed solitons which, we expect, will be of great importance in further development of this subject.

In the rest of the paper we proceed as follows. Section \ref{review} reviews briefly the main ingredients involved in the web-based formalism of \cite{Gaiotto:2015aoa}. It also discusses how the category of boundary conditions is constructed in this framework, and illustrates the idea with a simple example. Section \ref{twistedmasses} then provides a more extensive overview of the twisted mass construction and the new phenomena one encounters in this more general setting. In section \ref{subsec:GeneralStrategy} we discuss the basic issues one encounters when trying to build a web-based formalism for such theories and outline a general strategy to address these issues. In Section \ref{toyexample} we illustrate the general strategy explicitly in a simple toy example and work out the categories of boundary conditions for both the left and right half-planes, showing their Koszul duality. Section \ref{koszul} then generalizes this result to an arbitrary theory with a single vacuum and a single twisted mass, showing how one can recover the cobar complex and therefore Koszul duality of left and right thimbles using webs. We conclude in Section \ref{conc} and also include a collection of illustrative examples of Landau-Ginzburg theories with non-trivial twisted masses in Appendix \ref{examples}. In Appendix \ref{App:groupactions} we describe briefly some aspects of an equivariant generalization of the web formalism which is needed in our discussion. In Appendix \ref{multivackoszul} we initiate the discussion of how Koszul duality generalizes in the presence of multiple vacua. This discussion even applies without twisted masses. 

There is some overlap of the results here with those presented in \cite{Khan:2021hve}, but we stress that the present paper reports on some important progress we have made since the appearance of  \cite{Khan:2021hve}.

\subsection*{Acknowledgements} We thank Tudor Dimofte, Davide Gaiotto and Edward Witten for valuable discussions. We thank Davide Gaiotto for some very useful comments on a preliminary draft. AK also thanks Semon Rezchikov for useful discussions and Justin Hilburn for encouragement. AK is supported by the Bershadsky Fund and the Sivian Fund at the Institute for Advanced Study along with the National Science Foundation under Grant No. PHY-2207584. The work of GM is supported by  the US Department of Energy under grant DE-SC0010008 to Rutgers University.  Furthermore, this work was in part written while GM was visiting the Institute for Advanced Study in Princeton, and he thanks the IAS for its excellent hospitality. At the IAS he was in part supported by the IBM Einstein Fellow Fund.  GM also thanks the Aspen Center for Physics, where this work was completed, for hospitality. The ACP is 
supported by National Science Foundation grant PHY-2210452.

\section{Review of the Web Formalism} \label{review}

\subsection{Infrared Data and Bulk Local Operators}
As stated in the introduction the starting point of the web-based formalism consists of specifying a finite set of vacua $\mathbb{V}$ where a typical element is denoted by a lowercase Latin letter such as $i$ or $j$. To each of these vacua we associate a complex number $i \rightarrow z_i \in \mathbb{C}$ known as the vacuum weights. We assume that the vacuum weights are in general position (and also away from ``exceptional walls'' as defined in \cite{Gaiotto:2015aoa}). In particular, for every distinct triple $(i,j,k)$ the $z_i,z_j,z_k$ are the vertices of a non-degenerate triangle. The central charge $z_{ij}$ in the $ij$-sector is expressed as a simple difference of vacuum weights 
\begin{equation}
 z_{ij} := z_j - z_i
\end{equation} so that in the complex plane we draw the central charge as a vector pointing from the $i$th vacuum weight to the $j$th vacuum weight. 

\begin{figure}
    \centering
    \includegraphics[width=0.15\linewidth]{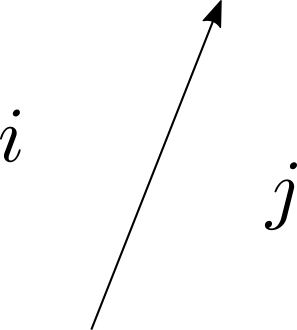}
    \caption{The oriented edge is parallel to $z_{ij} = z_j - z_i$. In Landau-Ginzburg models once we choose a phase $\zeta$ we have $z_i = \zeta^{-1}W_i$ where $W_i$ is a critical value of $W$.}
    \label{edge}
\end{figure}

Given the vacuum weights, one can define the notion of a plane web. A plane web simply consists of a graph drawn on the plane $\mathbb{R}^2$, where the faces are labeled by vacua, and edges are subject to the \textit{slope constraint}. The slope constraint says that if an edge, when oriented, has a face labelled $i \in \mathbb{V}$ to its left, and a face labelled $j \in \mathbb{V}$ to its right, it must be parallel to the complex number $z_{ij}$. See Figure \ref{edge}. 
A web naturally comes in a moduli space carrying an action of the group $\mathbb{R}_+ \ltimes\mathbb{R}^2$ consisting of scale transformations (through a choice of origin), which simply scale up every edge length, and translations which act by translating the position of each vertex by a fixed vector. Rotations do not act as they would result in a graph that violates the slope constraint.

\begin{figure}
    \centering
    \includegraphics[width=0.4\textwidth]{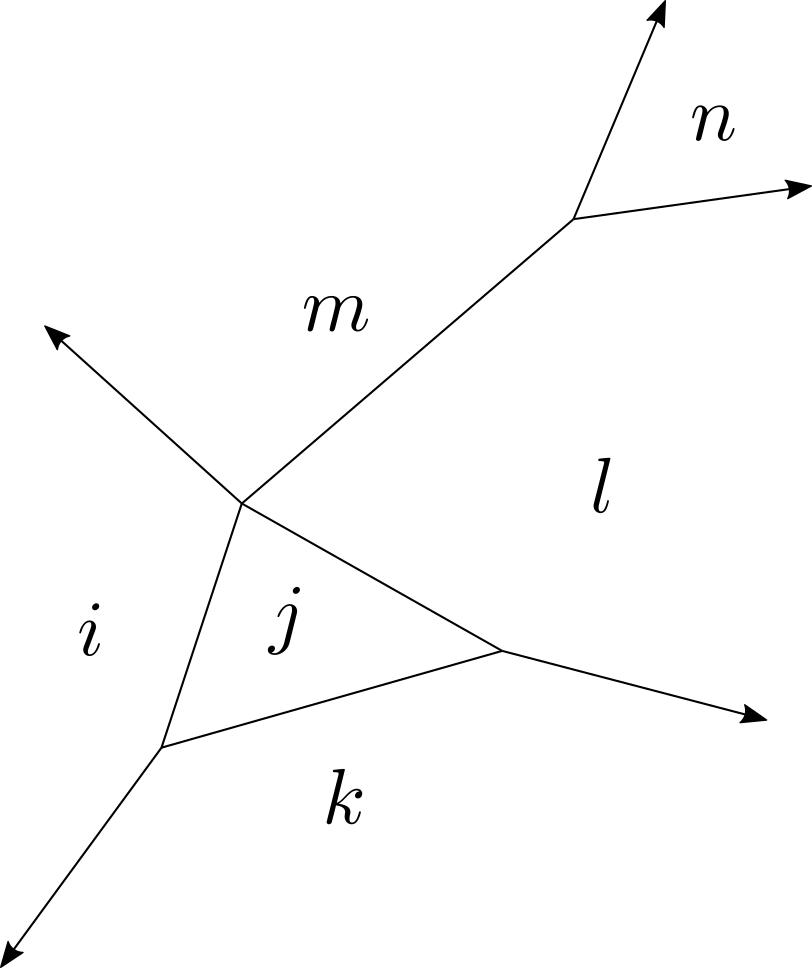}
    \caption{A planar web with four vertices and four internal edges. Its faces are labeled by vacua, elements of $\mathbb{V}$ denoted by letters such as $i,j,k,\dots$ and edges satisfy the slope constraint.}
    \label{randomweb}
\end{figure}

We can form the reduced moduli space of a web by dividing with the action of $\mathbb{R}_+ \ltimes \mathbb{R}^2.$ If this reduced moduli space for a given web is just a point, so that the web has no moduli other than translations and an overall scaling, it is called a \textit{taut} web. If the reduced moduli space of a web is one-dimensional, it is called \textit{sliding}. The web of Figure \ref{randomweb} for instance is a sliding web. If we shrunk the edge separating the face labelled $m$ and $l$ to a point, it would become taut.

A given vertex of a web determines a \textit{cyclic fan of vacua}, which consists of a cyclically ordered set 
\be 
\{i_0, i_1, \dots, i_n\} 
\ee of vacua such that the corresponding central charges 
\be 
\{z_{i_0i_1}, z_{i_1 i_2}, \dots, z_{i_n i_0} \} 
\ee
have slopes that are clockwise-ordered. 

\begin{remark}
We remark on the following degenerate cases. First we consider the set consisting a single vacuum $\{i \}$ as a cylic fan of vacua, calling it a zero-valent fan. Second, we consider the (unordered) set consisting of two distinct vacua $\{i,j\}$ also as a cyclic fan, calling it a bivalent fan. 
\end{remark}

\begin{remark} Cyclic fans of vacua are precisely in one-to-one correspondence with convex polygons (considered as subsets of $\mathbb{C}$) with vacuum weights as vertices. \end{remark}

The web formalism assumes we're given a collection of vector spaces $\{ R_{ij} \}$ graded by an integral\footnote{More precisely the spaces are graded by a $\mathbb{Z}$-torsor because of the well-known charge fractionalization effect.} ``cohomological degree", one for each ordered pair $(i,j)$ along with non-degenerate pairings 
\be 
K_{ij}: R_{ij}\otimes R_{ji} \rightarrow \mathbb{C}
\ee 
carrying cohomological degree $-1$ that are symmetric in the sense that 
\begin{equation} K_{ij}(r_{ij}, s_{ji}) = K_{ji}(s_{ji}, r_{ij}) 
\end{equation} for every $r_{ij} \in R_{ij}$ and $s_{ji} \in R_{ji}.$ 
\footnote{In many cases one sets $R_{ji} = R_{ij}^{\vee}[1]$, where as usual $V^{\vee}$ denotes the linear dual of the graded vector space $V$, and the pairing $K_{ij}$ is then given by the pairing between a $R_{ij}$ and $R_{ij}^{\vee}[1].$}
 
For a Landau-Ginzburg model defined by a K\"{a}hler target space $X$ and a Morse superpotential $W$, the vacuum set $\mathbb{V}$ consist of the set of critical points of $W$, which we assume to be a finite set. We choose a phase $\zeta$ on the unit circle which is not in the set $\{\frac{W_i - W_j}{|W_i - W_j|} \}_{i \neq j}$
of critical phases. The choice of $\zeta$ determines a choice of supercharge we wish to preserve. Letting $W_i = W(\phi_i)$ be the critical value of the $i$th critical point, the vacuum weights are then given by
\begin{equation}
z_i = \zeta^{-1}W_i.
\end{equation} 

We construct the space  $R_{ij}$  from the space of classical $ij$-BPS solitons. The latter are maps $\phi: \mathbb{R} \rightarrow X$ with 
\be 
\lim_{x \to -\infty} \phi(x) = \phi_i, \,\,\,\,\,\,\, \lim_{x \to +\infty} \phi(x) = \phi_j
\ee 
that solve the $\zeta$-soliton equation
\begin{equation}\label{eq:zeta-soliton}
    \frac{\text{d}\phi^a}{ \text{d}x} = \zeta g^{a\bar{b}} \frac{\partial \overline{W}}{ \partial \overline{\phi}^{\bar{b}}},
\end{equation} 
where, here, the phase $\zeta$ should be set to the critical value  
\be 
\zeta = \zeta_{ij} := \frac{W_j - W_i}{|W_j - W_i|}.
\ee 
(The phase $\zeta$ in  \eqref{eq:zeta-soliton} is not to be confused with the phase determining the vacuum weights above.) If $\zeta$ in equation \eqref{eq:zeta-soliton} is not the critical phase $\zeta_{ij}$, there are no solutions with the above boundary conditions at the infinite ends of $\mathbb{R}$. One can anticipate this using Morse theory. The $\zeta$-soliton equation is the gradient flow equation for the K\"{a}hler metric on $X$ and the real function $\text{Re}(\zeta^{-1}W)$. The Morse indices of critical points of the real part of a holomorphic function are all equal, being the complex dimension of the target space, 
\be
\mu(\phi_i) = \text{dim}_{\mathbb{C}}X, \,\,\,\,\,\, \phi_i \in \text{Crit}(W)
\ee 
and so the expected dimension of the reduced moduli space of solutions $\mathcal{M}(\phi_i, \phi_j)$ to the gradient flow equation   
\be 
\text{dim} \mathcal{M}(\phi_i, \phi_j) = \mu(\phi_j) - \mu(\phi_i) -1 = -1 
\ee 
is negative and therefore generically empty.
\footnote{We can also see this in a more geometric fashion by projecting a soliton to the $W$-plane.
See, for example, the classic discussion in \cite{Cecotti:1992qh}.  }
Introducing a parameter $\zeta$ increases the expected dimension by one, and so by tuning it to a particular value, we may get a solution.

The way we can build well-defined graded vector spaces from solutions of the $\zeta$-soliton equation uses the fact that the $\zeta$-soliton equation arises as the critical point equation of an actional functional on the space of maps from $\mathbb{R}$ to $X$. This happens because, by using holomorphicity of $W$ and the fact that $g$ is a K\"{a}hler metric, the $\zeta$-soliton equation may also be viewed as the Hamiltonian flow equation for the K\"{a}hler form $\omega$ and the Hamiltonian $H= \text{Im}(\zeta^{-1}W)$. Hamiltonian flow equations in turn arise as critical points of the symplectic action functional. 

To be more precise, we let $\mathfrak{X}_{ij}$ be the space of maps from $\mathbb{R}$ to $X$ with $ij$ boundary conditions 
\be 
\mathfrak{X}_{ij} = \{\phi: \mathbb{R} \rightarrow X| \lim_{ x\rightarrow -\infty} \phi(x) = \phi_i, \,\,\, \lim_{x \rightarrow + \infty} \phi(x) = \phi_j \},
\ee 
equipping it with the Riemannian metric induced from the K\"{a}hler metric on $X$: 
\be
||\delta \phi||^2 = \int_{\mathbb{R}} g_{a \bar{b}}(\phi(x))\delta \phi^a  \delta \ov{\phi}^{\ov{b}}.
\ee
Consider the function $h_{\zeta}: \mathfrak{X}_{ij} \rightarrow \mathbb{R}$ given as follows. Let $\lambda$ be a Liouville form for the symplectic form on $X$ so that $\text{d}\lambda = \omega$ (we assume $X$ is an exact symplectic manifold). $h_{\zeta}$ is given by 
\footnote{Strictly speaking, $h_{\zeta}$ is infinite with the boundary conditions under consideration.
However, this infinity is unimportant. It can be removed by an additive infinite constant that does not affect the variation. Alternatively, one can simply work from the start with the one form $\delta h_\zeta$. }
\be 
h_{\zeta}[\phi] = \int_{\mathbb{R}} \phi^*(\lambda) - \text{d}x \,\text{Im} \big(\zeta^{-1}W \big),
\ee
which is nothing but the symplectic action functional on $(X, \omega)$ with Hamiltonian 
\be\label{eq:Ham-zetaW}
H= \text{Im}(\zeta^{-1}W). 
\ee The $\zeta$-soliton equation is the critical point equation for this symplectic functional 
\be 
\delta h_{\zeta} = 0.
\ee This perspective tells us another crucial property of the $\zeta$-soliton equation: along a solution the value of $\text{Im}(\zeta^{-1}W)$ is fixed to be a constant. Combining this with the previous interpretation in terms of gradient flows gives us another reason as to why there can only be non-trivial solutions with $ij$ boundary conditions for $\zeta= \zeta_{ij}.$

One would like to define $R_{ij}$ to be something like the Morse-Smale-Witten complex for the functional $h_{\zeta}$ with $\zeta = \zeta_{ij} $, however that is complicated by the fact that $h_{\zeta}$ is not Morse: critical points are not isolated due to the translational invariance
\footnote{This is also related to the fact that the real function $\text{Re}(\zeta_{ij}^{-1}W)$ is not Morse-Smale: the thimbles do not intersect transversally.}
of $h_{\zeta}.$ One way to get around this is to break the translational invariance in the following way. We let $\zeta = e^{i \theta}$, choose $\theta_{ij}$ so that $\zeta_{ij} = e^{i \theta_{ij}}$, and promote $\theta$ to be function $\theta(x): \mathbb{R} \rightarrow S^1$ such that
\begin{align}
    \lim_{x \to -\infty} \theta(x) = \theta_{ij} - \epsilon, \\ \lim_{x \to +\infty} \theta(x) = \theta_{ij} + \epsilon, \\ \frac{d \theta(x)}{ d x} \geq 0.
\end{align} for a small positive constant $\epsilon >0 .$ The function 
\be 
h = \int_{\mathbb{R}} \phi^*(\lambda) - \text{d}x \,\text{Im}(e^{-i\theta(x)} W)
\ee 
is then a non-degenerate Morse function on $\mathfrak{X}_{ij}$ (assuming $W$ itself is a generic Morse function), and we define $R_{ij}$ to be the MSW complex of $h$. It will be generated by solutions of the \textit{forced} flow equation 
\be
\frac{\text{d}\phi^a}{ \text{d}x} = e^{i \theta(x)} g^{a \bar{b}} \frac{\del \ov{W}}{ \del \ov{\phi}^{\bar{b}}}.
\ee
Because $\mathfrak{X}_{ij}$ is infinite dimensional one must define the fermion degree of a soliton with care. In finite dimensional Morse theory one can define fermion degree as minus one-half of the sum of signs of the 
eigenvalues of the Hession. In infinite dimensions this is generalized to an $\eta$ invariant. 
The  Hessian of $h_{\theta(x)}$ around a soliton $\phi$ can be identified with a one-dimensional Dirac operator \cite{Gaiotto:2015aoa,Khan:2020hir}, and one uses (minus) the $\eta$ invariant of that operator.

Moreover, on this graded vector space, there is a differential given by counting solutions of the $e^{i \theta(x)}$ instanton equation, obtained as the gradient flow equation for $h_{\theta(x)}$ 
\be 
\frac{\del \phi^I}{\del \tau} = g^{IJ} \frac{\delta h_{\theta(x)}}{ \delta \phi^J}
\ee 
with the following boundary conditions:
\bea \lim_{\tau \rightarrow -\infty} \phi(\tau, x) = \phi^a_{ij}(x), \,\,\,\,\,\,\,\, \lim_{\tau \to +\infty} = \phi^{b}_{ij}(x), \\ \lim_{x \to -\infty} \phi(\tau,x) = \phi_i, \,\,\,\,\,\,\, \lim_{x \to +\infty} \phi(\tau,x) = \phi_j.\eea
By using the usual Morse theory arguments one can show that the chain complex $R_{ij}$ is invariant up to homotopy under small enough changes of the function $\theta(x)$. The cohomology of the complex $R_{ij}$ is closely related to the space of quantum BPS states with boundary conditions $ij$. 

One may relate the complex $R_{ij}$ to what we would obtain by performing semi-classical quantization of an $ij$-soliton. A classical $ij$ soliton has a bosonic zero mode corresponding to the translational symmetry of the $\zeta$-soliton equation, and the broken supersymmetries give rise to two fermionic zero modes which are its superpartners. This leads to a doublet of states in the quantum Hilbert space with fermion numbers $f_{ij}$ and $f_{ij}+1$. By the argument provided in 16.3.2 of \cite{Gaiotto:2015aoa}, $R_{ij}$ coincides with the space formed by taking the ``upper" soliton number state for each classical soliton, namely the state with fermion number $f_{ij}+1.$ 
The gradation $f_{ij}$ of an $ij$-soliton $\phi_{ij}$ in turn satisfies 
\begin{equation}
    f_{ij} \, \text{mod} \, \mathbb{Z} = \frac{1}{2\pi} \Big( \text{Arg}\big(W''(\phi_j) \big) -\text{Arg}\big(W''(\phi_i) \big) \Big).
\end{equation}
The pairings $K_{ij}$ pair up a $\zeta_{ij}$-soliton with a $\zeta_{ji}$-soliton obtained from reversing the direction $x \rightarrow -x$ of space.

It is also useful to relate the complex $R_{ij}$ with the intersection points of Lefschetz thimbles. Recall that a left Lefschetz thimble of phase $\zeta$ is given as the the set of points of $X$ which flow to the critical point $\phi_i$ in the infinite past under the flow map 
\be 
f^{\zeta}_x : X \rightarrow X, \,\,\,\,\, x \in \mathbb{R}
\ee 
given by the gradient flow of $\text{Re}(\zeta^{-1}W)$: 
\be\label{eq:LeftLefshetz-def}
L_i(\zeta) := \{p \in X | \lim_{x \to -\infty} f^{\zeta}_{x}(p) = \phi_i\}. 
\ee
Similarly a right thimble of a critical point $\phi_i$ is given by 
\be\label{eq:RightLefshetz-def}
R_i(\zeta) := \{p \in X | \lim_{x \to +\infty} f^{\zeta}_{x}(p) = \phi_i\}
\ee which in fact coincides with $L_i(-\zeta).$ Writing $\zeta = e^{i \theta}$ we will sometimes also use the notation $L_i(\theta)$ and $R_i(\theta)$. It is often very useful to 
note that under the flow $f^{\zeta}$ the quantity ${\rm Im}(\zeta^{-1}W)$ is conserved. 
(See equation \eqref{eq:Ham-zetaW} above.) 
Taking $\theta(x)$ to be a step function going between $\theta_{ij} - \epsilon$ to $\theta_{ij} + \epsilon$, (where $\epsilon$ can have either sign, but is small and nonzero) it is easy to see that $R_{ij}$ coincides with the free vector space generated by each intersection point of left and right thimbles, slightly rotated away from the phase of an $ij$-soliton, namely generators of $R_{ij}$ are in 1-1 correspondence with the intersection points: 
\be\label{eq:IntersectionCriterion}
L_i(e^{i(\theta_{ij}-\epsilon)}) \cap R_j(e^{i (\theta_{ij}+ \epsilon)}).
\ee 
Our assumption is that this intersection is transversal, consisting of a finite set of points, so that $R_{ij}$ is a finite-dimensional vector space. 

Going back to the abstract discussion, let us now define our first non-trivial algebraic structure. For this, we associate a vector space to each cyclic fan of vacua. Letting $I = \{i_0, i_1, \dots, i_n\}$ be a cyclic fan of vacua, we define 
\begin{equation}
    R_I = R_{i_0 i_1} \otimes R_{i_1 i_2} \otimes \dots \otimes R_{i_{n}i_0}.
\end{equation} We may again deal with the degenerate cases by assigning to the one-element cyclic fan $\{i\}$ the one-dimensional vector space $\mathbb{C}$,\footnote{This has to do with the Morse nature of the superpotential, or more generally, the assumption that each vacuum $i$ of the theory is massive. If we assumed that $W$ is Morse-Bott, the right thing would be to assign it the deRham cohomology of the $i$th critical manifold} generated by an element $\phi_i$ carrying cohomological degree zero,
\begin{equation}
    R_{\{i\}} = \mathbb{C} \phi_i,
\end{equation}
and assigning to the two-element cyclic fan $\{i,j\}$ the space 
\begin{equation}
    R_{\{i,j\}} = R_{ij} \otimes R_{ji}.
\end{equation} 

Let us denote the set of cyclic fans of vacua as $\mathrm{Fan}$ and set: 
\begin{equation}
R_c := \bigoplus_{I \in \mathrm{Fan} } R_I.
\end{equation} 
A little more explicitly, we can organize things by the valency of the fan, so that the zero-valent fans contribute a space $\oplus_{i \in \mathbb{V}} \mathbb{C} \phi_i \cong \mathbb{C}^{|\mathbb{V}|}$, the bivalent fan $\{i,j\}$ contributes the summand $R_{ij} \otimes R_{ji}$, and so on, so that 
\begin{align}
R_c = \oplus_{i \in \mathbb{C}}\mathbb{C}\phi_i \bigoplus \oplus_{(i \neq j)} R_{ij} \otimes R_{ji} \bigoplus \oplus_{ \{i,j,k\} \text{ cyclic fan } } R_{ij} \otimes R_{jk} \otimes R_{ki} \oplus \dots.
\end{align}

Given a web $\mathfrak{w}$ there is a natural multilinear operation \begin{equation}
    \rho(\mathfrak{w}): R_{c}^{\otimes |V(\mathfrak{w})|} \rightarrow R_{c}
\end{equation} given by applying the contraction map $K$ associated to each internal edge of the web $\mathfrak{w}$. Let $T_n$ denote the set of (deformation classes of) taut
webs with $n$ vertices, and define an $n$-ary map $\lambda_n: R_c^{\otimes n} \rightarrow R_c$ given by 
\begin{equation}\label{eq:Bare-L-infinity}
    \lambda_n = \sum_{\mathfrak{w} \in T_n} \rho(\mathfrak{w}).
\end{equation} 
We will call this the \emph{noninteracting} $L_\infty$ algebra. It will presently be deformed by a Maurer-Cartan element to 
produce an \emph{interacting} $L_\infty$ algebra. 
For clarity we spell out the degenerate case of $\lambda_1$ for the noninteracting $L_\infty$ algebra. The taut webs with a single vertex can easily be classified as follows. They consist of a single edge parallel to a central charge $Z_{ij}$, along with a closed zero valent vertex either on the face labelled by $i$ or the face labeled by $j$. Letting $\phi_i$ be a generating element for the one-dimensional vector space $R_{\{i\}},$ by definition this leads to the map
\begin{equation} 
\lambda_1(\phi_i) = \sum_{j \neq i} K_{ij}^{-1}(1) \in \oplus_{j \neq i} R_{ij} \otimes R_{ji}.
\end{equation}

A key result of \cite{Gaiotto:2015aoa} is the 
\begin{theorem} The collection of maps $ \{\lambda_n \}_{n=1,2,\dots}$ equip $R_c$ with the structure of an $L_{\infty}$-algebra.
\end{theorem}

The proof uses the idea that webs can be convolved: given two webs $\mathfrak{w}, \mathfrak{w}'$ such that the fan at infinity of $\mathfrak{w}'$ coincides with a fan at one of the finite vertices $v$ of $\mathfrak{w}$, one may define their convolution $\mathfrak{w}*_{v} \mathfrak{w}'$ by gluing in a copy of $\mathfrak{w'}$ at the vertex $v$ of $\mathfrak{w}$. We may then consider the formal sum of all taut webs $\mathfrak{t}$, and by considering boundaries of the moduli space of sliding webs deduce the fundamental relation 
\be\label{eq:t star t}
\mathfrak{t}* \mathfrak{t} = 0. \ee 
This in turn leads to the $L_{\infty}$-relations on $R_c$. For more details see sections 2.1 and 4.1 of \cite{Gaiotto:2015aoa}. We also refer the reader to \cite{Kapranov:2014uwa} for an interpretation of the $L_{\infty}$-operations and $L_{\infty}$-relations using secondary polytopes. 

The main ingredients mentioned so far, namely vacua, central charges and ``Hilbert" spaces of BPS states are rather familiar from the study of BPS states in two-dimensional $\mathcal{N}=(2,2)$ quantum field theories. A less familiar, but nonetheless crucial, ingredient that enters the algebra of the infrared is a kind of interaction vertex of BPS particles. This ingredient is known as the interior amplitude. An interior amplitude $\beta \in R_c$ is an element carrying cohomological degree $+2$ that solves the $L_{\infty}$ Maurer-Cartan equation: 
\begin{equation}
    \lambda_1(\beta) + \frac{1}{2} \lambda_2(\beta, \beta) + \frac{1}{3!}\lambda_3(\beta, \beta, \beta) + \dots = 0.
\end{equation}

\begin{figure}
    \centering
    \includegraphics[width=\textwidth]{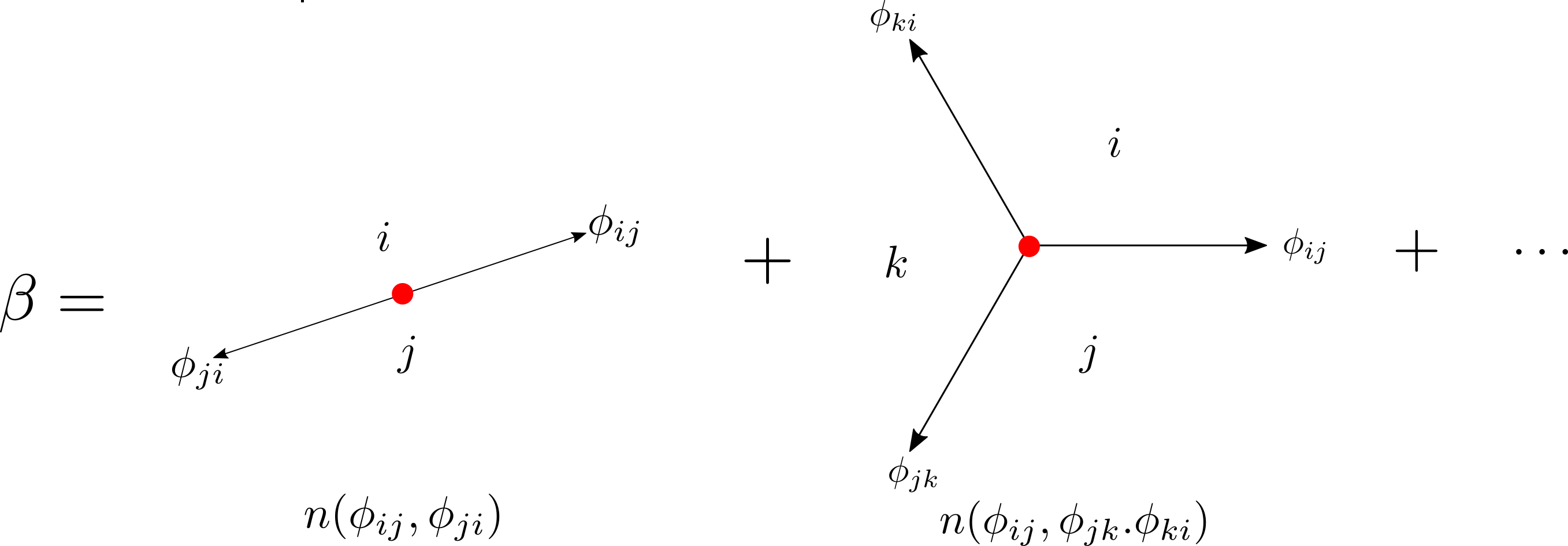}
    \caption{Schematic depiction of $\beta$. It decomposes as a sum of bivalent components, trivalent ones and so on.}
    \label{beta}
\end{figure}

One should think of the component of $\beta$ involving a cyclic fan of soltions  
\be 
\{\phi_{i_0 i_1}, \dots, \phi_{i_n i_0}\} 
\ee 
as an $(n+1)$-valent interaction vertex that couples the $(n+1)$ BPS particles involved in the fan. The Maurer-Cartan equation is then a the condition that such an interaction vertex is indeed supersymmetric. For a very useful analogy see Appendix \ref{app:BF-Analogy}. 
For a more detailed discussion of $\beta$ and its path integral interpretation we refer the reader to Section 14.6 in \cite{Gaiotto:2015aoa}. See also Figure \ref{beta} for a schematic depiction of $\beta$. The interior amplitude 
$\beta$ deforms  the noninteracting $L_\infty$ algebra following from \eqref{eq:t star t} to the interacting one. 

The \textit{infrared data} of a theory consists of the vacuum set $\mathbb{V}$, the vacuum weights $\{z_i\}_{i \in \mathbb{V}}$, the space of solitions along with their pairings $\{R_{ij}, K_{ij}\}_{i \neq j}$ and finally an interior amplitude $\beta \in R_{c}$.

\begin{figure}
    \centering
    \includegraphics[width=0.9\textwidth]{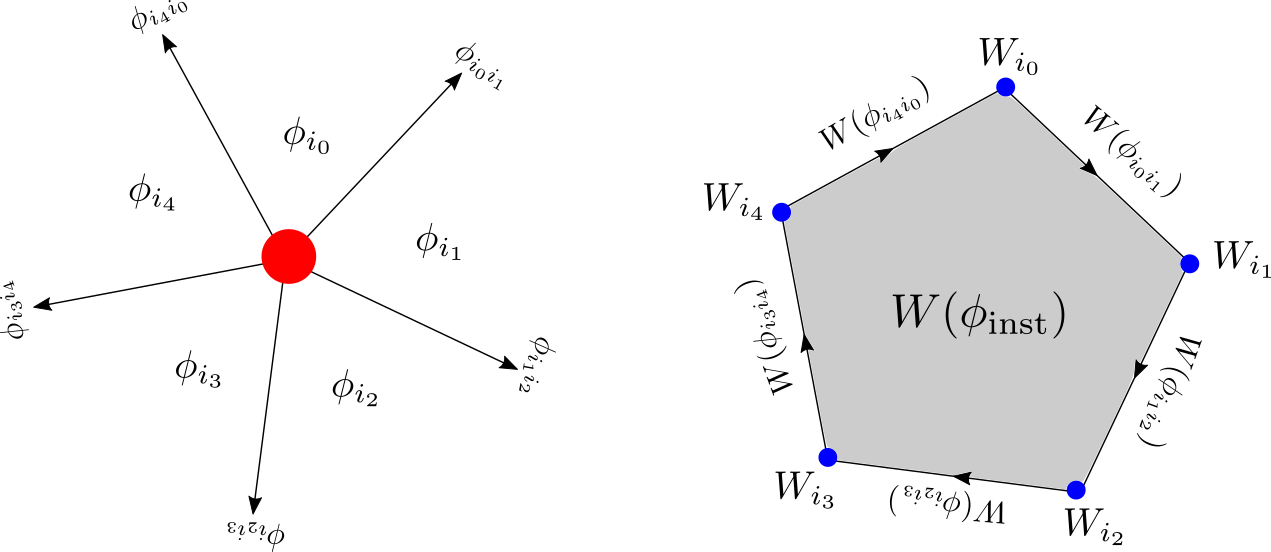}
    \caption{Depiction of fan boundary conditions and the image under $W$.}
    \label{fanbcs}
\end{figure}

In a Landau-Ginzburg model, the interior amplitude $\beta \in R_c$ is determined by counting solutions of the $\zeta$-instanton equation 
\begin{equation} \label{zetainst}
    \frac{\partial \phi^a}{ \partial \bar{z}} = \zeta g^{a\bar{b}} \frac{\partial \overline{W}}{\partial \overline{\phi}^{\bar{b}}}
\end{equation}
with fan boundary conditions, and no reduced moduli. Recall that by a fan boundary condition
\footnote{See also \cite{Gaiotto:2015aoa} Appendix E for a more extensive definition}
we mean a map $\phi: \mathbb{C} \rightarrow X$ obeying the $\zeta$-instanton equation \eqref{zetainst} with the following boundary conditions.
We begin by letting $\{i_0, \dots, i_n\}$ be a cyclic fan of vacua, and let 
\be 
\{\phi_{i_0 i_1}, \phi_{i_1 i_2}, \dots, \phi_{i_n i_0} \},
\ee
be a cyclic fan of solitons. We then require the following: We first recall that given a function of a complex variable $f(z,\bar{z})$ we may consider the limit\footnote{Being more pedantic, we're defining $f_{\theta}(w, \bar{w}) := f(e^{-i\theta} w, e^{i\theta} \bar{w})$ and then letting $w = x + i y$ and taking the limit $y \rightarrow \infty$, i.e $f_{\theta}(x) = \lim_{y \rightarrow \infty} f_{\theta}(w, \bar{w})$.} $$f_{\theta}(x) = \lim_{\text{Im}(e^{-i\theta} z) \rightarrow \infty} f(z,\bar{z})$$ assumed to exist, to obtain a function of a real variable $x$. We then impose the following boundary conditions. Letting $\theta_{ij} = \text{Arg}(\zeta_{ij})$, we suppose
\bea 
\lim_{ \text{Im} \big(\zeta_{k, k+1}^{-1} z \big) \to \infty } \phi(z, \bar{z}) = \phi_{i_k i_{k+1}}(x), \,\,\,\,\,\, \text{  for  } \,\,\,\,\,\,\, k =0, \dots, n \\  \lim_{\text{Im}(e^{-i \theta_k}z) \rightarrow \infty} \phi(z, \bar{z}) = \phi_k \,\,\,\,\,\,\,\,\,\,\,\,\,  \theta_{k-1,k }< \theta_{k} < \theta_{k,k+1}
\eea
The image of a $\zeta$-instanton satisfying fan boundary conditions  under the superpotential $W$ is conjectured to fill the interior of corresponding (convex) polygon. See Figure \ref{fanbcs} 
for a depiction of fan boundary conditions and their image in the $W$-plane. More details on the fan boundary conditions and why their signed counts satisfy the $L_{\infty}$ Maurer-Cartan equation is discussed at length in section 14 of \cite{Gaiotto:2015aoa}.  

In the next subsection \ref{subsec:ExampleQuarticLG} we detail the infrared data for a particularly simple Landau-Ginzburg model. It will have a nontrivial interior amplitude. 

\subsection{Example: Quartic Landau-Ginzburg Model}\label{subsec:ExampleQuarticLG}

\begin{figure}
    \centering
    \includegraphics[width=0.8\textwidth]{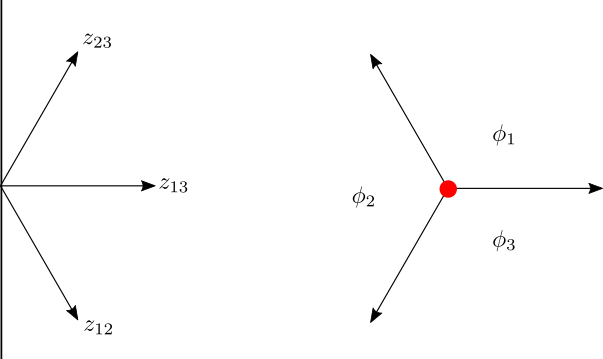}
    \caption{Left: Central charges in the right half-plane for the quartic, $\mathbb{Z}_3$-symmetric LG model with $\zeta = -\text{i}$. Right: Picture of basic interaction vertex of the model.}
    \label{quarticz}
\end{figure}

Consider the theory of a single chiral superfield $\phi,$ and superpotential
\begin{equation}
    W = \frac{1}{4}\phi^4 - \phi.
\end{equation} The model has vacua $\mathbb{V} = \{\phi_1, \phi_2, \phi_3\}$ where 
\be 
\phi_1 = e^{\frac{4\pi \text{i}}{3}}, \,\,\,\,\, \phi_2 = 1, \,\,\,\,\,\, \phi_3 = e^{\frac{2\pi i}{3}},
\ee 
and taking $\zeta = -\text{i}$, the corresponding vacuum weights are 
\be 
z_1 = -\frac{3 \text{i}}{4} e^{\frac{4\pi i}{3}}, \,\,\,\,\, z_2 = -\frac{3 \text{i}}{4}, \,\,\,\,\, z_3 = -\frac{3 \text{i}}{4} e^{\frac{2\pi i}{3}} .
\ee
The central charges are thus 
\begin{align}
z_{23} &= \frac{3}{4} \text{i}\Big(1-\text{exp}\big(\frac{2\pi \text{i}}{3} \big) \Big) = \frac{3\sqrt{3}}{8} + \frac{9}{8}\text{i}~ ,\\
z_{13} &= -\frac{3}{4} \text{i} \Big( \text{exp} \big(\frac{2\pi \text{i}}{3} \big) - \text{exp} \big(\frac{-2\pi \text{i}}{3} \big) \Big)
= \frac{3\sqrt{3}}{4} ~ ,\\ 
 z_{12} &= -\frac{3}{4} \text{i} \Big(1- \text{exp} \big(\frac{4\pi \text{i}}{3} \big) \Big)=  \frac{3\sqrt{3}}{8} - \frac{9}{8}\text{i} ~ .
\end{align}
and are depicted in Figure \ref{quarticz}.

\begin{figure}
    \centering
    \includegraphics[width=0.35\textwidth]{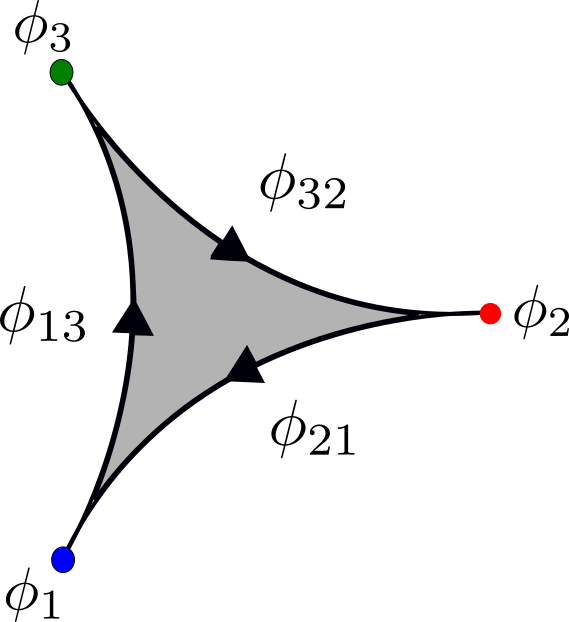}
    \caption{The three vacua, solitons and $\zeta$-instanton of the quartic Landau-Ginzburg model in the $\phi$-plane.}
    \label{quarticinst}
\end{figure}

Determining the spaces of solitons is also a simple matter. One can show that there is a single BPS soliton trajectory between any pair of distinct vacua. In order to determine the fermion degrees of these solitons we use the universal formula for the fractional part, along with the $\mathbb{Z}_3$-symmetry. We find that the classical $13$ soliton trajectory corresponds in the quantum theory to a doublet of BPS particles of charges 
\footnote{Using the freedom in the definition of the fermion number discussed in equation \eqref{eq:ShiftFermNumber} below we can shift away the integer part.}
\begin{equation}
(f_{13}, f_{13}+1) = \Big(-\frac{1}{3}, \frac{2}{3} \Big),
\end{equation} 
and by $\mathbb{Z}_3$-symmetry, these are also the fermion numbers for the $21$ and $32$ particles, 
\begin{equation}
f_{13} = f_{21} = f_{32}.
\end{equation}
In the formalism $R_{ij}$ is always defined to be the subspace corresponding to the particle with the bigger fermion number, so that we have 
\begin{equation} 
R_{13} = R_{21} = R_{32} = \mathbb{C}^{\big[\frac{2}{3} \big]}.
\end{equation}
The corresponding anti-solitons to these consist of the $31$, $12$ and $23$ solitons which therefore have fermion numbers 
\begin{align}
(f_{13}, f_{13} +1) &= \big(-\frac{2}{3}, \frac{1}{3} \big), \\ f_{13} &= f_{12} = f_{23}.
\end{align} 
and so the corresponding soliton spaces are
\begin{equation}
R_{21} = R_{31} = R_{32} = \mathbb{C}^{\big[\frac{1}{3} \big]}.
\end{equation}
The interior amplitude lives in the space $R_c$ which is also straight-forward to determine in this example. The cyclic fans are easy to enumerate: there are three zero-valent fans, with corresponding one-dimensional spaces in degree $0$, three bivalent fans each with corresponding one-dimensional space in degree $1$, for instance 
\begin{equation}
R_{12} \otimes R_{21} = \mathbb{C}^{\big[\frac{1}{3} \big]} \otimes \mathbb{C}^{\big[\frac{2}{3} \big]} \cong \mathbb{C}^{[1]},
\end{equation}
and a single trivalent cyclic fan given by $\{1,3,2\}$ with the corresponding space being one-dimensional in degree $2$: 
\begin{equation}
    R_{13} \otimes R_{32} \otimes R_{21} =  \Big(\mathbb{C}^{\big[\frac{2}{3} \big]} \Big)^{\otimes 3} \cong \mathbb{C}^{[2]}.
\end{equation}
Thus $R_c$ is a space spanned by three vectors in degree zero, three in degree $1$ and one and in degree $+2$. The Maurer-Cartan element can be taken to be
\begin{equation} \label{quarticamp}
    \beta = \phi_{13} \otimes \phi_{32} \otimes \phi_{21}.
\end{equation} 
This $\beta$ indeed trivially satisfies the Maurer-Cartan equation.
\footnote{The Maurer-Cartan equation produces an element in degree $+3$, whereas the space $R_c$ is only concentrated non-trivially in degrees $0$, $1$ and $2$.}
That this is the interior amplitude of the quartic Landau-Ginzburg theory would follow from the existence of a 
unique map $\phi: \mathbb{C} \rightarrow \mathbb{C}$ with trivalent fan boundary conditions that satisfies the $\zeta$-instanton equation. We claim that this is indeed the case, and the image of this map is as depicted in Figure \ref{quarticinst} (which already appeared in \cite{Khan:2020hir}). 
One can give numerical evidence for existence of such solutions. 
See figure 3 of \cite{Saffin:1999au}.
\footnote{For some simple superpotentials one can even exhibit exact solutions to the $\zeta$-instanton equation with fan boundary conditions \cite{Oda:1999az}. }

\begin{remark}\label{rmk:RiemannMapping} The existence of the $\zeta$-instanton above would follow immediately from an analogue of the Riemann mapping theorem for the $\zeta$-instanton equation, but such a statement is not available at present. In fact, we conjecture more generally that 
for ``nice'' superpotentials (which should certainly include polynomial superpotentials) of a single chiral field, there is an analog of the Riemann mapping theorem: With fan boundary conditions there is a solution of the $\zeta$-instanton equation mapping $\mathbb{C}$ in a 1-1 fashion to the interior of the region bounded by the ``polygon'' of boosted solitons. (This region in turn is mapped by $W$ to a polygonal region in the complex $W$-plane.) In contrast to the Riemann mapping theorem we do \underline{not} expect uniqueness (modulo obvious symmetries) of the solutions . See section 14.2 of \cite{Gaiotto:2015aoa} for an extended discussion of that point. Even when the target manifold is just the complex $\phi$-plane the problem is nontrivial and poses an interesting challenge in the theory of partial differential equations. Progress on this problem is being made by R. Mazzeo and M. Zimet.  \end{remark}

\subsection{Bulk Observables} Let us now briefly discuss how the infrared data allows us to construct the physical $L_{\infty}$-algebra of bulk observables of our $\mathcal{N}=(2,2)$ theory. Previously we noted the space $R_c$ has the structure of an $L_{\infty}$-algebra \eqref{eq:Bare-L-infinity}, which we called the noninteracting $L_\infty$ algebra. Given a   Maurer-Cartan element $\beta$ we can deform the multiplications $\lambda_n$ to define another   $L_{\infty}$-algebra   via the formula
\begin{equation}
    \lambda_n[\beta](a_1,\dots,a_n) = \sum_{k \geq 0} \frac{1}{k!} \lambda_{n+k}(a_1, \dots, a_n, \beta, \dots, \beta).
\end{equation} 
In terms of web diagrams, in the deformed $L_{\infty}$-algebra, in order to compute $\lambda_n[\beta]$, we consider all taut webs with 
\underline{at least} $n$ vertices, where the extra vertices are occupied by the interior amplitude $\beta$.
\footnote{This is analogous to an interaction vertex in a perturbative quantum field theory. This is our reason for calling the $L_{\infty}$-algebra with $\beta = 0$ the noninteracting $L_{\infty}$-algebra and the deformed one the interacting $L_{\infty}$-algebra.}
The space of bulk local operators associated to a theory with vacuum set $\mathbb{V}$, vacuum weights $\mathbb{W} := z( \mathbb{V})$ and spaces of BPS solitons $\{R_{ij} \}$ and interior amplitude $\beta$ is defined to be the $L_{\infty}$-algebra $R_c$ deformed by the Maurer-Cartan element $\beta$. In particular, the deformed $\lambda_1$ provides a differential \be d_{\beta}: R_c \rightarrow R_c \ee so that $(d_{\beta})^2 = 0,$ and the cohomology with respect to this differential are the on-shell local operators.

We leave it as a straightforward exercise to the reader to show that for the  quartic Landau-Ginzburg model the noninteracting $L_\infty$ algebra is $3+3+1=7$ dimensional with $3$-dimensional cohomology while the nontrivial interior amplitude $\beta$ deforms it so that the 
the cohomology of bulk local operators is one-dimensional and concentrated in degree $0$: 
\be H^*(R_c) \cong \mathbb{C},\ee generated by the operator corresponding to the "identity" observable.

\vspace{0.1pt}

\begin{remark} 
The ubiquitous example of Landau-Ginzburg models might mislead one into thinking that the web-based framework is only about the ``A-model with superpotential". However we stress that it applies to any massive $\mathcal{N}=(2,2)$ quantum field theory with an unbroken $U(1)_R$-symmetry. For instance the supersymmetric $\mathbb{CP}^{N-1}$ model has a dynamically generated mass gap, central charges and soliton sectors, so the formalism directly applies (without needing to use the well-known mirror Landau-Ginzburg model).
\end{remark} 

\begin{remark} \label{foot:E2} In the introduction we mentioned that although the bulk observables are expected to form an 
$E_2$ algebra the paper \cite{Gaiotto:2015aoa} only constructed an $L_\infty$ algebra as we have just described. 
An $E_2$-algebra can be viewed as an $L_{\infty}$-algebra with extra structure. 
For example, the cohomology of an $E_2$ algebra is a Gerstenhaber algebra (i.e. a shifted Poisson algebra), which has both a degree zero associative product and a degree $-1$ Lie bracket.
As far as we know, a construction of the full $E_2$-algebra structure on the space of bulk observables along the lines of \cite{Gaiotto:2015aoa}, has not appeared in the literature.   The $L_\infty$ structure constructed by webs gives the Gerstenhaber Lie bracket but does not give a construction of the Gerstenhaber product. Thus there is a gap to fill. 

One approach to defining the full $E_2$-structure, suggested to us by D.~Gaiotto, is as follows. Roughly speaking, an $E_2$-algebra assigns an operation $m_{C}$ to each chain $C$ in the configuration space of points on the plane (modulo the action of $\mathbb{R}_+ \ltimes \mathbb{R}^2$  ). There is a natural embedding of disjoint unions of web moduli spaces into the configuration space of points. The intersection of this embedding with a given chain $C$ is given by a disjoint union of subsets labelled by webs, and in order to define $m_C$ we simply sum over these webs, thus defining a possible collection of $E_2$-operations. Another approach to define the $E_2$-structure is to consider local observables along the identity interface and consider the $A_{\infty}$-products by looking at taut interface webs, along with webs that contribute to OPEs of the identity interface. This collection of operations would lead to a particular ``model" of an $E_2$-algebra. For either approach it is worthwhile to fill-in the details and obtain the full $E_2$-structure on bulk observables. However we must leave this for another occasion. 

\end{remark}

\subsection{Half-Plane Webs and $A_{\infty}$-Categories}
We now turn to a discussion of the $A_{\infty}$-category of boundary conditions. For this we have to make suitable modifications to the discussion of planar webs in order to accommodate half-planes. 

We begin by choosing a half-plane $H \subset \mathbb{R}^2$, so that the boundary of $H$ is not parallel to any of the central charges $\{z_{ij} \}.$ A half-plane web simply consists of a graph drawn on $H$ where again the faces are labeled by vacua, and the edges satisfy the slope constraint. The group acting on taut half-plane webs now consists of translations parallel to the boundary along with overall scalings through a choice of origin on the boundary.  A half-plane web will be called taut if the moduli space is rigid modulo the action of $\mathbb{R}_+ \ltimes \mathbb{R}$.

\begin{figure}
    \centering
    \includegraphics[width=0.85\textwidth]{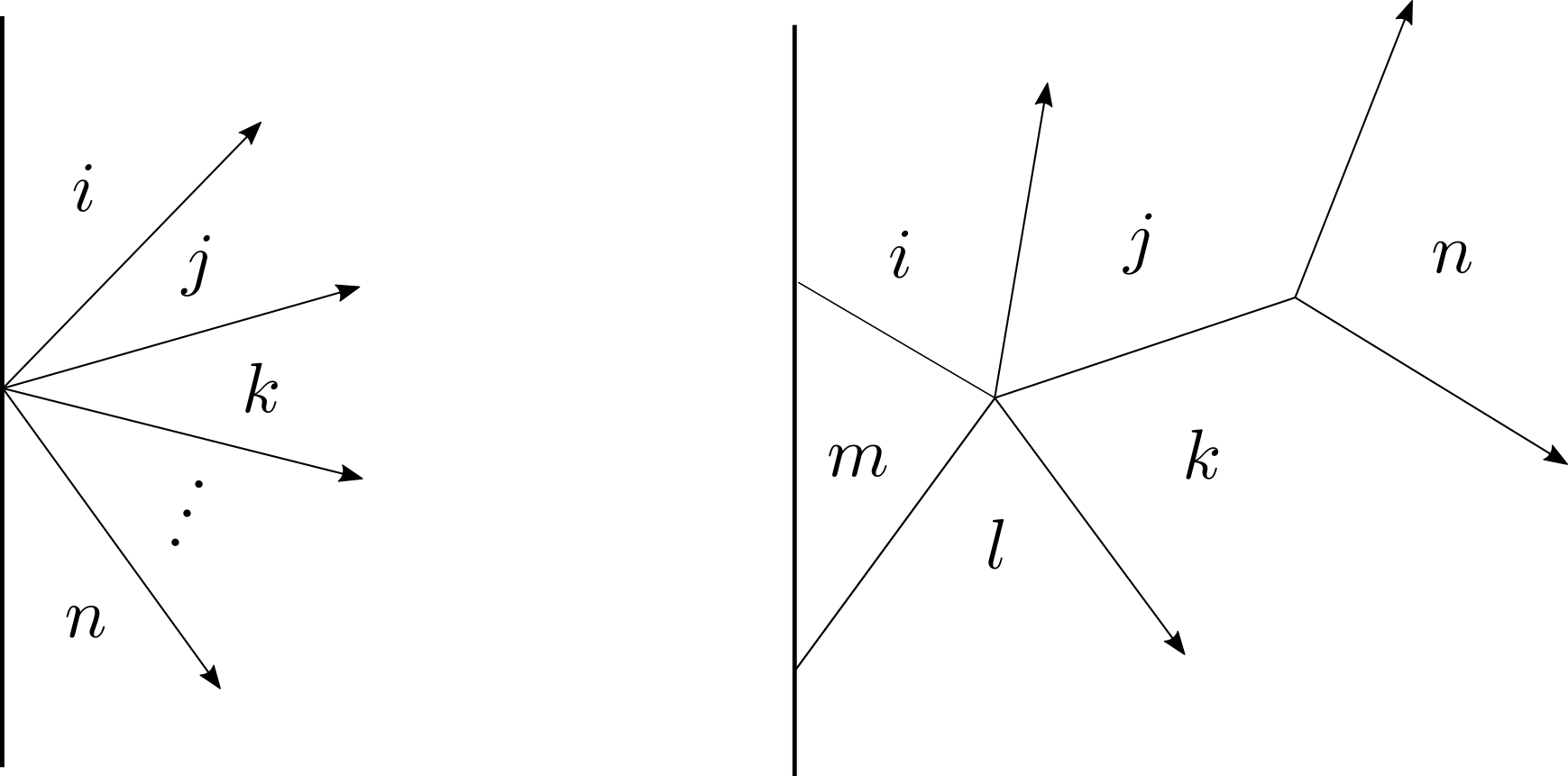}
    \caption{The left depicts a generic half-plane fan, whereas the right displays a half-plane web. Here $H$ is the right half-plane $x\geq 0$.}
    \label{hpweb}
\end{figure}

A vertex that lies along the boundary of a taut half-plane web determines a half-plane fan. A half-plane fan consists simply of an \textit{ordered} (as opposed to cyclically ordered) set of vacua 
\be 
\{i_0, i_1, \dots, i_n\}
\ee such that the corresponding central charges 
\be 
(z_{i_0 i_1}, z_{i_1 i_2}, \dots, z_{i_{n-1} i_n})
\ee 
are clockwise ordered complex numbers in the half-plane $H$. We denote the set of half-plane fans as $\mathrm{HFans}.$ Similar to before, we associate a vector space to a half-plane fan as follows: 
\begin{equation}
    R_{\{i_0, i_1, i_2, \dots, i_{n} \}} = R_{i_0 i_1} \otimes R_{i_1 i_2} \otimes \dots \otimes R_{i_{n-1} i_n}.
\end{equation}
The $A_{\infty}$   thimble category 
associated to the half-plane $H$ has a set of objects $\{\mathfrak{T}_i \}_{i \in \mathbb{V}}$ in one-to-one correspondence with the vacuum set $\mathbb{V}$. The space of morphisms
\begin{equation}
    \text{Hom}(\mathfrak{T}_i, \mathfrak{T}_j) =: \widehat{R}_{ji}
\end{equation} 
is defined by taking a direct sum of the spaces associated to all half-plane fans of the type $\{i, \dots, j\}$:
\begin{equation}
    \widehat{R}_{ij} = \bigoplus_{I = \{i, \dots, j\} \in \mathrm{HFan} } R_I.
\end{equation} 
The $k$th $A_{\infty}$-map 
\be
m_k: \widehat{R}_{i_0 i_1} \otimes \dots \otimes \widehat{R}_{i_{k-1} i_k} \rightarrow \widehat{R}_{i_0 i_k} 
\ee 
is obtained by summing over taut half-plane webs with $k$-boundary vertices, and where each bulk vertex is occupied by the Maurer-Cartan element $\beta.$ In our conventions, if the half-plane is the right half-plane then composition of morphisms should be read vertically upward.
As discussed in detail in \cite{Gaiotto:2015aoa}, one can show that these maps satisfy the $A_{\infty}$-associativity relations. The essential part of the argument is that if we consider the boundaries of the moduli space of sliding half-plane webs, we are lead to a relation of the form \be \mathfrak{t}_{H}* \mathfrak{t}_H + \mathfrak{t}_H* \mathfrak{t} = 0\ee where as before $\mathfrak{t}$ consists of the formal sum of taut planar webs, $\mathfrak{t}_H$ consists of the formal sum of taut half-plane webs, and $*$ denotes convolution of webs.
 
\begin{remark}
    $\widehat{R}_{ii}$ is taken to be one-dimensional in cohomological degree zero. The only half-plane fan of this type is just visualized to be the half-plane labelled by $i$ and a single vertex at the boundary. 
\end{remark}

\begin{remark}
    The objects of the thimble category carry a natural ordering: we say that 
    \be \mathfrak{T}_i< \mathfrak{T}_j  \,\, \text{ if } \,\, z_{ij} \in H .\ee We then have that if $\mathfrak{T}_i> \mathfrak{T}_j$ then $\widehat{R}_{ij} = \{0\},$ as there are no half-plane fans starting from $i$ and ending at $j$. The thimble category thus carries a semi-orthogonal decomposition. 
\end{remark}

\begin{remark}\label{Rmk:ZetaConv} We defined the Leftshetz thimble $R_i(\zeta)$ to be the thimble with gradient flow $Re(\zeta^{-1}W)$ that flows to 
$\phi_i$ as $x\to +\infty$. Unfortunately, in our conventions (following \cite{Gaiotto:2015aoa}) the supersymmetry preserved by the category with half plane $H$ and phase $\zeta$ has boundary conditions defined by gradient flow with $Re(-i \zeta^{-1}W)$. 
(See equation $(11.12)$ of \cite{Gaiotto:2015aoa}.)

\end{remark}

The web category built from thimbles actually allows us to describe a wider variety of other boundary conditions.
We can choose any Lagrangian submanifold of the target space as a boundary condition for the bosonic fields on the 
boundary of the half-plane subject to certain growth conditions. For example,  for the right half-plane and general $\zeta$ the allowed branes have $\text{Im}(\zeta^{-1}W) \rightarrow \infty$ as discussed in Section 12 of \cite{Gaiotto:2015aoa}.
These more general boundary conditions can be described using the notion of a \textit{twisted complex} of the thimble 
boundary conditions. A twisted complex consists of a multiplicity (or ``Chan-Paton'')  space $\mathcal{E}_i$ for each $i \in \mathbb{V}$ along with an element 
\be 
\mathcal{B} \in \widehat{R}(\mathcal{E}) := \bigoplus_{i, j \in \mathbb{V}} \mathcal{E}_i \otimes \widehat{R}_{ij} \otimes \mathcal{E}_j^{\vee}.
\ee 
of degree $+1$ that satisfies the $A_{\infty}$ Maurer-Cartan equation 
\be  \sum_{n  \geq 1} m_n(\mathcal{B}^{\otimes n}) = 0,
\ee
in the $A_{\infty}$-algebra\footnote{The $A_{\infty}$-structure on $\widehat{R}(\mathcal{E})$ is given simply by combining the $A_{\infty}$-structure on the thimble category with the evaluation map between the multiplicity spaces $\{\mathcal{E}_i\}$ and their duals.} $\widehat{R}(\mathcal{E}).$ 
For a Lagrangian brane $L$ as above, the spaces $\{\mathcal{E}_i(L) \}$ are determined by solving the $\text{i}\zeta$-soliton equation
\footnote{We use the terminology established in equation \eqref{eq:zeta-soliton} above, where $\zeta$ is the same $\zeta$ for which we are computing the thimble category}
for a map $\phi: \mathbb{R}_{+} \rightarrow X$ such that 
\be \lim_{x \to 0} \phi(x) \in L, \,\,\,\,\,\,\,\,\,  \lim_{x \to \infty} \phi(x) = \phi_i.\ee
The boundary amplitude $\mathcal{B}$ on the other hand is determined by solving the $\zeta$-instanton equation on $\mathbb{R}_+ \times \mathbb{R}$ with an appropriate generalization of fan boundary conditions in the presence of boundaries.

\subsubsection{Example: The Quartic Landau-Ginzburg Model}

We illustrate the ideas associated to the web construction of the $A_\infty$ category of boundary conditions in the specific case of the quartic LG model studied above. 

\begin{figure}
    \centering
    \includegraphics[width=0.25\textwidth]{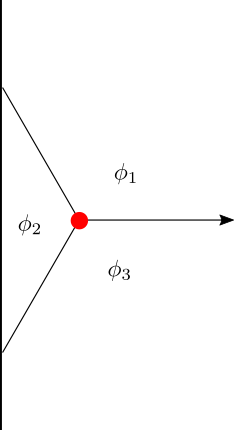}
    \caption{The taut half-plane web with the interior amplitude $\beta$ in \eqref{quarticamp} leads to the product expressed in \eqref{quarticproduct}. }
    \label{quarticweb}
\end{figure}

We take $H$ to be the right-half plane, and continue to take $\zeta = -i$. Since the central charges $\{z_{12}, z_{13}, z_{23}\}$ lie in the right-half plane, this orders the thimbles so that 
\be 
\mathfrak{T}_1< \mathfrak{T}_2< \mathfrak{T}_3.
\ee 
Each of the spaces $\widehat{R}_{12}, \widehat{R}_{13}, \widehat{R}_{23}$ has only a single half-plane fan that contributes to it, so that each of these spaces coincide with their unhatted counterparts, and are thus one-dimensional, with the following degrees
\begin{align}
    \widehat{R}_{12} &= \mathbb{C}^{[\frac{1}{3}]}, \\
    \widehat{R}_{23} &= \mathbb{C}^{[\frac{1}{3}]}, \\
    \widehat{R}_{13} &= \mathbb{C}^{[\frac{2}{3}]}.
\end{align}
In order to work out the $A_{\infty}$-maps, we must enumerate the taut half-plane webs. The only relevant taut half-plane web in this case simply consists of the one in Figure \ref{quarticweb}. This leads to a map 
\be 
m_2: \widehat{R}_{12} \otimes \widehat{R}_{23} \rightarrow \widehat{R}_{13}, 
\ee given by 
\be \label{quarticproduct} 
m_2(\phi_{12} \, ,  \,\phi_{23}) = \phi_{13}.
\ee

\begin{figure}
    \centering
    \includegraphics[width=\textwidth]{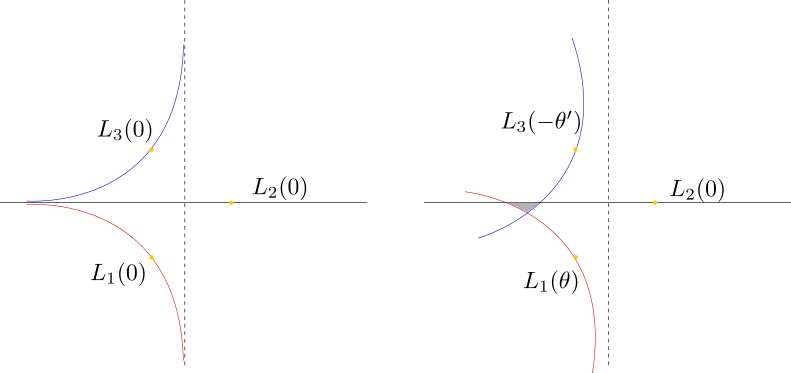}
    \caption{The left depicts the left thimbles at angle $\theta = 0$, in the $\phi$-plane, whereas the right depicts slightly rotated versions along with the image of the holomorphic triangle that leads to the non-trivial product in the Fukaya-Seidel category associated to the quartic superpotential. This figure, and similar figures below were created with 
    a drawing program and are not numerically precise. They are qualitative representations of the true situation. 
    }
    \label{quarticthimbles}
\end{figure}

It is instructive to verify that the category we just determined matches the one computed via methods more familiar to symplectic geometers. In the standard Fukaya-Seidel category we take objects to be slightly deformed thimble Lagrangians of the target space, morphism spaces are generated by intersection points, and the $A_{\infty}$-maps are determined by counting solutions of the holomorphic map equation. For this quartic Landau-Ginzburg model, the relevant thimbles are those defined by gradient flow of ${\rm Re}(W)$ (see Remark \ref{Rmk:ZetaConv} above.)  
These have deformation classes in   the first homology of $\mathbb{C}$ relative to the region where $\text{Re}(W) \rightarrow \infty,$ a region which consists of a union of four angular sectors: 
\be 
-\frac{\pi}{8} + \frac{\pi}{2} k <\theta < \frac{\pi}{8} + \frac{\pi}{2} k, \,\,\,\, k = 0,1,2,3.
\ee 
The thimble $L_2(0)$ of $\phi_2 = 1$ consists of the $x$-axis, being the part of the set of $\phi \in \mathbb{C}$ with 
\be 
\text{Im} \Big( \,\frac{1}{4} \phi^4 - \phi \Big) = 0,
\ee
which lies in the regions above. The thimble $L_1(0)$ of $\phi_1 = e^{-2 \pi i/3}$ consists a curve that asymptotes to the negative part of the real axis, and the negative part of the imaginary axis, and goes through the vacuum $\phi_1$, and the thimble $L_3(0)$ is obtained by reflecting $L_1(0)$ across the $x$-axis. These thimbles can be deformed slighlty so that each pairwise intersection consists of a unique point (and all three points are distinct points of the complex plane). This tells us that each morphism space is one-dimensional. Moreover, there is a non-trivial holomorphic triangle with Lagrangian boundary conditions, which gives us a non-trivial composition in this category. See Figure \ref{quarticthimbles} for a Figure of these thimbles, their intersection points and the image of the holomorphic triangle.

In either case we find that the algebra associated to the thimble category  coincides with the path algebra of the $A_3$ quiver (in order to get precisely this we must degree shift certain objects in the vacuum web category).

\begin{figure}
    \centering
    \includegraphics[width=\textwidth]{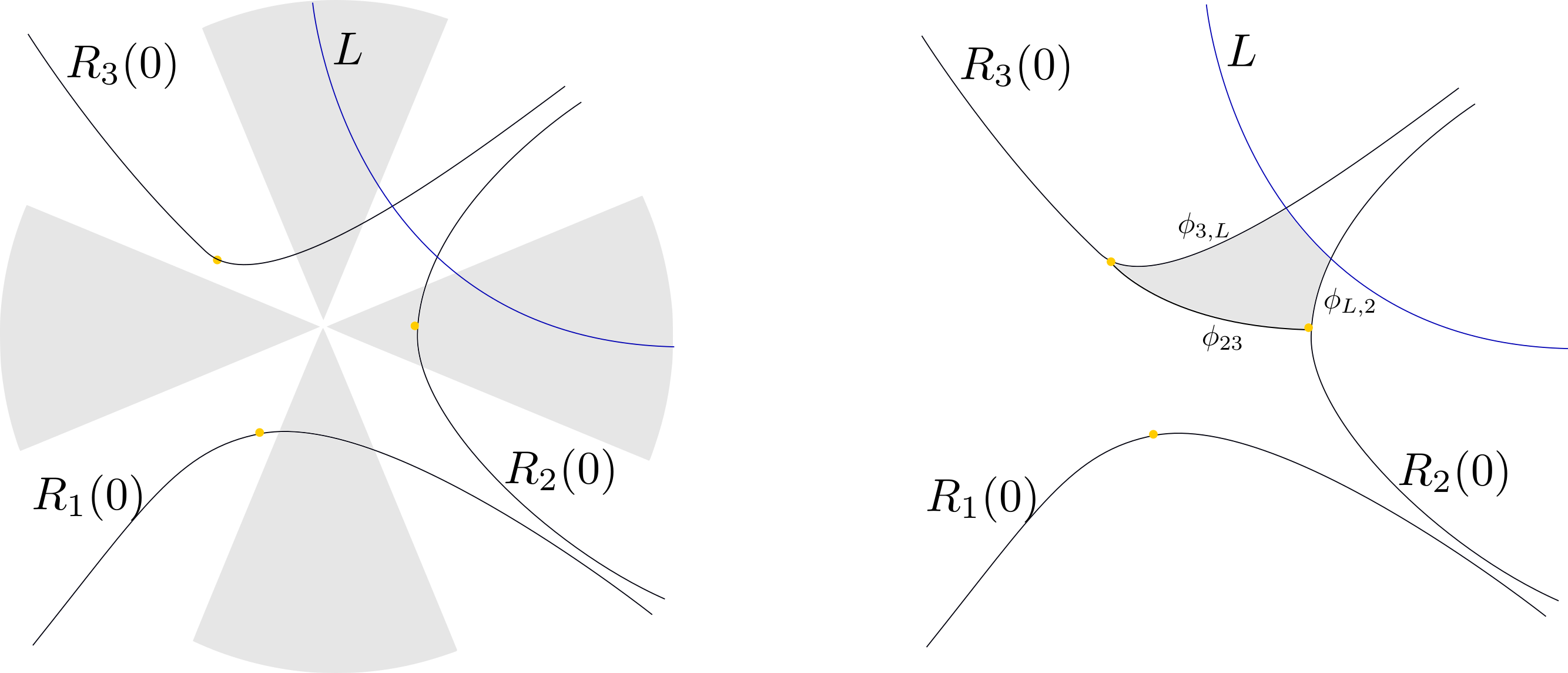}
    \caption{The left depicts the ``good" regions (shaded) where the support of a supersymmetric boundary condition for the quartic Landau-Ginzburg model can go off to infinity, and a particular Lagrangian $L$ which is not one of the left thimbles. It also depicts the right thimbles $R_i(0)$ for $i=1,2,3$, which go off to infinity in the complement of the shaded regions. $L$ intersects non-trivially with $R_2(0)$ and $R_3(0)$, with the intersection points corresponding to half-solitons that generate the one-dimensional spaces $\mathcal{E}_2(L)$ and $\mathcal{E}_3(L)$. The right of the image depicts the paths of the two half-solitons $\phi_{L,2}$ and $\phi_{3,L}$ along with the soliton $\phi_{23}$. The shaded region depicts the image of the $\zeta$-instanton on $\mathbb{R}_+ \times \mathbb{R}$ with half-fan boundary conditions that leads to a non-trivial boundary amplitude $\mathcal{B}$ for $L$.}
    \label{twistedcomplex}
\end{figure}

As mentioned above, one can choose boundary conditions determined by Lagrangian subspaces of $X$ that are not thimbles. 
In the quartic Landau-Ginzburg model, for instance a Lagrangian that asymptotes to the positive real and imaginary axis is one such example. See   Figure \ref{twistedcomplex}. This will be described by a twisted complex. 
In order to determine the spaces $\{\mathcal{E}_i(L) \}_{i=1,2,3}$, namely the complexes of half-solitons with $(L, i)$ boundary conditions, we can determine the intersection of $L$ with the \textit{right} thimbles at $\zeta=1$.
These thimbles are straightforward to work out and are depicted in Figure \ref{twistedcomplex}. From this we see that the multiplicity spaces are non-trivial (and one-dimensional) only for $\phi_2$ and $\phi_3$, 
\be 
\mathcal{E}_1(L) \cong \{0\}, \,\,\,\,\, \mathcal{E}_2(L) \cong \mathbb{C}^{[f_{2,L}]}, \,\,\,\,\, \mathcal{E}_3(L) \cong \mathbb{C}^{[f_{3,L}]} 
\ee for some degrees $f_{2,L}$ and $f_{3,L}.$ Moreover, we claim there is a non-trivial boundary $\zeta$-instanton which leads to a boundary element 
\be
\mathcal{B} \in \mathcal{E}_2(L) \otimes \widehat{R}_{23} \otimes \mathcal{E}_3^{\vee}(L) 
\ee 
given by a generator, which trivially satisfies the Maurer-Cartan equation. In particular we must have 
\be 
f_{2,L} -f_{3,L} + \frac{1}{3} = 1 
\ee
for this to be the case. See Figure \ref{twistedcomplex}
for the image of the $\zeta$-instanton in the $\phi$-plane. Having finished specifying the boundary condition with support $L$ as a twisted complex, one can then go ahead and work out morphism spaces, and show for instance that $\text{Hom}(L,L)$ has a one-dimensional cohomology concentrated in degree zero, 
\be 
H^*\big(\text{Hom}(L,L) \big) \cong \mathbb{C}.
\ee
The full category of boundary conditions is that of the representation category of the $A_3$ quiver. 

We thus see in this example that web categories allow us to give a self-contained construction of the category of A-type boundary conditions for Landau-Ginzburg models.

\section{General Discussion of Twisted Masses} \label{twistedmasses}

In the previous section we have seen how the web-based framework allows us to describe the space of bulk local operators along with the category of boundary conditions for massive Landau-Ginzburg models. The formalism relied rather strongly on having a finite number of central charges, and on their additivity property 
\begin{equation}
z_{ij} + z_{jk} = z_{ik}.
\end{equation}
Indeed the formalism from the very start expresses the central charge $z_{ij}$ as a difference of vacuum weights $z_{ij} = z_j - z_i$ so that not only does the additivity property hold, but we also have the property  that 
\begin{equation}
z_{ii} = 0.
\end{equation}
In particular there are no non-trivial BPS states in the $ii$-sector. One might wonder if it is a general propery of  $\mathcal{N}=(2,2)$ theories that there are no BPS states in the $ii$ sector. In fact, this is far from the case, as we will explain in this section. We can already see that it is not the case using  classical analysis of   $\mathcal{N}=(2,2)$ field theories defined by a sigma model with a K\"{a}hler target manifold $X$. Classically, one may make the sigma model massive by writing down potentials that preserve supersymmetry. There are two kinds of potentials one can add. 

\textbf{Closed One-Forms.} As we discussed previously, one kind of potential we can add to a supersymmetric sigma model is associated to a holomorphic function 
\begin{equation}
W: X \rightarrow \mathbb{C}.
\end{equation}
The supersymmetric Lagrangian defined by $W$ includes a potential energy for the bosonic fields of the form
\begin{equation} 
L_W = g^{a\bar{b}} \frac{\partial W}{ \partial \phi^a} \frac{\partial \overline{W}}{ \partial \overline{\phi}^{\bar{b}}} ~ . 
\end{equation}
    Here we note a central observation: $W$ enters the Lagrangian (and the supersymmetry transformations) only through its first (and second) derivative. This suggests that one can make the following generalization. Let $\alpha$ be a holomorphic one-form on $X$ and consider adding
\begin{equation}
L_{\alpha} = g^{a\bar{b}} \alpha_a \bar{\alpha}_{\bar{b}}
\end{equation} to the bosonic part of the sigma model Lagrangian. Then one can show that there is an appropriate supersymmetric completion of this action so that the model has $\mathcal{N}=(2,2)$ supersymmetry provided $\alpha$ is \textit{closed}  
\begin{equation}
\text{d} \alpha = 0.
\end{equation}  To get a feeling for the $d$-closure condition of $\alpha$, we can consider the operators that arise upon quantization of the one-dimensional system obtained from dimensionally reducing the $\mathcal{N}=(2,2)$ theory\footnote{Later we will discuss the generalization of these operators that arise in the two-dimensional theory.}. Given a holomorphic function $W$ one may deform the Dolbeault operator $\overline{\partial}$ to \begin{equation}
\overline{\partial}_W := \overline{\partial} + \partial W \wedge \end{equation} which is still nilpotent. More generally the operator 
\begin{equation}
\overline{\partial} + \alpha \wedge 
\end{equation}
for a holomorphic one-form $\alpha$ is nilpotent provided $\alpha$ is closed. The previous case of a global superpotential is recovered when $\alpha$ is exact $\alpha = \text{d}W$ for a holomorphic function $W$, but in general we need not assume $\alpha$ is exact. In summary, one kind of potential consistent with supersymmetry simply corresponds to the geometric data of a closed holomorphic one-form on $X$.

\textbf{Isometries.} Second let $A$ be the maximal abelian subgroup of the group of continuous isometries of $X$, and suppose that $A$ is rank $r$ so that it is generated by a set of $r$ holomorphic vector fields $V^1, \dots, V^r$ which are mutually commuting 
\begin{equation}
[V^i, V^j] = 0.
\end{equation}
Pick a collection of $r$ complex numbers \begin{equation} \vec{m} := (m_1,\dots, m_r).\end{equation} We can then add \textit{twisted masses} for the group $A$, which corresponds to adding the potential 
\begin{equation} 
L_{\vec{m}} = | \sum_{i=1}^r m_i V^i|^2 
\end{equation} 
to the bosonic part of the supersymmetric sigma model Lagrangian and making an appropriate supersymmetric completion.

In order to get a feeling for this kind of deformation, it is again useful to discuss the operators that act on the Hilbert space of the theory in one dimension lower. In that context the twisted mass deformation deforms the Dolbeault operator $\overline{\partial}$ to 
\begin{equation}
\overline{\partial}_{\vec{m}} := \overline{\partial} + \sum_{i=1}^r m_i \,\iota_{ V^i}
\end{equation} where $\iota_V$ refers to the operation that contracts a differential form with the holomorphic vector field $V$.
This is still nilpotent\footnote{Note that in real equivariant cohomology the operator $d + u\, \iota_X$ squares to a Lie derivative $u \,\mathcal{L}_X$ acting on forms. In the holomorphic context however, the operator $\overline{\partial}_{\vec{m}}$ is nilpotent on the nose, since $V$ is a holomorphic vector field. The Lie derivative $\mathcal{L}_V$ on the other hand arises as the anti-commutator of $\overline{\partial}_{\vec{m}}$ with 
\begin{equation} 
\partial_{\vec{m}} := \partial + \sum_{i=1}^r \overline{m}_i \iota_{\overline{V}^i}.
\end{equation}} provided $\{V^i \}$ are holomorphic vector fields which mutually commute.

We remark that adding this term to the potential corresponds to adding the derivative square of the moment map evaluated on $(m_1, \dots, m_N)$ i.e if 
\be
h_{\vec m} = \sum_{i=1}^N m_i h^i 
\ee 
where $h^i$ is the moment map satisfying \begin{equation} \frac{\partial h^i}{\partial \phi^I} = (V^i)^J\omega_{IJ},\end{equation} the bosonic potential we add can be rewritten as 
\begin{equation}
L_{\vec m }  = g^{IJ} \frac{\partial h_{\vec m}}{\partial \phi^I} \left( \frac{\partial h_{\vec m}}{\partial \phi^J} \right)^*  ~.
\end{equation} 
Here $(\phi^I)_{I=1, \dots, \text{dim}_{\mathbb{R}}X}$ denote a set of real coordinates on $X$.

We can add both kinds of potentials and still obtain a $\mathcal{N}=(2,2)$ supersymmetric field theory provided $\alpha$ is invariant under the maximal abelian subgroup of isometries, so that 
\begin{equation}
\mathcal{L}_{V^i} \alpha = 0
\end{equation}
for $i=1, \dots,r.$ 

Thus classically, the most general $\mathcal{N}=(2,2)$ sigma model is specified by the data of
\begin{itemize}
    \item A K\"{a}hler manifold $X$
    \item A set of complex numbers \footnote{More invariantly $\vec{m} \in \mathfrak{a}$, the Lie algera of $A$.} $\vec{m} := (m_1, \dots, m_r)$ where $r$ is the rank of the maximal abelian subgroup $A$ of isometries of $X$. 
    \item An $A$-invariant closed holomorphic one-form $\alpha$ on $X$.
\end{itemize}

The classical vacua of the theory specified by this data corresponds to the intersection points of the zeros of $\alpha$ and the fixed points of the $A$-action on $X$. We assume the vacuum set is compact
and moreover each fixed point is isolated, so that there is a finite number of isolated vacua. In order to obtain a massive theory we assume that the derivative of $\alpha$ and $V$ is non-degenerate at each fixed point. 

A simple example of a theory where we both have a non-zero  one-form $\alpha$ and a non-zero twisted mass associated with an isometry consists of taking $X = \mathbb{C}^2$ with its standard K\"ahler metric and considering the $U(1)$-isometry that  rotates 
\begin{equation}
\varphi_{\xi} : (\phi_1, \phi_2) \rightarrow (e^{i \xi} \phi_1, e^{-i \xi} \phi_2).
\end{equation} 
for $\xi \in \IR$. 
We can turn on a twisted mass $m$ corresponding to this isometry. Moreover, we can consider the $U(1)$-invariant superpotential 
\begin{equation}
W = \phi_1 \phi_2.
\end{equation}
The theory has a single isolated massive vacuum located at 
\begin{equation}
\phi_1 = \phi_2 = 0.
\end{equation}

\begin{remark}
    The above example actually generalizes to a much broader situation. A natural place where both kinds of potentials arise is in the context of hyperK\"{a}hler geometry. Here given a continuous hyperK\"{a}hler isometry on a hyperK\"{a}hler manifold $X$, and fixing a complex structure $I$ on $X$, we can write down a tri-holomorphic (in particular $I$-holomorphic) vector field $V$ that generates the isometry and turn on the corresponding twisted mass. Moreover, letting \begin{equation} \Omega = \omega_{J} + \text{i} \omega_K \end{equation} be the $I$-holomorphic symplectic form on $X$ we may add a potential corresponding to the closed $I$-holomorphic one-form \begin{equation}
        \alpha = \iota_V \Omega.
    \end{equation} This corresponds to taking the superpotential (when its defined) to be the complex moment map 
    \begin{equation}
    W = \mu_c, \,\,\,\,\,\, d\mu_c = \iota_V \Omega
    \end{equation}
    In such a situation we actually preserve $\mathcal{N}=(4,4)$ supersymmetry\footnote{There is also a natural $U(1)$ R-symmetry. The R-symmetry group of $\mathcal{N}=(4,4)$ supersymmetry in $d=2$ is $SU(2) \times SU(2)$ but turning on this kind of potential breaks it to a $U(1)$ subgroup.}. 
\end{remark}

Having specified the data to which we associate an $\mathcal{N}=(2,2)$ sigma model with potential, let us now study whether the conditions for the web formalism are satisfied. 

The first problem one encounters in a general sigma model is that when both kinds of potential deformations are present, there is no natural $U(1)_R$-symmetry. To see this, recall that a classical $\mathcal{N}=(2,2)$ supersymmetric sigma model without any potential terms has both $U(1)_V$ and $U(1)_A$ R-symmetries. Turning on a one-form $\alpha$ (generically) breaks $U(1)_V$ explicitly, whereas turning on a twisted mass breaks $U(1)_A$ explicitly as we will now explain.

It is instructive to see how the explicit breaking of the R-symmetries manifests itself in the dimensional reduction of the theory to supersymmetric quantum mechanics. Recall that in supersymmetric quantum mechanics the Hilbert space is the space of (complexified) square integrable differential forms on $X$ where the two $U(1)$ symmetry operators have eigenvalues 
\begin{align}
F_A &= q+p, \\ F_V &= q-p,
\end{align}
when acting on the space of (square-integrable) $(p,q)$-differential forms 
\be \Omega^{(p,q)}_{L^2}(X) \subset \Omega^*_{L^2}(X) \ee where as usual $p$ ($q$) denotes the number of holomorphic (anti-holomorphic) indices carried by the form. In this language, the fact that  the one-form $\alpha$ breaks the $U(1)_V$-symmetry corresponds to the fact that the twisted Dolbeault operator 
\begin{equation}
\overline{\partial} + \alpha \wedge: \,\, \Omega^{(p,q)} \rightarrow \Omega^{(p,q+1)} \oplus \Omega^{(p+1,q)}
\end{equation} 
has a definite $F_A$-degree being $+1$ but no definite $F_V$-degree. The operator \begin{equation}
\overline{\partial} + m\,\iota_V : \Omega^{(p,q)} \rightarrow \Omega^{(p,q+1)} \oplus \Omega^{(p-1,q)}
\end{equation}
on the other hand works in the opposite way: it has a definite $F_V$ degree of $+1$ but no definite $F_A$ degree. Turning on both kinds of deformations leads to a differential of the form 
\be \ov{\del} + m \,\iota_V + \alpha \wedge\ee which gives a cohomology theory with no natural integral grading. 

Though it is interesting to try and work at this level of generality, in this paper we will require our formalism to have definite integral degrees. We thus see that for that to be the case we must either have either $\alpha = 0$ or $\vec{m} = 0$.

Let us now examine the nature of the central charges in these two cases. 

First if $\vec{m} =0$ (so we could work with the A-twisted model with $Q_{\text{A}}^2=0$) 
the non-vanishing central charge comes from the square of the B-type supercharge, which by definition is given by 
\begin{equation}
Q_{\text{B}} = \overline{Q}_+ + \overline{Q}_-,
\end{equation}
so that 
\begin{equation}
Z = Q_{\text{B}}^2 = \{\overline{Q}_+, \overline{Q}_-\}. 
\end{equation}
The B-type supercharge in turn is identified as a deformation of the Dolbeault operator acting on the space of differential forms on the mapping space 
\begin{equation}
\mathfrak{X} =\text{Map}(\mathbb{R},X),
\end{equation}
viewed as a complex manifold with the natural complex structure induced from the complex structure on $X$. The forms are required to be square integrable with respect to the norm defined by the K\"{a}hler metric on $\mathfrak{X}$ induced from $X$. The deformation involves adding both the operation of taking an interior product of a form with a vector field and adding the operation of wedging with a holomorphic one-form, so that the B-type operator takes the form 
\begin{equation}
Q_{\text{B}} = \overline{\partial} + \iota_u + \widehat{\alpha} \wedge.
\end{equation} Here $u$ comes from the natural $\mathbb{R}$-action on $\mathfrak{X}$ which simply takes 
\begin{equation}
T_a: \phi(x) \rightarrow \phi(x+a)
\end{equation}
for $a \in \mathbb{R}$. The vector field that generates this action is thus given by   
\begin{equation}
u = \int \text{d}x \, \frac{\partial \phi^a}{ \partial x} \frac{\delta }{\delta \phi^a}
\end{equation}
Similarly, $\widehat{\alpha}$ is the one-form on $\mathfrak{X}$ obtained from pulling back $\alpha$ to $\mathfrak{X} \times \mathbb{R}$ via the evaluation map 
\begin{equation}
\text{ev}: \mathfrak{X} \times \mathbb{R} \rightarrow X
\end{equation}
and integrating along the $\mathbb{R}$-direction:
\begin{equation}
\widehat{\alpha} = \int_{\mathbb{R}} \text{ev}^*(\alpha) = \int_{\mathbb{R}} \text{d}x \, \alpha_a \delta \phi^a.
\end{equation}
The closure of $\widehat{\alpha}$ simply follows from the closure of $\alpha$. The square of an operator of the form 
\begin{equation}
\widetilde{\overline{\partial}} = \overline{\partial} + \iota_K + \xi \wedge
\end{equation}
for a holomorphic vector field $K$ and a closed holomorphic one-form $\xi$ is given by the degree zero operator which multiplies a form by the function $\iota_K \xi$: \begin{equation}
\widetilde{\overline{\partial}}^2 = \iota_K \xi.
\end{equation}
Specializing to $\mathfrak{X}$ with the above $\xi$ and $K$, we find that the central charge is thus given by 
\begin{equation}
Z = \iota_u \widehat{\alpha} = \int \text{d}x \frac{\text{d}\phi^a}{\text{d}x} \alpha_a.
\end{equation}
If $\alpha = \text{d}W$ and we choose critical points $i$ and $j$ as the boundary conditions, we find 
\begin{equation}
Z_{ij} = W(\phi_j) - W(\phi_i)
\end{equation}
precisely the formula for the central charge we had in the abstract web formalism discussed previously. However, if $\alpha$ is a one-form with a non-trivial cohomology class, the conversion to a boundary integral is not possible. Thus if $b_1(X) \neq 0$, the condition that $z_{ii} = 0$ will not hold in general. We will say that a Landau-Ginzburg model has twisted masses if the one-form $\alpha$ has a non-trivial periods.  

In the other situation if $\alpha =0$, (so we could work with the B-twisted model with $Q_{\text{B}}^2=0$) 
the non-vanishing central charge (sometimes called the \textit{twisted} central charge) is identified as the square of the A-type supercharge 
\begin{equation}
Q_{\text{A}} = \overline{Q}_+ + Q_-
\end{equation}
so that 
\begin{equation}
\widetilde{Z} = Q_A^2 =  \{ \overline{Q}_+, Q_-\}.
\end{equation}
The A-type supercharge can in turn be identified as a deformation of the deRham operator acting again on the space of differential forms on $\mathfrak{X}$, now viewed as a Riemannian manifold with the metric induced from the (K\"{a}hler) metric on $X$. The deformed operator takes the form \begin{equation}
Q_{\text{A}} = \text{d} + \widehat{\omega}\wedge + \iota_{\widehat{V}}  
\end{equation}
where $\widehat{\omega}$ is the one-form on $\mathfrak{X}$ given by 
\begin{equation}
\widehat{\omega} = \int \text{d}x \,\omega_{AB} \frac{\text{d} \phi^A}{ \text{d} x} \delta \phi^B 
\end{equation}
where $\omega$ is the symplectic form on $X$ and $\widehat{V}$ is the vector field on $\mathfrak{X}$ given by 
\begin{equation}
\widehat{V} = \int \text{d}x \,V^A \frac{\delta}{ \delta \phi^A}.
\end{equation} The form $\widehat{\omega}$ is closed because $\omega$ is a closed two-form and $\widehat{V}$ generates an isometry of $\mathfrak{X}$ precisely because $V$ generates an isometry of $X$. 
The square of the supercharge of a deformed deRham operator of the type 
\begin{equation}
\widetilde{d} = \text{d} + \iota_T + \eta \wedge
\end{equation}
for a vector field $T$ and a closed one-form $\eta$ on a space $M$ is given by the degree zero operator 
\begin{equation}
\widetilde{d}^2 = \mathcal{L}_T + \iota_T \eta.
\end{equation}
For the vector field and one-form above then, we have 
\begin{equation}
\iota_{\widehat{V}} \widehat{\omega} = \int \text{d}x \, \omega_{AB} V^B\frac{\text{d} \phi^A}{ \text{d}x} = \int \phi^*(\text{d}h) 
\end{equation}
where $h$ is the moment map, and $\mathcal{L}_{\widehat{V}}$ just consists of the operator $q$ obtained from the $U(1)$ flavor symmetry. Thus the central charge acting on the space of differential forms on $\mathfrak{X}$ takes the form
\begin{equation}
\widetilde{Z} = mq + \text{i}m \int dh 
\end{equation}
Putting $ij$ boundary conditions we find that 
\begin{equation}
\widetilde{Z} = \text{i}m \big( h(\phi_j) - h(\phi_i) \big) + mq.
\end{equation} We also see at this stage that the global form of the Lie group $A$ associated to the Lie algebra $\mathfrak{a}$ becomes important: for the charge $q$ to have quantized eigenvalues, $A$ must be compact. In any case, once again we find that with a non-zero twisted mass $\widetilde{Z}_{ii} \neq 0.$

In conclusion the conditions of the web formalism can hold for the example of sigma models only when $\vec{m} = 0$ and $\alpha = \text{d}W$. At a more general point in moduli space they cease to hold.

\begin{remark}
    The expression for the classical central charge $Z$ for potentials coming from holomorphic one-forms is in fact quantum exact. Indeed the B-type supercharge whose square is $Z$ can be written entirely in terms of holomorphic quantities on the target space $X$, and such quantities are protected from corrections under the renormalization group flow \cite{Seiberg:1994bp}. The central charge coming from twisted masses on the other hand is well-known to undergo non-trivial quantum corrections, both from perturbative and instanton effects \cite{Dorey:1998yh, Losev:2003gs}. Indeed the A-type supercharge whose square is $\widetilde{Z}$ depends explicitly on the K\"{a}hler form, which is well-known to undergo quantum corrections. The general form of the central charge as a sum of a boundary term and a conserved charge still persists however.
\end{remark}

From now on we focus on the case of a one-form with non-trivial periods. The discussion of non-trivial twisted masses coming from target space isometries will take a similar structural form (as the two deformations are in fact mirror duals to each other \cite{Hori:2000kt}).

Suppose then that we are studying a supersymmetric sigma model with target space $X$ and a potential determined by a closed holomorphic one-form $\alpha$ on $X$. We have seen that for such theories the central charge cannot be expressed as the difference of a set of well-defined vacuum weights. Let us explore the physical consequences of this, and also explore whether there is some workaround to this.

First we observe the fact that the non-vanishing of the central charge in a sector of field space where the field $\phi$ approaches the same vacuum $\phi_i$ at both ends of space can lead to non-trivial BPS states. With such boundary conditions the field $\phi$ will trace out a trajectory in $X$ whose closure is a loop based at $\phi_i$. Letting $\gamma$ be the homology class of this loop, the central charge of such a field configuration is given by 
\begin{equation}
Z(\gamma) = \int_{\gamma} \alpha.
\end{equation} 
The central charge thus restricted to the $ii$-sector in particular defines a homomorphism 
\begin{equation}
Z: H_1(X, \mathbb{Z}) \rightarrow \mathbb{C}.
\end{equation}
The different sectors in the $ii$ field space are thus labeled by the quotient group 
\begin{equation}
\Gamma := H_1(X, \mathbb{Z})/ \text{ker}Z.
\end{equation}
In particular the torsion subgroup of $H_1(X, \mathbb{Z})$ is contained in $\text{ker} \, Z$. Therefore $\Gamma$ is a finite rank free abelian group 
\begin{equation}
\Gamma \cong \mathbb{Z}^{r},
\end{equation} for some non-negative integer $r$ (which in fact is assumed to be positive). Letting $\gamma$ therefore be a class for which the central charge is non-vanishing the BPS bound says that the lowest energy state in this sector of field space satisfies
\begin{equation}
E(\gamma) = \big| \int_{\gamma} \alpha \, \, \big|.
\end{equation}
The condition for a classical field configuration to have this energy is equivalent to it solving the flow equation for $\alpha$:
\begin{equation}
    \frac{\text{d} \phi^a}{ \text{d} x} = \zeta(\gamma) \, g^{a\overline{b}} \overline{\alpha}_{\overline{b}}
\end{equation} where 
\begin{equation}
\zeta(\gamma) := \frac{Z(\gamma)}{|Z(\gamma)|}
\end{equation}
with $ii$ boundary conditions. We call a solution to this equation a \textit{closed soliton} of charge $\gamma$. The space of closed BPS solitons can indeed be non-vanishing and our formalism must therefore account for them. 

In addition we also have the usual $ij$ solitons where $i \neq j$, with the refinement that the $ij$ sector now again splits into disconnected components now labelled by the $\Gamma$-torsor $\Gamma_{ij}$ which consists of the set of $1$-chains $c$ with boundary 
\begin{equation}
\partial c = \phi_j - \phi_i
\end{equation}
modulo the addition of boundaries. A soliton in the sector labelled by the chain $\gamma_{ij} \in \Gamma_{ij}$ solves the flow equation with 
\begin{equation}
\zeta = \zeta(\gamma_{ij}) := \frac{Z(\gamma_{ij})}{|Z(\gamma_{ij})|}
\end{equation}
where $Z(\gamma_{ij})$ is the integral of $\alpha$ along the open chain $\gamma_{ij}$. 

The above discussion leads to a useful mathematical framework for discussing the central charges of BPS solitons in theories with nontrivial twisted masses: 
We now have a \textit{vacuum groupoid} $\mathbb{V}$ with objects labeled by zeroes of $\alpha$, and morphism space 
\begin{equation}
\text{Hom}(i,j) = \Gamma_{ij}
\end{equation}
the space of 1-chains with $ij$ boundaries (modulo the addition of boundaries to such 1-chains), and the composition law 
\begin{equation}
\circ: \Gamma_{ij} \times \Gamma_{jk} \rightarrow \Gamma_{ik} 
\end{equation}
is given by addition of $1$-chains. The central charge is a groupoid homomorphism given by integrating $\alpha$ along a $1$-chain. In addition we have (potentially)  BPS solitons for each morphism $\gamma_{ij} \in \Gamma_{ij}$ in the vacuum groupoid, both for $i \neq j$ and $i=j.$ 

In order to prevent the discussion from getting too abstract, we have included a collection of illustrative examples of LG models with cohomologically non-trivial one-forms $\alpha$, along with their deck groups $\Gamma$ and BPS spectra in Appendix \ref{examples}.

\section{General Strategy}\label{subsec:GeneralStrategy}

Now we would like to start generalizing the web formalism of section \ref{review} to the case of theories with twisted masses.
The generalization has a relation to the generalization of Morse theory to Morse-Novikov theory, and the reader familiar with
Morse-Novikov theory might think that the the generalization of the web formalism to include twisted masses can be completely 
solved by employing a familiar trick: Instead of working on $X$ we may pass to a covering space where things would resemble the usual web formalism more closely. We would then work equivariantly with respect to the deck group on this covering space. We will now explain that this viewpoint is only partially correct.

Let $\widetilde{X}$ be the (unique) holomorphic cover of $X$ obtained from the universal abelian cover
\footnote{The universal abelian cover in turn is obtained from the universal cover $\pi: \widehat{X} \rightarrow X$ by modding out the action of the commutator subgroup $[\pi_1(X), \pi_1(X)]$ of the fundamental group $\pi_1(X)$ of $X$ (we assume $X$ is connected   and we suppress the dependence on basepoint): $$X_{\text{ab}} = \widehat{X}/[\pi_1(X), \pi_1(X)].$$ Put differently one could mod out the universal cover by the kernel of $Z$  considered as a homomorphism with domain $\pi_1(X)$. } 
$X_{\text{ab}}$, the covering space with deck transformations $H_1(X, \mathbb{Z})$, by quotienting with $\text{ker}Z$, the kernel of the central charge homomorphism: 
\begin{equation}
    \widetilde{X} := X_{\text{ab}}/\text{ker} \,Z  ~ . 
\end{equation}
The result is the ``smallest'' holomorphic covering $p: \widetilde{X} \rightarrow X$ such that the pullback of the one-form $\alpha$ becomes exact: 
\begin{equation}
    p^*(\alpha) = \text{d}\widetilde{W}, \,\,\,\,\,\,\,\,\, \widetilde{W}: \widetilde{X} \rightarrow \mathbb{C}.
\end{equation}
We let  $\Gamma$ denote the covering group for $p$. 
A zero $\phi_i$ of $\alpha$   lifts to a set $p^{-1}(\phi_i)$ which is a set with a free $\Gamma$-action. There will be one such $\Gamma$-orbit for each critical point. The critical points of $\widetilde{W}$ are the union  of these $\Gamma$-orbits 
\begin{equation}
    \text{Crit}\big( \,\widetilde{W} \, \big) = \bigsqcup_{\phi_i \in \text{Zero}(\alpha)} p^{-1}(\phi_i),
\end{equation}
which follows from the fact that $p$ is a covering map. On the covering space we then have a well-defined notion of a critical value, and the central charge for any two critical points is then expressed again as a difference of critical values. Moreover, a closed $ii$ soliton on $X$ now lifts to an ordinary soliton on $\widetilde{X}$ that interpolates between some point in $a_i \in p^{-1}(\phi_i)$ and its image $\gamma \cdot a_i \in p^{-1}(\phi_i)$ under the element $\gamma \in \Gamma$. We therefore see that working on a covering space gives us both well-defined vacuum weights so that the central charge can be written as a difference, and closed solitons get identified as ordinary solitons. Can we then run the web formalism as usual?

Things are complicated by one additional factor. In the usual web formalism it is a crucial assumption that the vacuum weights are in general position so that no three vacuum weights $\{z_i, z_j, z_k \} \subset \mathbb{C}$ lie along a line in the complex plane. This ensures that one can make a meaningful distinction between an $ik$-soliton and a bound state of an $ij$ and $jk$ soliton (so that there is a gap between the multiparticle continuum spectrum and the one-particle BPS spectrum). It also ensures that the web combinatorics work out like we want to. For the present case not only is this condition violated, in fact the situation is actually much worse: fix a non-zero element $\gamma \in \Gamma$ and a point $a_i \in p^{-1}(\phi_i)$. We have \begin{equation}
    \widetilde{W}(\gamma \cdot p) = \widetilde{W}(p) + Z(\gamma),
\end{equation} moreover 
\begin{equation}
\widetilde{W}(n \gamma \cdot p) = \widetilde{W}(p) + n Z(\gamma)
\end{equation}
for any $n \in \mathbb{Z}$. This implies that the entire infinite set 
\begin{equation}
\{W(n \gamma \cdot p)\}_{n \in \mathbb{Z}}
\end{equation}
of complex numbers all lie along the common line parallel to the complex number $Z(\gamma)$ in the $W$-plane! 
 We are thus lead to a highly degenerate situation, and one cannot hope for a reasonable web formalism without resolving this degeneracy. 
\footnote{We have entertained many different modifications of the web formalism in the degenerate case and have found them all unsatisfying, for one reason or another.}
The central source of the difficulty is the infinite alignment of critical values associated to the infinitude of elements $\gamma \in \Gamma$.

 In order to get a reasonable web formalism we may attempt to break the degeneracy by perturbing the superpotential $\widetilde{W}$ and then 
taking a limit in which the perturbation vanishes. This was, in fact, our very first approach to the problem. However, there are too many perturbations, and taking different limits with different perturbations in general will give incoherent and inconsistent answers.
The progress that led to the present paper is the discovery that - at least in some examples -  one can perturb $\tilde W$ in a ``small'' way so that the vacuum weights of the perturbed superpotential are non-degenerate but the breaking is nevertheless ``soft'' in the following sense: 
We can still preserve the $\Gamma$-invariance in the soliton spectrum, so that the perturbed model satisfies: 
\begin{equation}\label{eq:DeckInvPertRab}
R_{a,b} \cong R_{\gamma \cdot a, \gamma \cdot b}    
\end{equation} 
for each pair of (distinct) critical points $a,b$ of $\widetilde{W}$ and each $\gamma \in \Gamma$. 
That such a perturbation exists for all models with twisted masses is not 
\emph{a priori} clear to us. In the next section we will nevertheless demonstrate how this can be done in the simplest example of a theory with a non-trivial twisted mass.

To summarize, our strategy for constructing the $A_\infty$ categories of boundary conditions in models with twisted masses will be as follows:

\begin{enumerate}
    \item Pass to a cover $p: \widetilde{X} \rightarrow X$ with deck transformation group $\Gamma$ so that one has well-defined superpotential $\widetilde{W}$ and hence well-defined vacuum weights. 
    \item Perturb the superpotential $\widetilde{W}$ on the cover so that the vacuum weights  become non-degenerate while still preserving the $\Gamma$-invariance of the soliton spectrum. 
    \item Apply the web formalism with the perturbed central charges and  perturbed soliton spaces. Upon choosing a half-plane one computes  an $A_{\infty}$-category with the action of the deck group $\Gamma$ by autofunctors.
    \item Pass to the orbit category with respect to the deck group $\Gamma.$
\end{enumerate}

As we will see in the subsequent sections, this strategy is deceptively simple: In fact, no single perturbation will work for all 
computations of soliton spaces and all interior amplitudes, rather, a family of perturbations must be employed. 
This in turn leads to  interesting orders of limits issues. Comparison of different choices of perturbation moreover leads to a   version of Stokes' phenomenon. Nevertheless, we will in fact successfully implement the above outlined strategy. 

\section{Web Formalism for Mirror to Free Chirals} \label{toyexample}

We now illustrate the strategy of section \ref{subsec:GeneralStrategy} in detail and work out the category of boundary conditions (for a given half-plane and a given $\zeta$) for the simplest Landau-Ginzburg model with a Morse critical point and a non-trivial twisted mass\footnote{Some previous discussion of this Landau-Ginzburg model can be found in Appendix C of \cite{Harlow:2011ny}, Appendix A of \cite{Aganagic:2021ubp} and Section 4.1.1 of \cite{Kontsevich:2024esg} }. 

Consider the Landau-Ginzburg model with target space the punctured complex plane $\mathbb{C}^*$, and  one-form 
\begin{equation}
\alpha = \Big(\frac{m}{\phi} -1 \Big) \text{d}\phi
\end{equation} 
for a fixed non-zero complex number $m \neq 0.$ There is a single Morse vacuum located at \begin{equation}
    \phi = m.
\end{equation} The central charge homomorphism $Z: H_1(\mathbb{C}^*, \mathbb{Z}) \rightarrow \mathbb{C}$ is injective 
\begin{equation} 
\text{ker} \,Z = \{0\} \,\,\, \text{ for } \,\,\,  m\neq 0 
\end{equation}
so that the covering space we pass to will have deck group 
\begin{equation}
\Gamma = H_1(\mathbb{C}^*, \mathbb{Z}) \cong \mathbb{Z}.
\end{equation}
The covering space $\widetilde{X}$ is simply the universal cover of $\mathbb{C}^*$, namely the complex plane $\mathbb{C}$. Equipping the latter with the complex coordinate $Y$ the covering map $p: \mathbb{C} \rightarrow \mathbb{C}^*$ is the exponential map 
\begin{equation}
p(Y)=e^Y.
\end{equation}
We thus see that the superpotential $\widetilde{W}$ on $\mathbb{C}$ is given by 
\begin{equation}\label{eq:Wtilde-FreeChiral}
\widetilde{W} = mY - e^Y.
\end{equation}
The critical points of $\widetilde{W}$ are located at 
\begin{equation}
Y_k = \text{log}\,m + 2 \pi \text{i}k, \,\,\,\,\, k \in \mathbb{Z},
\end{equation}
with corresponding critical values 
\begin{equation}
\widetilde{W}(Y_k) = m(\text{log}\, m -1) + 2\pi \text{i} k m, \,\,\,\,\, k \in \mathbb{Z},
\end{equation}
so that they all lie along a line parallel to $\text{i}m$ in the complex $\widetilde{W}$-plane.

In order to obtain a non-degenerate web formalism we must perturb the function $\widetilde{W}$ so that the critical values get deformed to a non-degenerate configuration. 

\def\CW{\mathcal{W}}
In general, if we are studying the Morse theory of a holomorphic function  $\CW(Z)$
of a single complex variable $Z$ we may perturb it by choosing a holomorphic function $f(Z)$ 
and a small $\epsilon$ and considering the  perturbed function
\begin{equation}
    \CW^{\epsilon}(Z) := \CW(Z) + \epsilon f(Z)
\end{equation}
for a given function $f(Z).$ If $Z_i$ is a critical point of $\CW$ we may expand around it so that 
\begin{equation}
    Z_i(\epsilon) = Z_i + \epsilon \delta Z_i + O(\epsilon^2).
\end{equation} 
This is a critical point to first order in $\epsilon$ provided 
\begin{equation}
    \delta Z_i = - \frac{f'(Z_i)}{\CW''(Z_i)}.
\end{equation} 
We are also interested in how the critical value is perturbed. To first order the perturbed critical value is simply 
\begin{equation}
    \CW^{\epsilon}(Z_i(\epsilon)) = \CW(Z_i) + \epsilon f(Z_i) + O(\epsilon^2).
\end{equation}
Thus to first order in $\epsilon$ we may use a suitable perturbation function $f(Z)$ to arrange the critical values to a configuration that suits us. 

Going back to our model we choose the perturbation function of \eqref{eq:Wtilde-FreeChiral}
to be
\begin{equation}
    f(Y) = \frac{1}{2} Y^2.
\end{equation}
To first order in $\epsilon$ then the critical points are 
\be 
Y_k(\epsilon) = Y_k + \frac{\epsilon}{m} Y_k + \mathcal{O}(\epsilon^2) 
\ee
while our critical values lie at 
\begin{equation}\label{eq:perturbed-wtilde-1}
\tilde W^{\epsilon}_k = m(\text{log}m -1) + 2\pi \text{i}mk + \epsilon \frac{1}{2}(\text{log}m+ 2\pi i k)^2 + \mathcal{O}(\epsilon^2) . 
\end{equation} 
By rescaling $\phi$ and $\alpha$ we may set $m=1$, in which case \eqref{eq:perturbed-wtilde-1} simplifies to
\begin{equation}\label{eq:tWkeps}
   \tilde W_k^{\epsilon}= -1 + 2\pi \text{i} k -\epsilon \, 2\pi^2 k^2 + \mathcal{O}(\epsilon^2) .  
\end{equation}
When ${\rm Re}(\epsilon)$ is nonzero   the critical values are perturbed away from a line parallel to the imaginary axis by an amount that is $k$-dependent. All of the critical values are now lying on a parabola and thus the critical values are in general position, 
and therefore we can try to analyze the soliton spectrum of the perturbed model.

Before analyzing the soliton spectrum of the perturbed model we must note one crucial complication: The perturbation $f(Y) = \frac{1}{2} Y^2$ cannot be considered as a ``small" perturbation of the original superpotential $\tilde{W} = Y - e^Y.$  Indeed if we go to infinity in the $Y$-plane along a ray with $\text{Re} Y >0$, the asymptotic behavior of $\tilde{W} $ 
is unchanged. However, for a ray with $\text{Re} Y < 0$ the perturbation function begins to dominate $\tilde{W} $ and the global behavior can change. Therefore there is an important order of limits question. 
It is not correct to choose a fixed but small $\epsilon$ and define all the quantities in the web formalism. There is no $\epsilon$ for which the change of the theory is uniformly small.  Rather, for a fixed set of vacua, solitons, and webs we will make $\epsilon$ sufficiently small that it is indeed a small perturbation of those chosen vacua, solitons, and webs. 
For example, when we consider solitons between two vacua with fixed values of $k$ and $l$   we can choose an $\epsilon$, that guarantees that the perturbation is small in a large enough region containing the vacua and the soliton trajectories we are interested in. No uniform $\epsilon$ will work for all values of $k$ and $l$. We may then consider ``stable" solitons, which are the ones which continue to exist for an arbitrarily small $\epsilon$. 
\footnote{An example of a perturbation $f(Y)$ which does not change the asymptotic behavior is  $f(Y) = e^{r Y}$, where $0<r<1$ is an irrational number. We expect that this perturbation will not lead to the crucial equation \eqref{eq:DeckInvPertRab}.  }

The precise statement is as follows. 
\footnote{At this point, for notational simplicity, we 
henceforth replace the notation $\widetilde W(Y)$ by $W(Y)$. This should not cause confusion 
since a single-valued superpotential is only defined on the cover.}
\begin{theorem} Let 
\begin{equation} \label{perturbedw} 
 W^{\epsilon} = Y - e^Y + \epsilon \frac{Y^2}{2} , 
\end{equation} 
and let 
\begin{equation}
Y_k(\epsilon) = 2\pi \text{i} k + O(\epsilon)
\end{equation}
be the perturbation series of the critical point around the critical point $Y_k = 2\pi i k$ of $ W^0.$ Then for each $k,l \in \mathbb{Z}$ there is an $\epsilon > 0$ such that the perturbation series for the critical points $Y_k(\epsilon)$ and $Y_l(\epsilon)$ converges, and there is a unique soliton between $Y_k(\epsilon)$ and $Y_l(\epsilon)$, namely a solution to the flow equation
\begin{equation}
   \frac{\text{d} Y}{\text{d} x} = \zeta_{k,l}(\epsilon) \frac{\partial \overline{W}^{\epsilon}}{ \partial \overline{Y}}
\end{equation}
with $(k,l)$ boundary conditions and $\zeta_{k,l}(\epsilon)$ is the phase of $W_l^\epsilon - W_k^\epsilon$. Moreover, the soliton is stable in the sense that for each $0 < \delta \leq  \epsilon$ one has a unique
\footnote{up to translations} 
solution to the flow equation for $\text{Re}(\zeta^{-1}_{k,l} (\delta) \,W^{\delta} )$ that interpolates between $Y_{k}(\delta)$ and $Y_{l}(\delta).$
\footnote{In fact one really expects more then just existence for each $\delta$. The family of solitons should be continuous in $\delta$.} 
\end{theorem}

The proof of this theorem involves the following steps. We will first prove a key Lemma regarding the intersection pattern of Lefschetz thimbles in the unperturbed, $\epsilon=0$ theory. Next we will claim that for each $k,l \in \mathbb{Z}$, there is an $\epsilon > 0$ such that the perturbation in a region containing $Y_k$ and $Y_l$ can be considered small, and the Lefschetz thimbles change in a small way in this given region\footnote{We will also see that their global behavior, namely the behavior outside this region, \emph{does} change.}. In particular the number of intersection points is unchanged. Finally we will show that, in the perturbed theory the slopes are arranged so that the intersection points of thimbles are identified as $(k,l)$ solitons.

Recall that a left Lefschetz thimble at a critical point $p$ of a superpotential $W$ consists of the collection of points in the target $X$ that can be obtained from flowing away from $p$ along the gradient flow $\text{Re}(\zeta^{-1}W)$ so that 
\begin{equation}
\text{Re}(\zeta^{-1}W) \rightarrow + \infty
\end{equation}
as we go to infinity along the thimble. (See equation \eqref{eq:LeftLefshetz-def}.) The image $W(L_p(\zeta))$ consists of a ray starting from the critical value $W_p$ having angle $\text{Arg}(\zeta)$ from the positive horizontal.  A right thimble of phase $\zeta$ satisfies: 
\begin{equation}
R_p(\zeta) = L_p(-\zeta).
\end{equation}


\begin{lemma} \label{intersectionlemma} Let $L_k(\zeta)$ be the left Lefschetz thimble for the critical point $Y_k = 2\pi \text{i} k$ of the superpotential $W = Y - e^Y.$ Choose angles $\theta, \theta' \in (0, \frac{\pi}{2}) $ and let 
\begin{align}
I_{k,l}(\theta, \theta') &= L_k(-e^{-i\theta'}) \cap L_l(-e^{i \theta}), \\
J_{k,l}(\theta, \theta') &= L_k(e^{-i\theta'}) \cap L_l(e^{i\theta})
\end{align} Then $I_{k,l}(\theta, \theta')$ consists of a single point with positive orientation for each $k \leq l$ and is empty otherwise, and $J_{k,l}(\theta, \theta')$ consists of a single point with negative orientation if $k = l+1$ and positive orientation if $k=l$ and is empty otherwise.
\end{lemma}

\begin{figure}
    \centering
    \includegraphics[width=0.85\textwidth]{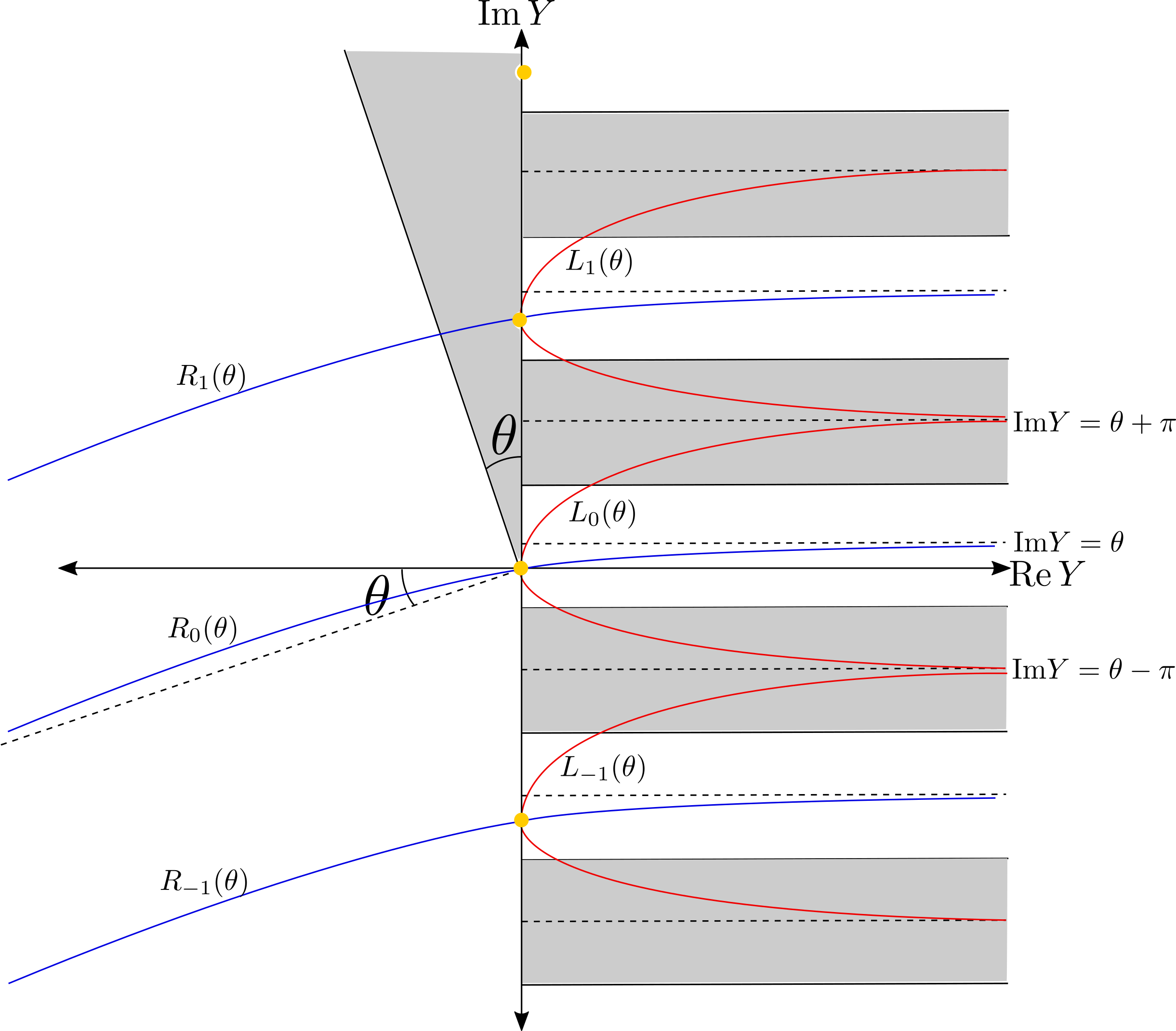}
    \caption{Left and right thimbles for $W=mY-e^Y$ in the $Y$-plane for $m=1$ and an angle $\theta \in (0, \frac{\pi}{2})$. The shaded regions are the ones in which a left thimble of angle $\theta$ is allowed to end. The yellow dots denote the critical points $Y_0 = 0, \,\,Y_1 = 2\pi i \,\, , Y_{-1}= -2\pi i.$ The dashed lines denote lines to which the thimbles asymptote.}
    \label{thimbles}
\end{figure}

\begin{figure}
    \centering
    \includegraphics[width=0.8\textwidth]{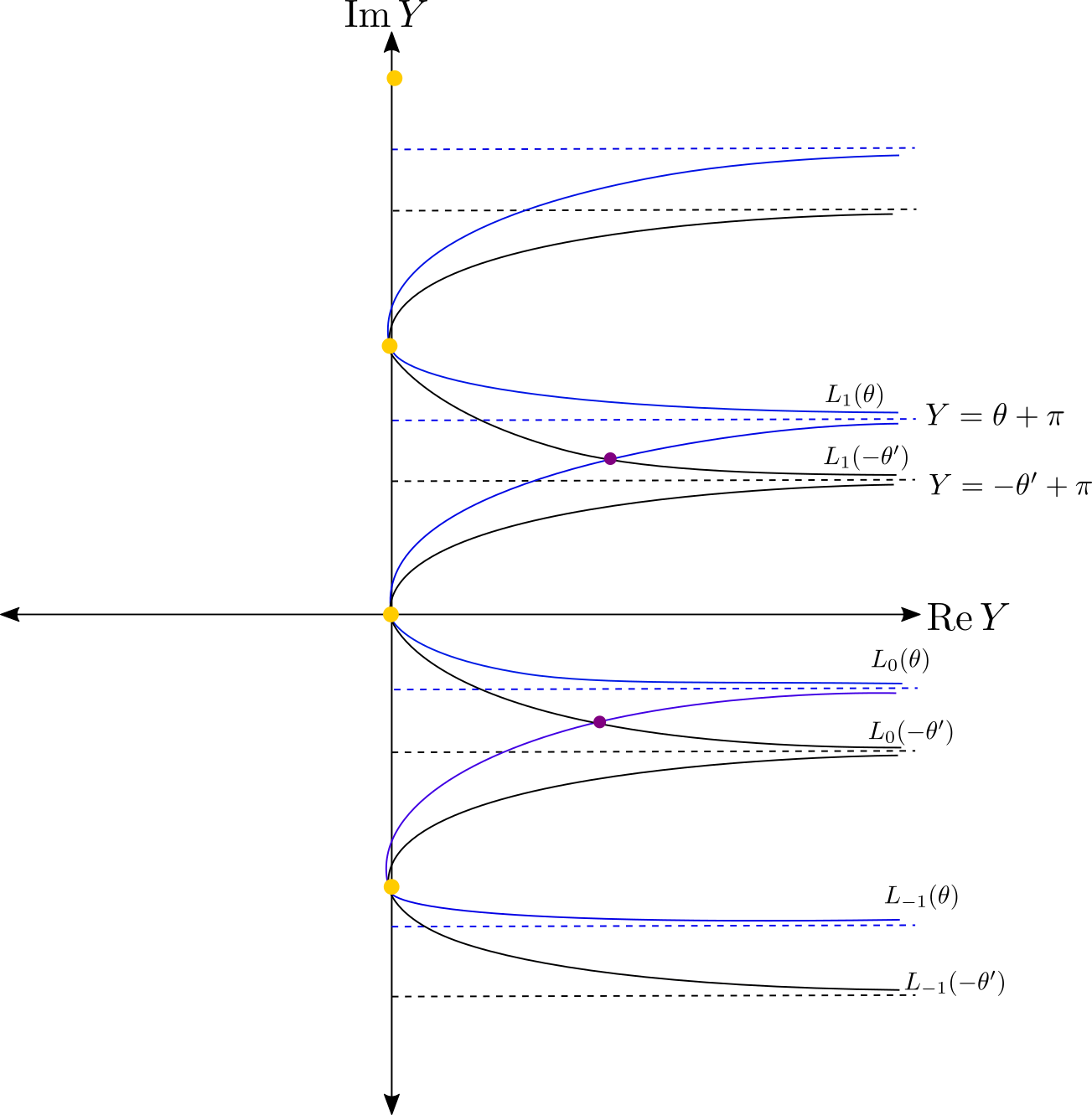}
    \caption{Intersection pattern of left thimbles $L_k(\theta)$ and $L_l(-\theta')$ for angles $\theta, \theta' \in (0,\frac{\pi}{2}).$}
    \label{intersectingleft}
\end{figure}

\begin{myproof}
    We begin by working out the possible regions in the complex $Y$-plane where the thimbles of phase $\zeta$ can go off to $\infty$. For this we look at the regions where as $Y \rightarrow \infty$, we have 
    \be 
    \text{Re}(\zeta^{-1}W) \rightarrow +\infty.
    \ee Let us work out the regions where this is satisfied for the left-half $Y$-plane and the right-half $Y$-plane separately. If $\text{Re} Y >0$ and large, the exponential term $-e^Y$ dominates dominates the linear term in $W$. Letting $\zeta = e^{i \theta}$ and $Y = x + \text{i}y$ we have then have 
    \begin{equation}
    \text{Re}(-\zeta^{-1}e^{Y}) = -e^x \text{cos}(y - \theta)
    \end{equation}
    which tells us that for this go to $+\infty$ as $x \rightarrow \infty$, $y$ must be such that 
    \be 
    \text{cos}(y-\theta)<0.
    \ee Thus we must have 
    \begin{equation} \label{regionpositive}
    y \in \bigcup_{k \in \mathbb{Z}} \big(\theta + \frac{\pi}{2} + 2\pi k, \, \theta + \frac{3 \pi}{2} + 2\pi k \big)
    \end{equation}
    This gives us a good set of regions where a thimble can end in the right half-plane. For the left-half plane and $|\text{Re}(Y)|$ large, the linear term dominates. Letting $Y = r e^{i \varphi}$ we have 
    \begin{equation}
    \text{Re}(\zeta^{-1}Y) = r \text{cos}(\varphi- \theta)
    \end{equation}
    and so we must have 
    \begin{equation}
    \varphi \in (\theta- \frac{\pi}{2}, \theta + \frac{\pi}{2})
    \end{equation} in order to have $\text{Re}(\zeta^{-1} W) \rightarrow \infty$ as $r \rightarrow \infty$. On the other hand in the left half-plane we have 
    \begin{equation} 
    \varphi \in \big(\frac{\pi}{2}, \frac{3\pi}{2} \big). 
    \end{equation}
    If $\theta \in [0, \pi)$ 
    %
    %
    the intersection of these two regions is the angular sector 
    \begin{equation} \label{regionnegative}
    \varphi \in \big(\frac{\pi}{2}, \frac{\pi}{2} +\theta \big). 
    \end{equation} 
    Thus in summary the regions in the $Y$-plane where $\text{Re}(e^{-i\theta}W) \rightarrow \infty$ as $Y \rightarrow \infty$ consist of the following. For $\theta \in [0, \pi)$ the regions consist of a union of the regions given in $\eqref{regionpositive}$ with $x>0 $ with the angular sector consisting of \eqref{regionnegative}. See the grey regions in Figure 
    \ref{thimbles}. It suffices to determine the regions for $\theta \in [0, \pi)$ because the regions for $\theta \in [\pi, 2\pi)$ are precisely the complements of the regions for $\theta' =  \theta- \pi.$

    We now come to the actual thimbles within these regions. It suffices to determine a single thimble $L_0(e^{\text{i} \theta})$ for $k=0$ as the rest can be determined by the action of the deck group. Our claim is that for $\theta \in (-\frac{\pi}{2}, \frac{\pi}{2})$ the thimble $L_0( e^{i \theta})$ is described as follows. The thimble $L_0(e^{i\theta})$ is a subset of the locus in the $Y$-plane such that 
    \begin{equation}
        \text{Im}(e^{-i\theta} W(Y)) = \text{Im}(e^{-i\theta} W(0)) ~ .
    \end{equation}
    Denoting real and imaginary parts of $Y$ by $Y = x + \text{i}y$ this becomes the 
    real in real variables: 
    \begin{equation}
        y \,\text{cos}(\theta) - x \text{sin}(\theta) - e^x \text{sin}(y-\theta) = \text{sin}(\theta). 
    \end{equation}
    This region for large $|Y|$ will asymptote to certain lines. For instance if we take $x$ to be large and positive, the exponential term dominates, and the left hand side can only be a bounded constant if the contribution from this term vanishes, which is only possible if 
    \be 
    y = \theta + \pi k, \,\,\,\, k \in \mathbb{Z}, \,\,\,\,\, x>0.
    \ee 
    Thus these are the lines which the region asymptotes to for $x$ large and positive. On the other hand, if $x$ is large and negative, the exponential term dies off and the linear term is bounded if and only if we have 
    \be y = \text{tan}(\theta)  \,x, \,\,\,\, x<0\ee so that we have for $x$ large and negative the region simply asymptotes to a line of angle $\theta$. We worked out the regions the thimbles may lie in in the previous paragraph. The thimble must also pass through the critical point $Y=0$. Given the asymptotes above, and the regions along with the fact that $L_k$ and $L_l$ cannot intersect for $k \neq l$, we find that the thimble $L_0(e^{i\theta})$ is a curve that is approximately the union of the lines $y = \theta - \pi$ and $y = \theta + \pi$ when $x$ is large and connects these two half-lines by passing through the origin when $x$ is small. The right thimble $L_0(-e^{i\theta})$ on the other hand lies in the complement of this region and thus is a curve passing through the origin which asymptotes to $y = \theta$ when $x$ is large and positive, passes through the origin, and then asymptotes to a line with slope $\text{tan}\,\theta$ for $x$ being large and negative. See Figure \ref{thimbles} for an illustration of these thimbles. 

    That this is the configuration of thimbles immediately implies the claimed result about their intersections. In order to work out the intersection points one simply looks at the relevant lines the thimbles asymptote to. 
     $L_k(e^{i\theta})$ will intersect with $L_l(e^{-i\theta'})$ for any $\theta, \theta' \in (0, \pi/2)$ as long as $k = l$ or $k = l+1$. 
$L_0(e^{i\theta})$ consists of a curve connecting $y= \theta-\pi$ to $y = \theta + \pi$ by going through $Y_0$, whereas $L_1(e^{-i\theta'})$ is the curve connecting $y= -\theta'+ \pi $ and $y = -\theta' + 3\pi$ by going through $Y_1$. Such curves will necessarily intersect exactly at a point, see Figure \ref{intersectingleft}, that is: 
 \begin{equation}
    L_0(e^{i\theta}) \circ L_1(e^{-i\theta'}) = -1.
    \end{equation}
On the other hand there are no other intersections (besides the trivial one).

    The right thimbles on the other hand have many more intersection points. For instance, $R_k(e^{i\theta})$ in the region with $x$ large and negative asymptotes to the half-line of slope $\theta$ going through the point $Y_k = 2\pi \text{i} k$. Thus we see that $R_k(e^{i\theta})$ and $R_l(e^{-i\theta})$ will intersect at a single point as long as $k  \geq l$: 
    \begin{equation}
    L_k(-e^{i\theta}) \circ L_l(-e^{-i\theta'}) = +1, \,\,\,\, k \geq l.
    \end{equation}
    See Figure \ref{intersectingright}. 
This proves the claim of Lemma \ref{intersectionlemma}. 
\end{myproof}

\begin{figure}
    \centering
    \includegraphics[width=0.72\textwidth]{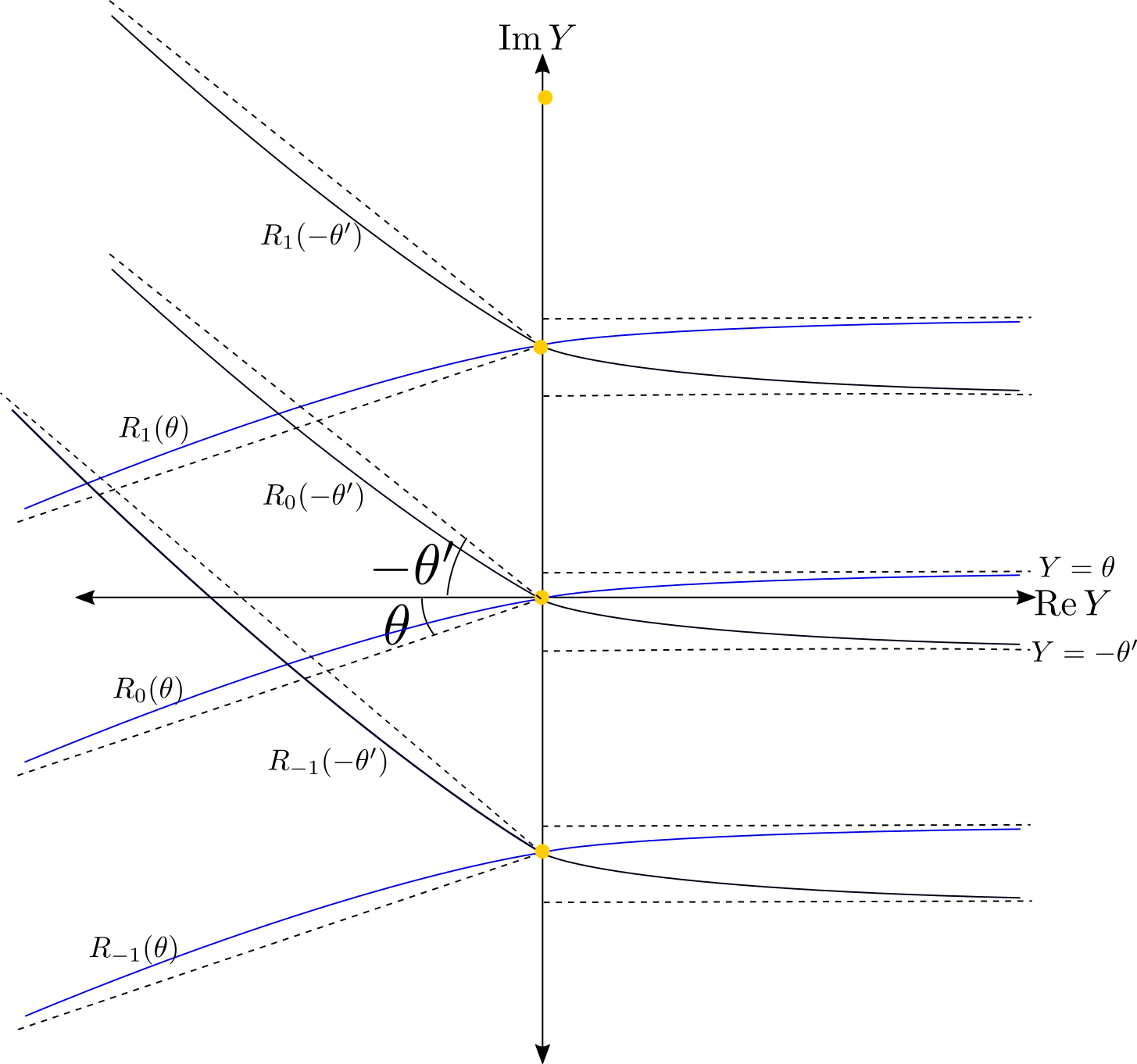}
    \caption{Intersection patterns for right thimbles $R_k(\theta)$ and $R_l(-\theta')$ for $\theta, \theta' \in (0, \frac{\pi}{2}).$}
    \label{intersectingright}
\end{figure}

We would now like to show that the intersection pattern of thimbles summarized in Lemma \ref{intersectionlemma} above implies that there is a stable $(k,l)$ soliton for each pair of distinct integers $(k,l)$. Recall that by a stable $(k,l)$ soliton we mean the following. We consider the superpotential $W_{\epsilon}(Y)$ given in \eqref{perturbedw}, with critical points $Y_k(\epsilon)$ and $Y_l(\epsilon)$. Suppose that there is an $\epsilon >0$ such that the perturbative series for $Y_k(\epsilon)$ and $Y_l(\epsilon)$ converge. Again, 
we will say that the perturbed model has a stable soliton if there is an $\epsilon>0$ such that there is a soliton between the (limiting values of) $Y_k(\epsilon)$ and $Y_l(\epsilon)$ given by the flow of $W^{\epsilon}$ and that for each $0< \delta < \epsilon$ the soliton varies continuously (as a function of $\delta$)  i.e there is a family of trajectories 
\begin{equation}
\{\gamma_{\delta}(x) | \text{lim}_{x \rightarrow -\infty} = Y_k(\delta), \text{lim}_{x \rightarrow + \infty} = Y_l(\delta),  \}_{0< \delta < \epsilon} 
\end{equation}
in $\mathbb{C}$ which solve the flow equation for $W^{\delta}(Y)$, interpolating between $Y_k(\delta)$ and $Y_l(\delta).$

Suppose (without losing generality) that $l > k.$ We will now show that for each such $k,l$ there is a (unique) stable soliton. The basic criterion for the existence of a soliton in the $(l,k)$ sector is the existence of an intersection point of the slightly perturbed left- and right- Lefshetz thimbles 
$L_l(\zeta_{lk}(\epsilon) e^{-i \epsilon'})$  and $R_k(\zeta_{lk}(\epsilon)e^{i \epsilon'})$ for 
some small positive $\epsilon'$ (not related to $\epsilon$). 
From equation \eqref{eq:tWkeps} we compute that $\zeta_{lk}(\epsilon)$, which is the phase of $\tilde W^\epsilon_k - \tilde W^\epsilon_l$ , is (for $l>k$): 
\be 
\zeta_{lk}(\epsilon) = e^{-i\pi/2+ i \pi \epsilon(k+l) + \mathcal{O}(\epsilon^2)} ~ . 
\ee
Using equation \eqref{eq:IntersectionCriterion} we can write the required intersection as 
\be 
L_l\left( - e^{ i\frac{\pi}{2} - i \epsilon' + i\pi \epsilon(k+l)+\mathcal{O}(\epsilon^2)} \right)
\cap L_k \left( - e^{-i \frac{\pi}{2} +i\epsilon' + i \pi \epsilon(k+l)+\mathcal{O}(\epsilon^2) } \right)
\ee

For a fixed $k,l$ we can make $0< \epsilon \ll \epsilon'$ so that we can apply  Lemma 4.2, which states 
that for each pair of angles $\theta$ and $\theta'$ between $0$ and $\frac{\pi}{2}$, 
the thimbles $L_k(-e^{-i\theta'})$ and $L_l(-e^{i \theta})$ intersect at a unique point $p_{k,l}(\theta, \theta')$
for $l>k$. It is clear that for each $\theta, \theta'$ and each $k,l$ one can choose an $\epsilon$ so that in a domain large enough to contain $Y_k, Y_l$ and $p_{k,l}(\theta, \theta')$ the perturbation of $W$ by $\epsilon \,f(Y)$ can be considered small\footnote{For simplicity we can take it to be a disk, so it will also contain the other critical points between $k$ and $l$.}. In particular this means that in this domain there are no new critical points introduced, and the original critical points $Y_k, Y_l$ have appropriate perturbations $Y_k(\epsilon), Y_l(\epsilon)$ by small amounts to critical points of $W_{\epsilon}$. Moreover, within this region the thimbles only change by a small amount and thus the perturbed thimbles $L^{\epsilon}_l(-e^{i\theta})$ and $L_k^{\epsilon}(-e^{-i\theta'})$ defined by $W_{\epsilon}$ still have a unique intersection point $p_{k,l}^{\epsilon}(\theta, \theta')$ in the region $D$, coming from perturbing the point $p_{k,l}(\theta, \theta').$ Note however the global behavior of the thimbles can and in fact does change (they go off to infinity in a totally different way), but this is not relevant to the problem of determining the stable soliton spectrum. This proves Theorem 4.1. 

One can describe the final step in the proof slightly differently: 
In the unperturbed case the critical values are 
\begin{equation}
W_n = -1 + 2\pi i n,
\end{equation}
for $n \in \mathbb{Z}$ and the thimbles simply map to half rays at angles $\theta$ and $-\theta'$ emanating from $W_k$ and $W_l$ respectively, which intersect. Upon perturbing, for $\epsilon$ small enough the critical values are approximately 
\begin{equation}
W_n^\epsilon \approx - 1 + 2\pi i n - \epsilon(2\pi^2 n^2)
\end{equation}
for $l \leq n \leq k$. Now under $W_{\epsilon}$ the domain $D$ in which the perturbation was small and which contains the perturbed critical points and intersection point of thimbles, maps to a domain (use open mapping theorem) containing the perturbed critical values which approximately lie on a parabola. Moreover, the perturbed thimbles map to rays at angles $\theta$ and $-\theta'$ that intersect. Now since all critical values in this domain in the $W_{\epsilon}$-plane lie along a parabola, the rays going from $W_l(\epsilon)$ at angle $-\theta'$ and then at angle $\theta$ to $W_k(\epsilon)$ is homotopic to the straight line that connects them.  See Figure \ref{homotopic}.

The $\epsilon \to 0$ limit of the $l,k$ solitons have an important property illustrated in  Figure \ref{perturbedsolitons}, namely, the solitons 
for $\vert l - k \vert > 1$ can be thought of as ``composites'' of the 
``elementary solitons'' with $\vert l - k \vert = 1$. In the $\phi$-plane 
the soliton winds multiple times. We will not attempt to make this observation extremely mathematically precise, but we regard it as important, and it will be the origin of certain Fock spaces which will emerge below.  It will also be used in our next step of determining the 
fermion numbers of the solitons.

Let us consider the fermion number of these $(k,l)$ solitons. 
As explained in 
\cite{Gaiotto:2015aoa} the multiplet of $(k,l)$ solitons has fermion 
number 
\be 
\big(\eta_{k,l} -\frac{1}{2} \, , \, \eta_{k,l} + \frac{1}{2} \big),
\ee 
where $\eta_{k,l}$ is the 
$\eta$-invariant (defined by summing over the nonzero eigenvalues) of the 
Dirac operator associated with the $Y_{k,l}$ soliton. The soliton 
complex $R_{k,l}$ will be one-dimensional with fermion degree 
$\eta_{k,l}+\frac{1}{2}$. CP tells us that 
\be 
\eta_{k,l} + \eta_{l,k}=0.
\ee
Moreover, the property that $(k,l)$ solitons are composites of  
elementary solitons tells us that as   $\epsilon \to 0$ 
\be
\begin{split}
\eta_{k,l} & = (l-k) \eta_{0,1} \\
\eta_{l,k} & = (l - k ) \eta_{1,0} \\
\end{split}
\ee
where $l>k$ and $\eta_{1,0}+ \eta_{0,1}=0$.  
Now we must take into account an important subtlety, already noted in 
\cite{Gaiotto:2015aoa}, concerning the definition of the fermion number: We can always 
add $dg$ to the fermion current where $g$ is an arbitrary function. This changes the fermion 
number of the soliton multiplet of $Y_{k,l}$ for $l>k$ to be 
\be\label{eq:ShiftFermNumber}
( (l-k) \eta_{0,1} + g_l - g_k -1/2, (l-k) \eta_{0,1} + g_l - g_k +1/2) 
\ee
and that of $Y_{l,k}$ to 
\be 
( (l-k) \eta_{1,0} + g_k - g_l -1/2, (l-k) \eta_{1,0} + g_k - g_l +1/2) 
\ee
Using this freedom we can set the fermion degree of $R_{kl}$ to be $0$ for 
$l>k$ and then $R_{lk}$ must have degree $+1$, hence 

\begin{equation}\label{eq:Rkl-k-less-l} 
R_{k,l} \cong \mathbb{C} 
\end{equation}
for each $k < l$ and  
\begin{equation}\label{eq:Rkl-k-more-l} 
R_{k,l} \cong \mathbb{C}^{[1]} 
\end{equation}
for each $k > l.$ Finally, we also have, trivially, $R_{kk} = \mathbb{C}$.

\begin{figure}
    \centering
    \includegraphics[width=0.4\textwidth]{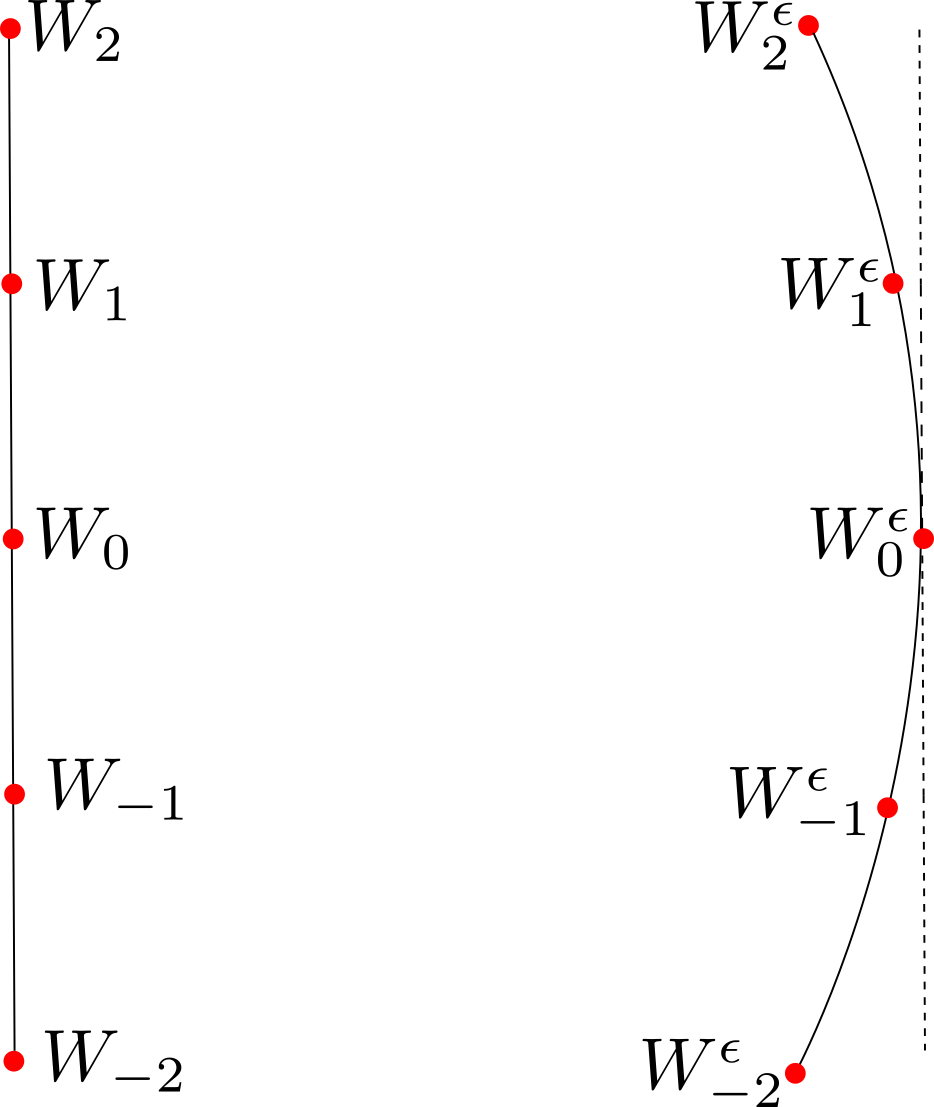}
    \caption{Critical values lying on a straight line parallel to the imaginary axis (left) are then perturbed to a convex configuration (right). }
    \label{perturbed}
\end{figure}

\begin{figure}
    \centering
    \includegraphics[width=0.6\textwidth]{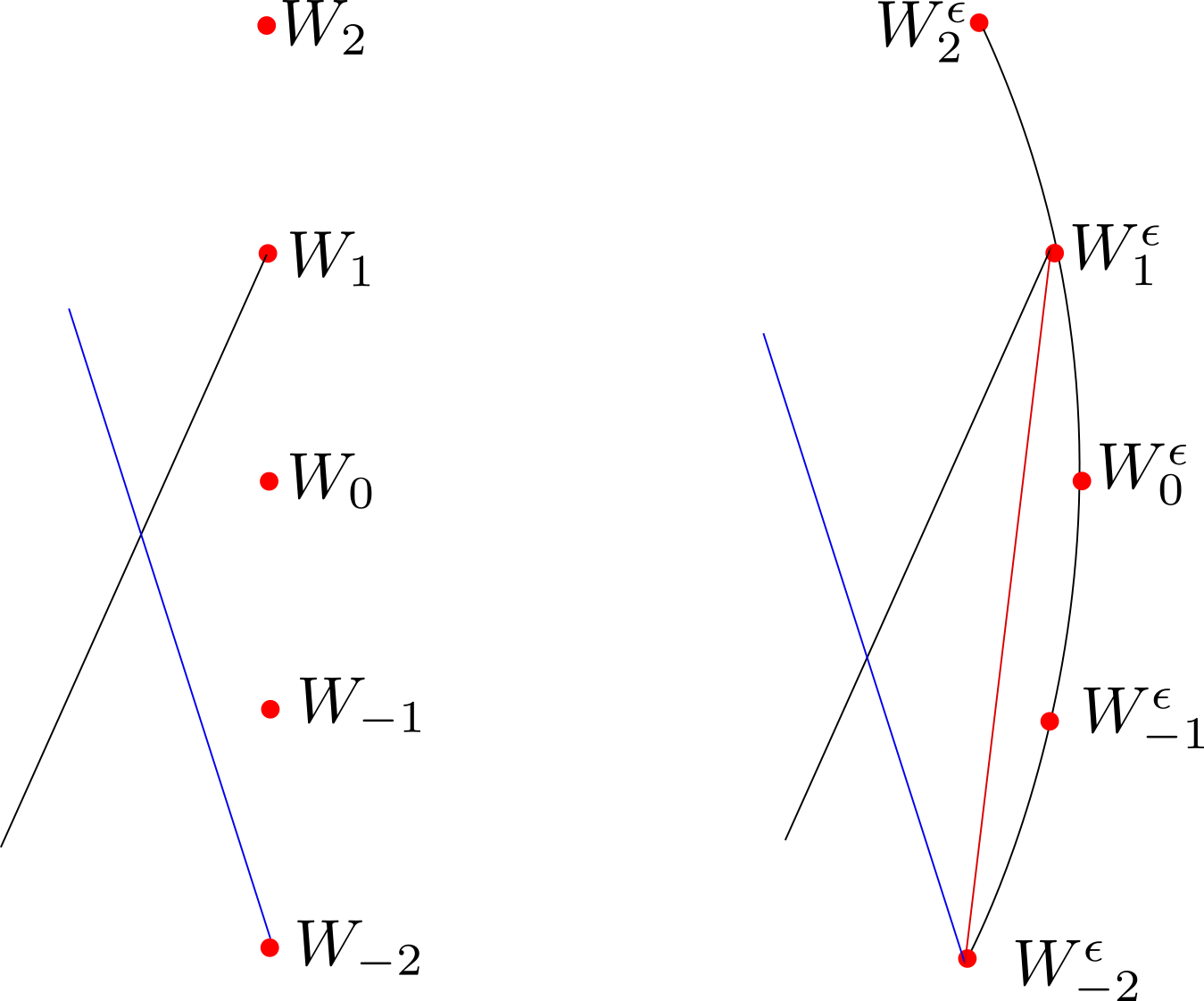}
    \caption{The left depicts the $W$-images of right thimbles for the unperturbed theory rotated by some angles $\theta$ and $-\theta'$. In the right of the image, after perturbation we see that the intersection point of the perturbed thimbles becomes homotopic to the image a soliton from the perturbation of the vacuum $Y_{-2}$ to the perturbation of the vacuum $Y_1$. }
    \label{homotopic}
\end{figure}

\begin{figure}
    \centering
    \includegraphics[width=0.6\textwidth]{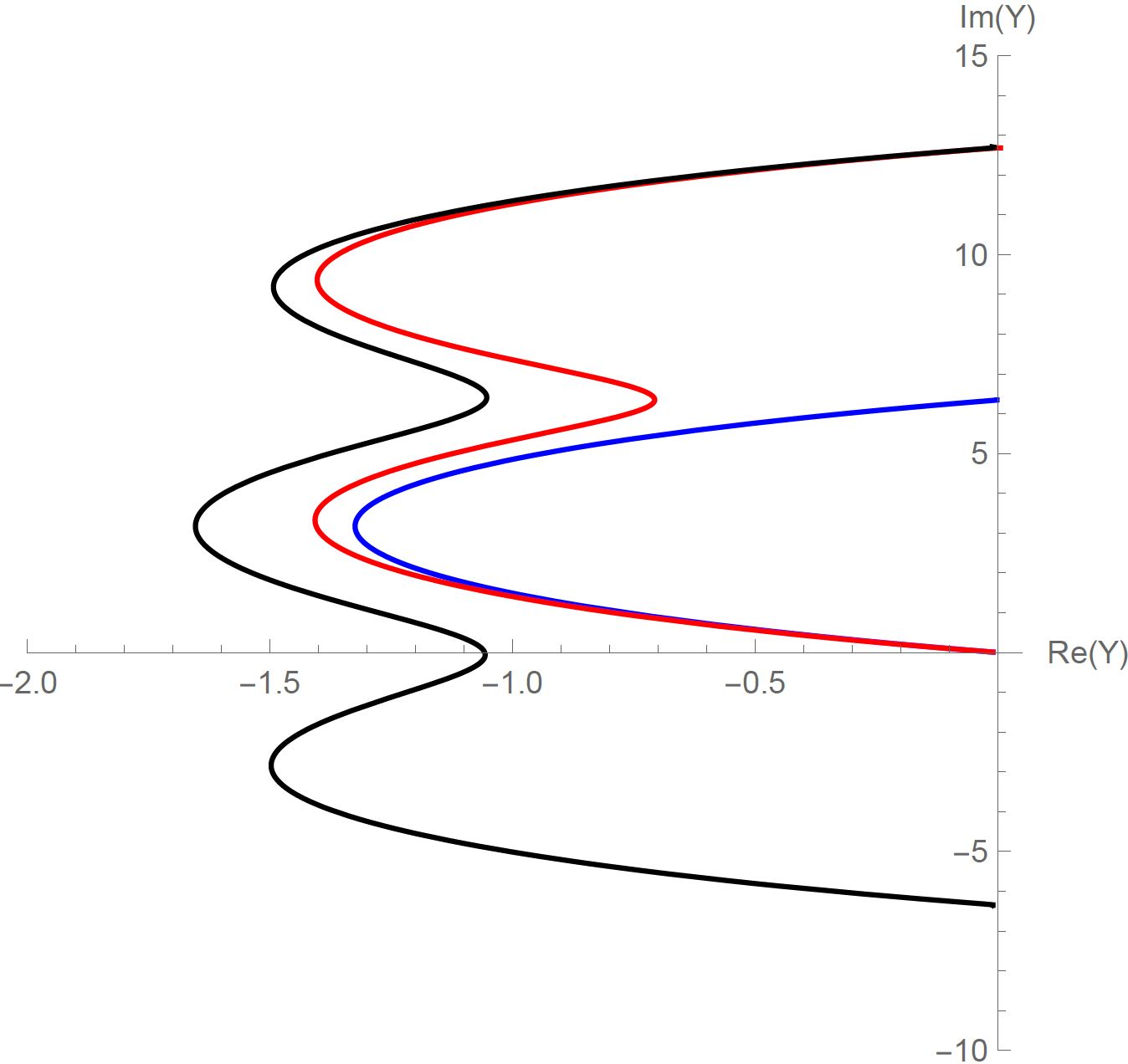}
    \caption{Some explicit soliton trajectories in the perturbed model with $\epsilon = 0.01$ are depicted. The trajectory in blue is a perturbation of the trajectory we identify as an elementary closed soliton of the original unperturbed model. The others should be regarded as single particle bound states which become  ``multiparticle states'' in the $\epsilon \to 0$ limit. 
    This picture is the  origin of the Fock spaces that play an important role in our subsequent discussion. In particular the 
    multiparticle states appearing in the $\epsilon\to 0$ limit  are governed by Fock space combinatorics. }
    \label{perturbedsolitons}
\end{figure}

\begin{figure}
    \centering
    \includegraphics[width=0.6\textwidth]{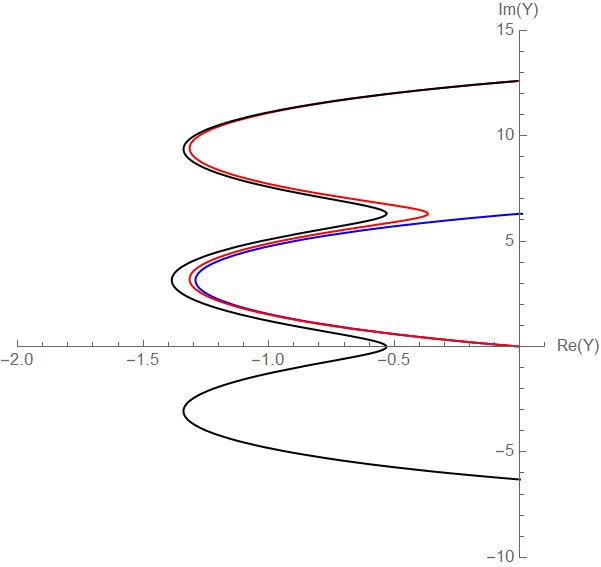}
    \caption{Some explicit soliton trajectories for the perturbed model with $\epsilon=0.003$. We see that as we make $\epsilon$ smaller, the trajectories between vacua that are more than one unit apart can be viewed as ``multiple gluings" of the closed soliton in the original unperturbed model.}
    \label{more perturbed solitons}
\end{figure}

In order to proceed with the calculation of the category of boundary conditions, we also need to discuss the notion of a stable $\zeta$-instanton. This is necessary in order to formulate the interior amplitude. For this we first need to discuss the notion of a stable cyclic fan of vacua and a stable cyclic fan of solitons. We will call a collection of vacua 
\begin{equation}
\{Y_{i_0}(\epsilon), \dots, Y_{i_n}(\epsilon)\}
\end{equation}
a stable cyclic fan of vacua if there is an $\epsilon>0$ such that for the central charges 
\begin{equation} 
\{z_{i_0 i_1}(\epsilon), z_{i_1 i_2}(\epsilon), \dots, z_{i_{n} i_0}(\epsilon) \} 
\end{equation}
the series in $\epsilon$ converges, and that for this $\epsilon$ the arguments of these central charges are clockwise ordered.
Moreover, this is stable in the sense that for each $0< \delta \leq \epsilon$ the central charges 
\begin{equation}
\{ z_{i_0 i_1}(\delta), z_{i_1 i_2}(\delta), \dots, z_{i_n i_0}(\delta) \}
\end{equation}
are also clockwise ordered. Given a stable cyclic fan, a stable fan of solitons is a collection 
\begin{equation}
(\phi_{i_0 i_1}(\epsilon), \dots, \phi_{i_n i_0}(\epsilon)) 
\end{equation}
of stable solitons for $W^{\epsilon}$. This collection of stable solitons defines a boundary value problem for the $\zeta$-instanton equation 
\begin{equation}
\frac{\partial \phi}{\partial \overline{z}} = \zeta \frac{\partial \overline{W^{\epsilon}}}{ \partial \overline{\phi}}
\end{equation}
for $W^{\epsilon}.$  As before, it will be called stable if the solution continues for each $0< \delta \leq \epsilon.$ 

\begin{figure}
    \centering
    \includegraphics[width=0.6\textwidth]{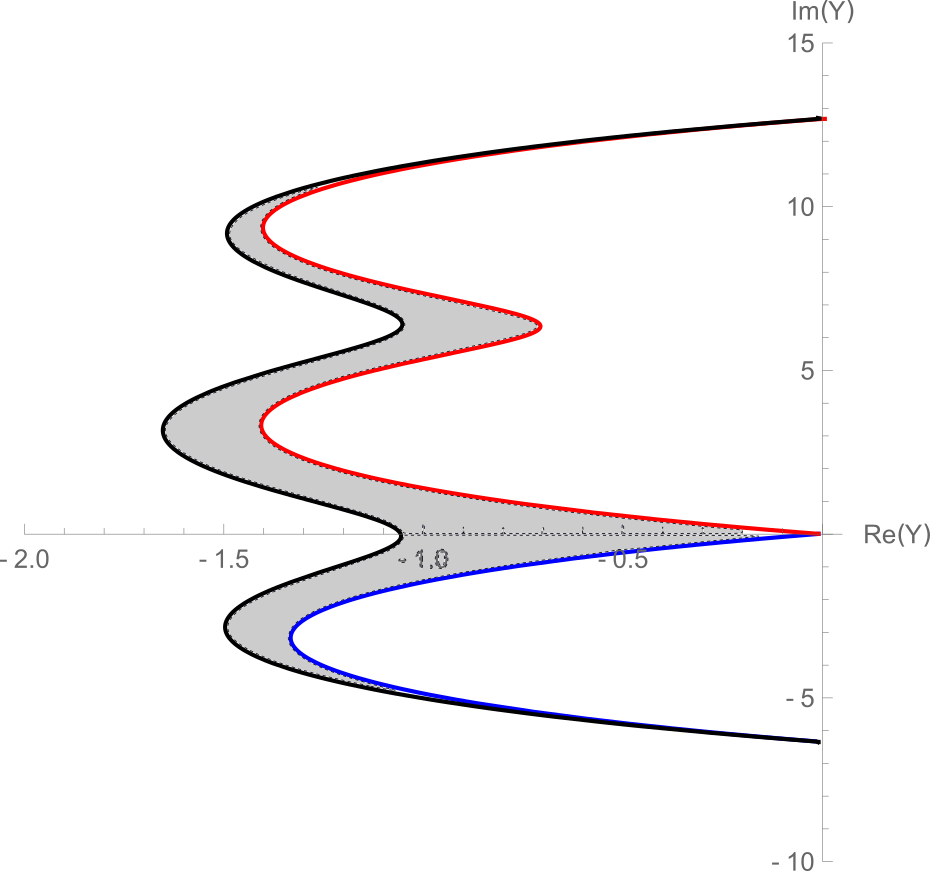}
    \caption{The image of a (stable) $\zeta$-instanton in the $Y$-plane for boundary conditions labelled by the triple $(k,l,m) = (2,0,-1)$ for $W^{\epsilon}$ with $\epsilon=0.01$. It contributes an element of the form $\beta_{(2,0,-1)}= \phi_{2,0} \otimes \phi_{0,-1} \otimes \phi_{-1,2}$ to the interior amplitude.}
    \label{zetatriangle}
\end{figure}

By the analogue of the conjectured Riemann mapping theorem for $\zeta$-instantons, there is a unique rigid $\zeta$-instanton for every triple of integers $k > l > m$. In other words, choosing $\epsilon$ small enough so that there is a triangle consisting of a $(k,l)$, an $(l,m)$ and an $(m,k)$ stable solitons, so that we may define the boundary value problem specified by fan boundary conditions for the solitons at this value of $\epsilon$ for the $\zeta$-instanton equation for $W^{\epsilon}.$ These form a triangle which may be filled in. See Figure \ref{zetatriangle} for an example. Thus we can write the formal expression: 
\be\label{eq:FreeChiralInteriorAmp}
\beta = \sum_{k> l > m} \phi_{k,l} \otimes \phi_{l,m} \otimes \phi_{m,k}.
\ee 
The expression is formal because, for each summand, we must choose an $\epsilon$ that is small enough to allow the triple of solitons to exist. 
No value of $\epsilon$ works uniformly for all triples.   Note that, thanks to \eqref{eq:Rkl-k-less-l} and \eqref{eq:Rkl-k-more-l},  two solitons in each triangle have degree $+1$ and the third one has degree $0$, so $\beta$ indeed carries degree $+2$, as it must. 

We have therefore finished specifying the analogue of the infrared data for this model: stable fans, stable solitons and stable instantons. 

We can now turn to the category of boundary conditions. Fix a half-plane $H \subset \mathbb{C}$. We will call a collection of vacua $(Y_{i_0}(\epsilon), \dots, Y_{i_n}(\epsilon))$ a stable half-fan if there is an $\epsilon > 0$ such that $(z_{i_0 i_1}(\epsilon), \dots, z_{i_{n-1} i_n}(\epsilon)) $ are all phase-ordered rays that lie in $H$. Moreover for any $0<\delta \leq \epsilon$ the corresponding rays for $\delta$ are still phase-ordered elements of $H$. The space $\widehat{R}_{k,l}$ is obtained by summing over stable half-fans in $H$ that start at $k$ and end at $l$, and assigning to it the tensor product of spaces of stable solitons.

We now choose $\zeta = -\text{i}$ and discuss the phase orderings of the central charges. 
For $\epsilon$ small we can say that  
\be\label{eq:znm-phaseordering}
z_{n,m}^\epsilon = 2\pi (n-m) + 2\pi^2 i \epsilon (n^2 - m^2)  \cong 2\pi (n-m)e^{i \pi \epsilon (n+m) } ~ . 
\ee 
For $n\not=m$ we note that $z_{n,m}$ is in the \underline{left}-half-plane for $n<m$ and in the \underline{right}-half-plane for $n>m$. 
Now it is useful to understand the phase ordering of two central charges of the form $z_{a,b}$ and $z_{b,c}$. When $a<b$ and $b<c$ so 
they are both in the \underline{left}-half-plane then for $\epsilon>0$, $z_{bc}$ has a (slightly) larger phase than $z_{ab}$.  The clockwise phase ordering is $z_{bc}$ then $z_{ab}$. On the other hand if $a>b$ and $b>c$ then they are both in the  \underline{right}-half-plane and now for $\epsilon>0$
the phase of $z_{ab}$ is (slightly) greater than $z_{bc}$, so the clockwise ordering is $z_{ab}$ then $z_{bc}$. If we change the sign of $\epsilon$, as we will in section \ref{subsec:ChangingSingEps}, then the above phase orderings are reversed.

Let us now choose  $H$ to be the \textbf{left}-half plane (LHP). Then  $z_{ij}\in H$ for $i<j$ so the ordering on thimbles is 
$\mathfrak{T}_i < \mathfrak{T}_j$   if the labels have the corresponding order $i<j$ as integers. 
For a given pair $i<j$ then the space $\widehat{R}^{\text{L}}_{ij}$ is given as follows (here the $\text{L}$ superscript denotes that these are the morphism spaces in the category associated to the left half-plane).  

\begin{figure}
    \centering
    \includegraphics[width=0.3\textwidth]{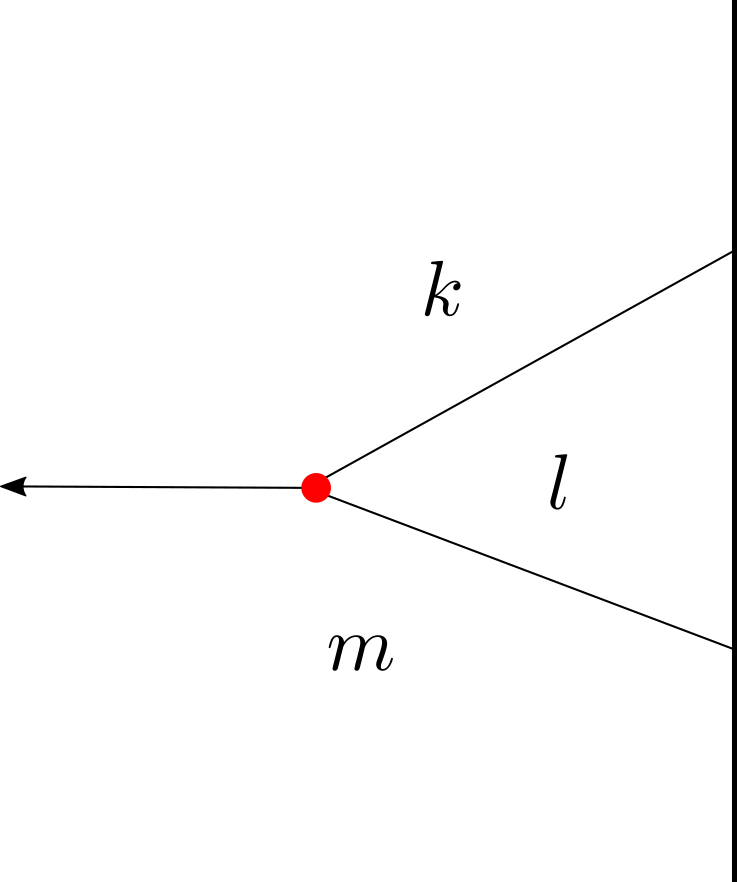}
    \caption{Web contributing to a non-trivial product in the category associated to the left half-plane. This gives the product in the polynomial algebra upon passing to coinvariants.}
    \label{leftproduct}
\end{figure}

It follows from the discussion under equation \eqref{eq:znm-phaseordering} that in the left half-plane the only LHP fan contributing 
to $\widehat{R}^{\text{L}}_{nm}$ has a single edge. Thus we simply have 
\begin{equation}
    \widehat{R}^{\text{L}}_{n,m} = R_{n,m} \cong \mathbb{C}
\end{equation} 
generated by a soliton $\phi_{n,m}(\epsilon)$ with an $\epsilon$ small enough.  The $A_{\infty}$-maps on this space are also easy to determine. As everything is concentrated in degree zero, the only possible non-trivial $A_{\infty}$ matrix element is of the form 
\begin{equation}
m_2: \widehat{R}^{\text{L}}_{k,l} \otimes \widehat{R}^{\text{L}}_{l,m} \rightarrow \widehat{R}^{\text{L}}_{k,m}
\end{equation}
for a triple $k<l<m$ which is determined as follows. For an $\epsilon$ small enough the critical values $k<l<m$ are such that the taut left half-plane web depicted in the Figure \ref{leftproduct}
exists and is stable. Moreover, as we discussed previously, there is indeed a unique stable $\zeta$-instanton for each such triple. This gives rise to the product 
\begin{equation}
    m_2(\phi_{k,l}(\epsilon), \phi_{l,m}(\epsilon)) = \phi_{k,m}(\epsilon).
\end{equation} This product is there for $\epsilon$ arbitrarily small (but positive), so we call it a stable product. This finishes the calculation of the category in the ``upstairs" theory on $\mathbb{C}$. Note that this category has an action of $\mathbb{Z}$ in a sense that is discussed in Section \ref{App:groupactions}. We indeed have that the vacuum set has a $\mathbb{Z}$-action, and there are natural isomorphisms 
\be
F_n: \widehat{R}_{k,l} \rightarrow \widehat{R}_{k+n, l+n}
\ee 
for each $n \in \mathbb{Z},$ which sends 
\be 
F_{n}(\phi_{k,l}(\epsilon)) = \phi_{k+n, l+n}(\epsilon)
\ee for an $\epsilon$ small enough that both solitons exist. Moreover, the product is preserved under this map. In order to obtain the category associated to the physical theory on $\mathbb{C}^*$, we must pass to the orbit category. 

The orbit category for the LHP simply has a single object $i$ corresponding to the single vacuum downstairs. By working with the formula given by fixing a reference element and summing over the orbits of the other one, the morphism space is given by 
\begin{equation}
\text{Hom}(\mathfrak{T}_i^{\text{L}}, \mathfrak{T}_i^{\text{L}}) \cong  \bigoplus_{n \geq 0} \widehat{R}_{0,n}.
\end{equation}
This is an infinite-dimensional space generated by $\phi_{0,n}$ for each $n \geq 0.$ Moreover the product in the orbit category is given as follows. Letting $x_n := \phi_{0,n}$ we simply have 
\begin{equation}
m_2^{\text{L}}(x_n, x_m) = x_{n+m}.
\end{equation}
We thus see that the orbit category associated to the left-half plane and phase $\zeta = -\text{i}$ is simply a category with a single object with endomorphism space given by a symmetric algebra in one variable: 
\be\label{eq:FreeChiralLHP-poseps}
\text{Hom}(\mathfrak{T}_i^{\text{L}}, \mathfrak{T}_i^{\text{L}}) \cong S^*(\mathbb{C}) ~ . 
\ee

It is also instructive to work out the category associated to the \textbf{right}-half plane and same phase $\zeta = -\text{i}.$ Here the induced ordering on the vacua is the opposite one from the integers that label them. Therefore the non-trivial spaces one has to determine consist of $\widehat{R}^{\text{R}}_{n,m}$ for $n>m.$ We note that for \textit{any} sequence of decreasing integers 
\be 
n > k_1 > \dots > k_{r} > m 
\ee 
there is an $\epsilon>0$ such that the fan $(z_{n,k_1}(\epsilon), \dots, z_{k_r m}(\epsilon))$ is a right-half fan. This simply follows from the fact that the corresponding critical values approximately lie on a parabola which is curved ``outwards" with respect to the relevant half-plane as long as $\epsilon>0$ (really $\text{Re}(\epsilon) > 0$). Thus the space $\widehat{R}_{n,m}^{\text{R}}$ consists of a summand for each such decreasing integer sequence 
\begin{equation}
\widehat{R}_{n,m}^{\text{R}} = \bigoplus_{n>k_1 > \dots > k_r > m} R_{n,k_1} \otimes \dots \otimes R_{k_r, m}. 
\end{equation} For each such sequence the corresponding summand is one-dimensional given by the tensor product of stable solitons 
\bea 
\phi_{n,k_1}(\epsilon) \otimes \dots \otimes \phi_{k_r, m}(\epsilon)
\eea 
for a small enough $\epsilon.$ Since $\phi_{a,b}(\epsilon)$ carries cohomological degree $1$ for each $a>b$ the summand associated to a sequence $n>k_1> \dots> k_r>m$ has cohomological degree $r+1$.

\begin{figure}
    \centering
    \includegraphics[width=0.85\textwidth]{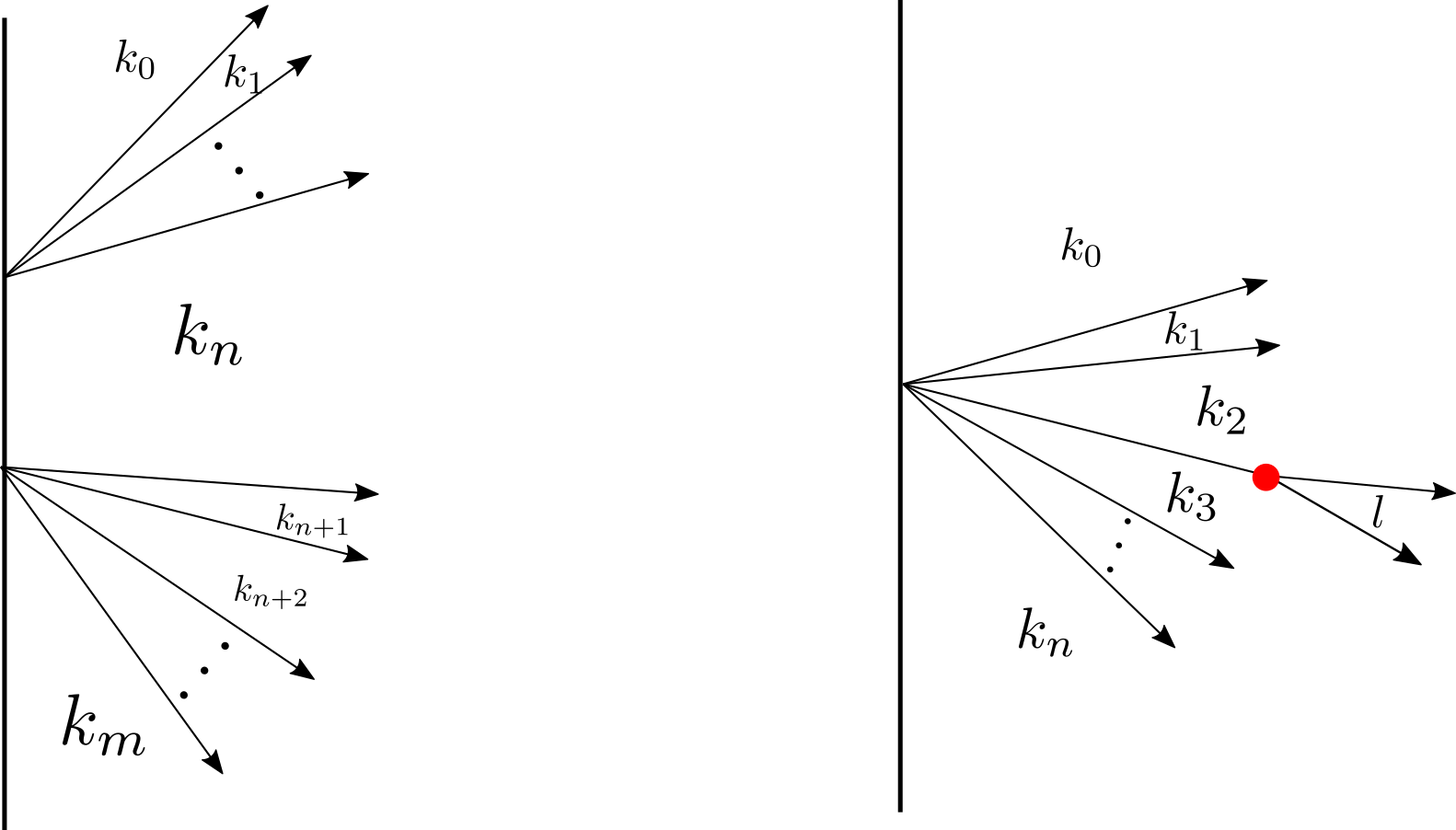}
    \caption{Left: a typical web that contributes to the product in the right half-plane web category. Right: a web that contributes to a differential.}
    \label{differential}
\end{figure}

Let us now determine the $A_{\infty}$-maps. It is not too difficult to see that there are only two kinds of \underline{taut} stable webs that can contribute: the first kind involves a single boundary vertex and a single bulk vertex, and the second kind involves two boundary vertices and no bulk vertex. Because the interior amplitude \eqref{eq:FreeChiralInteriorAmp} is trivalent only trivalent vertices can contribute. See 
Figure \ref{differential}.

A general element in the first class of webs consists of a boundary vertex associated to a sequence $k_0 > k_1 > \dots > k_r$. Moreover, it involves picking an edge associated to $(k_i, k_{i+1})$ and splitting it by inserting an integer $l_i$ such that $k_i > l_i > k_{i+1}$ in the middle, so that the bulk vertex is associated to the cyclic trivalent fan \be 
( l_i, k_{i+1}, k_{i} ).
\ee
(For models with more complicated interior amplitudes there will also be   splittings involving inserting more than one intermediate vacuum.)   By using the fact that we have a stable $\zeta$-instanton for each triple $k>l>m$, we see that this collection of webs gives rise to a differential of the form 
\begin{align}
\begin{split}
d^{\text{R}}(\phi_{k_0, k_1} \otimes \dots \otimes \phi_{k_{r-1}, k_r}) = \,\,\,\,\,\,\,\,\,\,\,\,\,\,\,\,\,\,\,\,\,\,\,\,\,\,\,\,\,\,\,\,\,\,\,\,\,\,\,\,\,\,\,\,\,\,\,\,\,\,\,\,\,\,\,\,\,\,\,\,\,\,\,\,\,\,,\,\,\,\,\,\,\,\,\,\,\,\,\,\,\,\,\,\, &\\  \sum_{\substack{0<i<r \\ k_i > l_i > k_{i+1}} } (-1)^i\phi_{k_0 k_1} \otimes \dots \otimes \phi_{k_i, l_i} \otimes \phi_{l_i, k_{i+1}} \otimes \dots \otimes \phi_{k_{r-1}, k_r}. \end{split} 
\end{align} In particular we have 
\begin{equation}
d^{\text{R}}(\phi_{n,m}) = \sum_{n>k>m} \phi_{n,k} \otimes \phi_{k,m}.
\end{equation}
One can explicitly check that this differential is nilpotent $d^2 = 0$.  (Recall \eqref{eq:Rkl-k-more-l}.) 

The second class of taut right half-plane webs leads to a product $m_2^{\text{R}}$ which simply says the multiplication of two elements associated to sequences $k_0 > \dots > k_r$ and $l_0> \dots > l_q$, supposing that $k_r> l_0$ is given by the element associated to the sequence $k_0 > \dots > l_q$: 
\begin{align}
    m_2^{\text{R}}(\phi_{k_0 k_1} \otimes \dots \otimes \phi_{k_{r-1}, k_r}\, , \, \phi_{l_0,l_1} \otimes \dots \otimes \phi_{l_{q-1}, l_q}) = \phi_{k_0, k_1} \otimes \dots \otimes \phi_{l_{q-1} l_q}.
\end{align}
The differential and the product are compatible so that the $A_{\infty}$-relations are satisfied. What we have discovered here is a 
special case of the cobar complex, discussed in detail below.

The cohomology of the differential $d^{\text{R}}$ is straightforward to compute: the chain complex $\widehat{R}_{k,l}$ for $k>l$ is simply \textit{acylic} provided $k > l+1$. 
\begin{equation}
H^*(\widehat{R}^{\text{R}}_{k,l}, d^{\text{R}}) = 0, \,\,\,\, k > l+1.
\end{equation}
This follows from the fact that one can write down a homotopy inverse to $d^{\text{R}}$ if $k> l+1$. The homotopy inverse 
\begin{equation}
\delta: \widehat{R}^{\text{R}}_{k,l} \rightarrow \widehat{R}^{\text{R}}_{k,l}
\end{equation} is a map of degree $-1$ given by summing over all ways to forget an element in the sequence 
\begin{equation}
\delta(\phi_{k_0, k_1} \otimes \dots \otimes \phi_{k_{r-1}, k_r}) = \sum_{1<j<r-1} \phi_{(k_0, \dots, \widehat{k_j}, \dots, k_r) }. 
\end{equation} It is then straightforward to check that 
\be 
\delta d + d \delta = 1.
\ee
Thus the only morphism spaces with non-vanishing cohomology are of the form $\widehat{R}^{\text{R}}_{k,k}$ which is one-dimensional in degree zero \begin{equation}
\widehat{R}^{\text{R}}_{k,k} \cong \mathbb{C},
\end{equation}
and $\widehat{R}^{\text{R}}_{k+1,k}$ which is one-dimensional in degree $+1,$ 
\begin{equation}
\widehat{R}^{\text{R}}_{k+1,k} \cong \mathbb{C}^{[1]}.
\end{equation}

The category associated to the right half-plane again has an action of the deck group $\mathbb{Z}$ by autofunctors. Let us pass to the orbit category with respect to this action. It consists of again a single object $\mathfrak{T}_i^{\text{R}}$ associated to the unique vacuum. The endomorphism algebra of this object is given by 
\begin{equation}
\text{Hom}(\mft_{i}^{\text{R}}, \mft_{i}^{\text{R}}) = \bigoplus_{n \geq 0} \widehat{R}_{n,0},
\end{equation}
which has the structure of a differential graded algebra.  The cohomology calculation in the previous paragraph tells us that only the summands associated to $n=0$ and $n=1$ are non-trivial in cohomology. By taking into account the product, this amounts to the statement that the endomorphism algebra for the unique object associated to the right half-plane is quasi-isomorphic to the exterior algebra in one variable
\begin{equation}\label{eq:FreeChiralRHP-poseps}
\text{Hom}(\mathfrak{T}^{\text{R}}_i, \mathfrak{T}^{\text{R}}_i) \simeq \Lambda^{*}(\mathbb{C}).
\end{equation}

The boundary algebras of observables \eqref{eq:FreeChiralLHP-poseps} and \eqref{eq:FreeChiralRHP-poseps}  associated to the left and right half-planes, respectively  are thus related by Koszul duality. The remaining examples of this section will exhibit the same phenomenon. 
In section \ref{koszul} we will prove that this holds more generally.

\subsection{Changing the sign of $\epsilon$}\label{subsec:ChangingSingEps}

At this point it is natural to ask how sensitive our results are to the choice of perturbing superpotential.  We now show that had we chosen to perturb with $\epsilon f(Y)$ where $\epsilon<0$ (really $\text{Re}(\epsilon)<0$) then the stable soliton spectrum
in fact changes rather dramatically. Nevertheless, the final result for the left and right boundary algebras will still be the same up to quasi-isomorphism!

The basic reason that the soliton spectrum changes dramatically is that if we continue $\epsilon$ across the wall in the $\epsilon$-plane
separating $\text{Re} \, \epsilon > 0$ from  $\text{Re} \, \epsilon < 0$ then the first order configuration of critical values switches from lying on a parabola with ``positive" curvature to "negative" curvature. As a result the collection of stable fans, solitons and instantons jumps discontinuously. In particular, the calculation we performed for the soliton spectrum that appears in Theorem 4.1, which involved looking at the intersection sets of perturbed right thimbles, no longer computes the soliton spectrum, as the path that connects the vacua by going along the right thimbles via an intersection point, when projected to the $W$-plane is no longer homotopic to the straight path between critical values. Instead we have the following statement.

\begin{theorem}
    The spectrum of stable solitons for the perturbed model 
    \begin{equation}
    W^{-\epsilon}(Y) = Y - e^{Y}- \epsilon \frac{Y^2}{2} 
    \end{equation} 
    with $\epsilon>0$ consists of a unique soliton for each pair of perturbed vacua that are one unit apart and is trivial otherwise. \bea R_{k+1,k} &\cong& \mathbb{C}^{[1]}, \\ R_{k, k+1} &\cong& \mathbb{C}.\eea  Moreover, there are no stable $\zeta$-instantons.
\end{theorem}

\begin{myproof}
    This follows from a similar calculation to the one that was performed for the case $\epsilon>0$. The crucial part is the second part of Lemma 4.2, for $J_{kl}(\theta,\theta')$ which states that the left thimbles of the unperturbed theory intersect non-trivially if they are at most one unit apart. We can then perform a small perturbation so that this remains true in a given open set. After the perturbation we may identify the intersection points of thimbles as solitons, proving the claim above. As there are no non-trivial triangles anymore (since a triangle would require a soliton between vacua that are perturbations of vacua which are at least two units apart), there are no stable rigid $\zeta$-instantons.
\end{myproof}

Let us work out the category of boundary conditions with this perturbation. We choose $H$ to be the right half-plane and $\zeta = -\text{i}.$ The vacuum weights for $\epsilon<0$ are now such that the only half-plane fans consist of single rays ${k,l}$ with $k>l,$ so that we now have 
\begin{equation}
    \widehat{R}_{k,l} = R_{k,l}  
\end{equation} which means that 
\begin{equation}
    \widehat{R}_{k,l} \cong \begin{cases}
        \mathbb{C} &  k=l \\
        \mathbb{C}^{[1]} & k = l+1 \\
        0 & \text{ otherwise}.
    \end{cases}
\end{equation} 
The generator of $\widehat{R}_{k,k}$ acts as the identity, and the map 
\be 
m_2: \widehat{R}_{l+2,l+1} \otimes \widehat{R}_{l+1, l} \rightarrow \widehat{R}_{l+2,l}
\ee 
vanishes as the target is simply zero-dimensional. We thus see that we reproduce $\widehat{R}^{\text{R}}_{k,l}$ up to quasi-isomorphism. Upon passing to the orbit category this again reproduces the exterior algebra in one variable as the endomorphism algebra of the unique generating object. 

We thus see that there is an $A_{\infty}$-equivalence between the right half-plane categories for $\epsilon>0$ and $\epsilon<0$. In the former case the calculation was more elaborate and relied on the existence of non-trivial $\zeta$-instantons. In the latter we reproduced the final answer in a more direct fashion. 

Finally, for the LHP and $\zeta=-i$ since 
\be\label{eq:z-freechir-minuspert}
z^{-\epsilon}_{ab} \cong  2\pi(a-b) e^{-i \pi \epsilon (a+b)} \ee there are many LHP fans based on 
sequences of vacua with $k_1> k_2> \cdots > k_n$, but since the only nonvanishing space $R_{a,b}$ with $a<b$  has $b=a+1$ we have
\be 
{\rm Hom}(\mathfrak{T}^{L,-},\mathfrak{T}^{L,-})  \cong \bigoplus_{n\geq 0 } \hat R_{0,n}
\ee
in the quotient category. (The superscript minus reminds us that we take the minus perturbation.)  We can identify 
\be
\hat R_{0,n} \cong R_{0,1}\otimes R_{1,2} \cdots \otimes R_{n-1,n} \cong \mathbb{C} x^n 
\ee
where $x \sim \phi_{0,1}$. 
The multiplication is induced by taut webs with two boundary vertices and no bulk vertices. The differential is zero. Once again 
we get a more direct derivation of the symmetric algebra $S^*\mathbb{C}$.

The entire discussion is summarized in table \ref{table} below.  

\subsection{Multiple Free Chirals}\label{subsec:MultipleChirals}

We finally include a brief discussion of multiple free chirals. The model we are now interested in is the Landau-Ginzburg model mirror to the free theory of $N$ free chiral fields with standard kinetic term and a single twisted mass. More precisely the free theory with target space $\mathbb{C}^N$ has the maximal abelian group of isometries given by $A = U(1)^N$, with corresponding complexified Lie algebra $\mathfrak{a}\otimes \mathbb{C} \cong \mathbb{C}^{N}$ and we pick the twisted mass to be the element in the dual space given by 
\begin{equation}
\vec{\widetilde{m}} = (\widetilde{m}, \dots, \widetilde{m}) 
\end{equation}
for some $\widetilde{m} \neq 0.$ We choose the same mass in all entries in order to keep the 
Deck group of rank one when passing to the cover. 
Letting $2\pi i m = \widetilde{m},$ this model is mirror dual to the Landau-Ginzburg model with target space $X=(\mathbb{C}^*)^N$ and Landau-Ginzburg one-form 
\begin{equation}
\alpha = \sum_{i=1}^N \Big(\frac{m}{\phi_i} -1 \Big) \text{d}\phi_i.    
\end{equation}
The model has a single Morse zero located at 
\begin{equation}
\phi_1 = \dots = \phi_N = m.
\end{equation}

Let us pass to the minimal cover where $\alpha=dW $ and $W$ is single-valued as follows.  $H_1((\mathbb{C}^*)^N, \mathbb{Z})$ is a rank $N$ free abelian group generated by $\gamma_1, \dots, \gamma_N$ where $\gamma_i$ is the cycle in which we wind around once in the counter-clockwise direction around the origin in the $\phi_i$-plane with all other coordinates kept at some constant value, and the central charge homomorphism $Z$ maps this to the complex numbers via 
\be 
Z(\gamma) = \int_{\gamma} \alpha.
\ee 
The kernel of $Z$ is non-trivial and given by elements 
\begin{equation}
a_1 \gamma_1 + \dots + a_N \gamma_N
\end{equation}
where $a_i \in \mathbb{Z}$ are integers that satisfy 
\begin{equation}
\sum_{i=1}^N a_i = 0.
\end{equation}
$\text{ker}Z$ is therefore of rank $N-1$ with one possible set of generators being given by 
\begin{equation}
c_i = \gamma_{i}- \gamma_{i+1},
\end{equation}
for $i=1, \dots, N-1.$ The universal cover is $\mathbb{C}^N$, with standard  coordinates $(Y_1, \dots, Y_N)$ such that   $\phi_i = e^{Y_i}$. 
The   $n$th generator $\gamma_i$ of the deck group $H_1((\mathbb{C}^*)^N, \mathbb{Z})$ acts via 
\begin{equation}
\gamma_n:(Y_1, \dots, Y_n, \dots, Y_N) \rightarrow (Y_1, \dots, Y_n + 2\pi \text{i}, \dots, Y_N).
\end{equation}
The minimal cover $\widetilde{X}$ can be identified with the quotient  space $\mathbb{C}^N/\mathbb{Z}^{N-1}$ consisting of equivalence classes $(Y_1, \dots, Y_N)$ of complex numbers where we make the identification 
\be\label{eq:Z-N-1-tmn}
(Y_1, \dots, Y_N) \sim (Y_1 + 2\pi \text{i} a_1, \dots, Y_N + 2\pi \text{i} a_N)
\ee
where 
\be 
a_1+ \dots + a_N=0. 
\ee 

\bigskip
\begin{remark} 
The space $\widetilde{X}$ is biholomorphic to $(\mathbb{C}^*)^{N-1} \times \mathbb{C}$. One choice of isomorphism is given by defining
$U_i = Y_i - Y_{i+1}$, $1\leq i \leq N-1$ and $V = Y_1 + \cdots + Y_N$. Then $V$ is invariant under \eqref{eq:Z-N-1-tmn}.  $U_i$ is not invariant but 
\be 
w_i = \exp[ \sum_{j=1}^{N-1} M_{i,j} U_j ]\in \mathbb{C}^*, \,\,\,\, \,\,\, 1\leq i \leq (N-1)
\ee is invariant, where $M$ is a matrix such that  
\be 
\begin{pmatrix} Y_1 \\  Y_2 \\ \vdots  \\ Y_N \\ \end{pmatrix}= M   \begin{pmatrix} U_1 \\ \vdots \\ U_{N-1} \\ V \\  \end{pmatrix}.
\ee
Explicitly, 
\be 
M= N^{-1} \begin{cases}   (N-j)  & i \leq j \qquad {\rm and} \qquad j \leq (N-1) \\ 
-j &  i > j \qquad {\rm and} \qquad j \leq (N-1) \\ 
1 &  j=N. \\  \end{cases}
\ee
Thus the superpotential is 
\be
W = m V - e^{\frac{1}{N} V}  \sum_{i=1}^{N-1} w_i - e^{\frac{1}{N} V}  (\prod_{i=1}^{N-1} w_i )^{-1}
\ee
The matrix $M$ is closely related to the weights of $SU(N)$. Indeed the theory has a hidden $SU(N)$ symmetry which is manifested in the mirror dual consisting of $N$ free chirals with a potential energy $\sim \vert m \vert^2 \sum_{i=1}^N \vert \varphi_i \vert^2 $. We expect the interior amplitude to be $SU(N)$ invariant. \end{remark}

\bigskip
The critical points on $\mathbb{C}^N$ are given by the set 
\be 
Y_{(k_1, \dots, k_N)} = (\text{log}m + 2\pi i k_1, \dots, \text{log}m + 2\pi i k_N)
.\ee Two critical points $Y_{\vec{k}}$ and $Y_{\vec{l}}$ project to the same point on $\widetilde{X}$ 
\be
\pi(Y_{(k_1, \dots,k_N)}) = \pi(Y_{(l_1, \dots, l_N)}) 
\ee if and only if 
\be 
\sum_{i=1}^N k_i = \sum_{i=1}^N l_i.
\ee
 Thus the projection of the critical points to $\widetilde{X}$ can be identified as a $\mathbb{Z}$-torsor so that on $\widetilde{X}$ the critical points $x_k$ are uniquely labeled by an integer $k \in \mathbb{Z}$, corresponding to, for instance,
 \be 
 Y_{(k,0,\dots, 0)} := (\text{log} m + 2\pi i k, \text{log}m , \dots \text{log}m).
 \ee 
 The corresponding critical values are given by 
 \be 
 W(x_k)   = N m (\text{log}m-1) + 2\pi i k m,
 \ee 
 which all lie on a line as expected.

Before perturbing and working out the soliton spectrum, it will be useful for us to work out the intersection pattern of (left) thimbles for this superpotential. We are interested in the thimble $L_k(e^{i\theta})$ supported at $x_k$ in $\widetilde{X}$ and the intersection sets $L_k(e^{i\theta}) \cap L_l(e^{-i\theta'})$ as oriented manifolds for $\theta, \theta' \in (0, \frac{\pi}{2}).$ Our strategy to do this will be familiar: we simply lift a thimble from $\widetilde{X}$ to the universal cover, look at the intersection pattern there and then project down.

Set $m=1.$

The gradient flow equation on the universal cover separates into $N$ independent flows so the thimble $L_{(k_1, \dots, k_N)}(\theta)$ associated to the point $Y_{(k_1, \dots, k_N)}$ on the universal cover $\mathbb{C}^N$ is given simply by 
\begin{equation}
L_{(k_1, \dots, k_N)}(\theta) = L_{k_1}(\theta) \times \dots \times L_{k_N}(\theta) \subset \mathbb{C} \times \dots \times \mathbb{C}
\end{equation}
where $L_k(\theta)$ is the thimble worked out in the previous section on $\mathbb{C}$ going through $Y_k = 2\pi i k$. We therefore have 
\be 
L_{(k_1, \dots, k_N)} (\theta) \cap L_{(l_1, \dots, l_N)}(-\theta') = \prod_{i=1}^N L_{k_i}(\theta) \cap L_{l_i}(-\theta').
\ee

For instance if $N=2$ and $\theta,\theta' \in (0, \frac{\pi}{2})$  we have that the non-trivial intersecting thimbles are 
\be L_{(n,m)}(\theta) \cap L_{(n,m)}(-\theta') = \{Y_{(n,m)}\} \ee 
which is a set consisting of just the critical point $Y_{n,m}$, (which is positively oriented as usual) and moreover 
\begin{align}
L_{(n,m)}(\theta) \cap L_{(n+1,m)}(-\theta'), \\ L_{(n,m)}(\theta) \cap L_{(n,m+1)}(-\theta'), \\ L_{(n,m)}(\theta) \cap L_{(n+1, m+1)}(-\theta') \end{align}
are all sets which consist of a single point. The first two intersection points carry a negative orientation whereas the third one is positively oriented. All other intersection sets are empty.

Thus we see that $L_{(0,\dots,0)}(\theta)$ intersects non-trivially with $2^N$ thimbles which are associated to the vacua $(j_1, \dots, j_N)$ where $j_i \in \{0,1\}$: the general formula is thus 
\begin{equation}\label{eq:MultiLeftIntersect}
  \vert   L_{(k_1, \dots, k_N)}(\theta) \cap L_{(k_1+j_1, \dots, k_N + j_N)}(-\theta') \vert 
  = \begin{cases}  1  & \text{ if } j_i \in \{0,1 \}, \\ 0  & \text{ otherwise } \end{cases}  
\end{equation} 
Moreover, the orientation of the intersection point between the two thimbles above is given by $(-1)^{j_1 + \dots + j_N}.$

Let  $\mathcal{L}_k(\theta)$ be the left thimble on $\tilde X$ through $x_k$, and let 
\be
\pi: \mathbb{C}^{N} \rightarrow \widetilde{X} 
\ee
be the projection. Then we have 
\be\label{eq:ProjectMultiLeft}
\pi(L_{(k_1, \dots, k_N)}(\theta)) = \mathcal{L}_{k_1 + \dots + k_N}(\theta)
\ee 
Since $\pi$ is a covering map we can determine the intersection sets $\mathcal{L}_k(\theta) \cap \mathcal{L}_l(-\theta')$ of thimbles on $\widetilde{X}$ by studying the intersections on the cover. In this way we see that the cardinality of 
\be\label{eq:CurlyLIntersect}
\mathcal{L}_{0}(\theta) \cap \mathcal{L}_{k}(-\theta')
\ee 
is given by counting  a set of points all of which carry an orientation of $(-1)^k$. The points can be enumerated by collections of $N$ numbers drawn from $\{0,1\}$ whose total sum is $k$. Thus the intersection set will be nonzero for $k=0,1 \dots, N$ with the intersection number given by the binomial coefficient \be \mathcal{L}_{0}(\theta) \circ \mathcal{L}_{k}(-\theta') = (-1)^k\binom{N}{k}.\ee 

Let us now perturb the superpotential by adding a perturbation function 
\be 
W \rightarrow W - \epsilon f(V)
\ee 
where as before 
\be
V = \frac{1}{N}(Y_1+ \dots + Y_N) 
\ee 
and 
\be \label{perturbationfunction}
f(V) = \frac{1}{2} V^2 
\ee 
We will only write out a few details for $\text{Re}(\epsilon)>0$. Choosing $\zeta = - \text{i}$, the analog of equation \eqref{eq:z-freechir-minuspert}
is now 
\begin{equation}
z^{-\epsilon}_{a,b} \cong 2\pi (a-b) e^{-\text{i} \pi \epsilon (a+b)/N^2}
\end{equation}
so that all the phase-ordering considerations of the $N=1$ case carry over unchanged.

Now the discussion before applies, in that the spectrum of stable solitons can be read off from the intersection numbers of unperturbed thimbles. This allows us to conclude that the spectrum of stable solitons for the model perturbed as above is given by 
\be
R_{k,l} \cong (\mathbb{C}^{\binom{N}{k-l}})^{[k-l]} \cong \Lambda^{k-l}(V) 
\ee
for $k\geq l$ where $V$ is an $N$-dimensional vector space carrying cohomological degree $+1.$ The CP conjugate of this space is given by 
\be 
R_{k,l} \cong (\Lambda^{l-k}(V))^{\vee}[-1]
\ee 
for $k<l.$

We finally conjecture that the interior amplitude of the model is given by finitely many triangles as follows. 
\footnote{Again this is an interesting statement in the theory of partial differential equations. It would be quite interesting 
to have a rigorous proof of this statement.}
For each triple $k,l,m$ which determines a stable cyclic fan $(x_k, x_l, x_m)$ where $k>m>l$, the corresponding amplitude 
\be 
\beta_{k,l,m} \in R_{k,l} \otimes R_{l,m} \otimes R_{m,k} 
\ee 
is given as follows. The space in which it lives is 
\be 
\Lambda^{(k-l)}(V) \otimes \big(\Lambda^{(m-l)}(V) \big)^{\vee}[-1] \otimes \big(\Lambda^{(k-m)}(V)^{\vee} \big)[-1],
\ee 
so specifying an element $\beta_{k,l,m}$ of degree +2 is the same as specifying a map 
\be 
\Lambda^{(k-m)}(V) \otimes \Lambda^{(m-l)}(V) \rightarrow \Lambda^{(k-l)}(V) 
\ee 
of degree zero. We take it to be given by the map which coincides with the multiplication map of elements in the exterior algebra, and claim that this is precisely the element determined by counting $\zeta$-instantons with trivalent boundary conditions dictated by the triple $(k,l,m)$. 
Verifying that this is indeed the case presents an interesting challenge in the theory of partial differential equations. 

We have thus finished specifying the spectrum of stable solitons and instantons for the mirror LG model to $N$ chiral superfields. We may now work out the category of thimbles for $H$ being the right half-plane and $\zeta = -\text{i}$. Again, the ordering induced on the vacua is opposite the natural ordering on the integers, and since the perturbation with a negative sign is such that the only half-plane fans carry a single ray, so that we have 
\begin{equation}
    \widehat{R}_{k,l} \cong \Lambda^{k-l}(V)
\end{equation} for $k>l$. The only non-trivial web is of the form appearing in the bottom left of Figure \ref{koszulidea} with $n=2$.
and involves the trivalent amplitude discussed above occupying the bulk vertex. This gives rise to the only non-trivial matrix elements which are of the form $m_2: \widehat{R}_{k,l} \otimes \widehat{R}_{l,m} \rightarrow \widehat{R}_{k,m} $ and coincide with multiplication in the exterior algebra.

Passing to the orbit category we thus find a single object $\mathfrak{T}_i^{\text{R}}$ with morphism space given by \begin{equation}
    \text{Hom}(\mft_i^{\text{R}}, \mft_i^{\text{R}}) = \Lambda^{*}(V)
\end{equation} where $V$ space carries cohomological degree $+1.$ Thus the endomorphism algebra is the exterior algebra in $N$ variables. 

We leave it to the reader to work out the category associated to the left-half plane and check that the left and right boundary algebras are, up to quasi-isomorphism, Koszul dual. In particular: 
\be \text{Hom}(\mft_i^{\text{L}}, \mft_i^{\text{L}}) \simeq S^*(V^{\vee}[1]), \ee so that the endomorphism algebra consists of the symmetric algebra in $N$ variables (all of which carry vanishing cohomological degree).
In order to get a feeling for the kind of computation involved, let us work out the charge 2 part (namely the morphism space between vacua that are two units apart) of the endomorphism algebra of the left half-plane. The space $\widehat{R}_{0,2}$ is given by two summands, one coming from a left half-plane fan with a single arrow parallel to $z_{0,2}^{-\epsilon}$, and the other coming from the half-plane fan with arrows $(z_{0,1}^{-\epsilon}, z_{1,2}^{-\epsilon})$ so that \be \widehat{R}_{0,2} = R_{0,2} \oplus ( R_{0,1}\otimes R_{1,2}) .\ee As a vector space we can write 
\be \widehat{R}_{0,2} \cong \big(\Lambda^{2}(V^\vee) \big)[1] \oplus (V^{\vee}[1])^{\otimes 2} \ee 
so that the first component is an $\binom{N}{2}$-dimensional vector space carrying cohomological degree $-1$ (recall that $V$ carries degree $1$, so the second exterior power of its dual carries degree $-2$, then we perform an overall shift by $+1$) whereas the second component is an $N^2$-dimensional vector space in cohomological degree $0$. The space $\widehat{R}_{0,2}$ may be given a basis by giving a basis of the form
\be 
\{f_{(i,j)} \mid i < j, \; i, j \in \{1, \dots, N \} \} ~ , 
\ee 
with each element carrying degree $-1$, for the first summand, along with a basis of the form 
\be
\{f_i \otimes f_j \mid i, j = 1, \dots, N \},
\ee
with each element carrying degree $0$ for the second summand. The interior amplitude component $\beta_{(2,0,1)}$ discussed previously, when inserted into the bulk vertex of the taut web depicted in Figure \ref{koszulidea} with  $n=2$ \footnote{with $(k_2,k_1,k_0) = (0,1,2)$} gives rise to the differential 
\begin{align} \label{diffcharge2}
d^{L}(f_{(i,j)}) = f_i \otimes f_j - f_j \otimes f_i.
\end{align}
We thus arrive at a two-term chain complex 
\begin{align}
\begin{CD}
0 @>>> \mathbb{C}^{\binom{N}{2}} @>d^L>> \mathbb{C}^{N^2} @>>> 0 
\end{CD}
\end{align}
where $d^L$ is given as above. Since the map $d^L: \mathbb{C}^{\binom{N}{2}} \rightarrow \mathbb{C}^{N^2}$ is injective, the cohomology of the above complex lies in degree zero and is of dimension \be N^2 - \frac{N(N-1)}{2} = \frac{N(N+1)}{2}.\ee This is precisely the dimension of $S^2(V^{\vee}[1]),$ the charge two component of the symmetric algebra of the $N$-dimensional vector space $V^{\vee}[1]$. Moreover, the form of the differential \eqref{diffcharge2} tells us that the symmetric algebra relations, namely the commutativity of $f_i$ and $f_j$ hold in cohomology. More generally, as we will discuss below, the left half-plane recovers the cobar complex of $\Lambda^{*}(V)$ which is well-known to be quasi-isomorphic to $S^*(V^{\vee}[1])$ (see for instance the classic paper of Priddy \cite{priddykoszul}). What we have worked out above is a special case of this quasi-isomorphism when the charge is restricted to be 2. 

\begin{remark}
    As we mentioned above the Landau-Ginzburg model we are discussing in the present section is the mirror dual to the theory of $N$ chiral superfields $(\Phi^1, \dots, \Phi^N)$ with a non-trivial twisted mass associated to the $U(1)$-action \be (\Phi^1, \dots, \Phi^N) \to (e^{i \theta} \Phi^1, \dots, e^{i \theta} \Phi^N). \ee The category of boundary conditions of a variety $X$ in the presence of a twisted mass associated to a $G$-action is given as follows. We consider the $G$-equivariant derived category of $X$, which carries an action of the dual group $G^{\vee}$. The category of boundary conditions is then obtained by passing to the orbit category with respect to $G^{\vee}.$ For $X= \mathbb{C}^N$ with the $U(1)$-action as above, both the structure sheaf and the skyscraper sheaf are objects of the equivariant derived category and upon passing to the orbit category their self morphism spaces are precisely given by the symmetric algebra and exterior algebras of $\mathbb{C}^N$, respectively. Thus we see that the thimble of the Landau-Ginzburg model for the right half-plane is mirror dual to the skyscraper sheaf at the origin whereas the left half-plane thimble is mirror to the structure sheaf. 
\end{remark}

It is very interesting to consider how the above considerations are altered if we change the sign of $\text{Re}(\epsilon)$. 
As in the previous section we expect equivalent results for the other sign of $Re(\epsilon)$, although the details will be more complicated. 
We now make some brief remarks, leaving details to the reader. 
One key step in the computation is the analog of the intersection computation of equations \eqref{eq:MultiLeftIntersect} to 
\eqref{eq:CurlyLIntersect}. Now we find that 
\begin{equation}\label{eq:MultiRightIntersect}
    R_{(k_1, \dots, k_N)}(\theta) \cap R_{(k_1-j_1, \dots, k_N - j_N)}(-\theta') = \begin{cases}  \{\text{pt} \} & \text{ if } j_i \geq 0 , \\ \emptyset & \text{ if } j_i < 0 \end{cases}  
\end{equation} 
Moreover, the orientation of the intersection point between the two thimbles above is given by $+1$.
Once again we consider the projection  $\pi: \mathbb{C}^{N} \rightarrow \widetilde{X} $ of $R_{(k_1, \dots, k_N)}(\theta)$ 
to $\mathcal{R}_{k}$ with $k=k_1 + \cdots k_N$. Now the cardinality of the intersection 
\be\label{eq:CurlyLIntersect}
\mathcal{R}_{0}(\theta) \cap \mathcal{R}_{k}(-\theta')
\ee 
is given by computing the number of ways we can add $N$ nonnegative integers to get $k$. This is, manifestly, the degeneracy of the particle number $k$ subspace of a bosonic Fock space for $N$ bosonic oscillators, so that the soliton spaces in the perturbed model now form bosonic Fock spaces. Moreover a natural choice of interior amplitude along with enumerating the possible half-plane webs allows us to derive that the left half-plane web algebra coincides with the symmetric algebra of $\mathbb{C}^N$, whereas the right half-plane will give the cobar complex of $\mathbb{S}^*(\mathbb{C}^N)$. The latter is well-known to be quasi-isomorphic to $\Lambda^*(\mathbb{C}^N)$, thus giving us agreement with the other sign of the perturbation.

The calculations of this section may by succinctly summarized in a table.
\begin{center} \label{table}

\begin{tabular}{|c|c|c|} 
\hline
\, & $W+\epsilon f(V)$ & $W-\epsilon f(V)$ \\
\hline
LHP & $S^{*}(\mathbb{C}^N)$ & $\text{Cobar} \big(\Lambda^{*}(\mathbb{C}^N) \big)$\\
\hline
RHP & $\text{Cobar} \big(S^*(\mathbb{C}^N) \big)$ & $\Lambda^{*}(\mathbb{C}^N)$ \\

\hline
\end{tabular}

\end{center}

Here $\text{Re}(\epsilon)>0$ and $f(V)$ is the perturbation function given in \eqref{perturbationfunction}. Exchanging the rows (i.e interchanging the half-plane) for a given perturbation gives us Koszul dual algebras whereas interchanging the columns (i.e interchanging the sign of $\epsilon$) gives us quasi-isomorphic algebras.

\section{General Remarks On Rank One Models With A Single Vacuum } \label{koszul} 

In the previous section we found that in the simplest Landau-Ginzburg model with a non-trivial twisted mass and unique vacuum, the boundary algebras computed via the web formalism associated to opposite half-planes are Koszul dual to each other. Moreover in section 
\ref{subsec:MultipleChirals} we saw the appearance of certain ``Fock spaces" built from spaces of closed solitons. In this section we remark on the generality of these results. We begin by showing the Koszul duality between the left and right half-plane web algebras holds for a general rank one model with a unique vacuum  
and is independent of the details of the (stable) soliton spectrum. 
\footnote{In fact we expect an appropriate ``categorical" version of Koszul duality between $A_{\infty}$-categories of opposite half-planes to hold in complete generality. Namely for an arbitrary Landau-Ginzburg model defined by a one-form $\alpha$, and indeed a general $\mathcal{N}=(2,2)$ quantum field theory satisfying the assumptions we have mentioned above. Statements somewhat along these lines are familiar from the literature on derived categories of sheaves, see for instance \cite{Auroux:2008xno}. For more on the categorical version of Koszul duality see Appendix \ref{multivackoszul}.}
We do this by observing that if $A$ is the $A_{\infty}$-web algebra for a particular half-plane, the web formalism reproduces the cobar complex of $A$ as the $A_{\infty}$-web algebra for the opposite half-plane. 
Moreover we conjecture that the spectrum of stable solitons always takes the form of Fock spaces built from the spectrum of closed solitons of the unperturbed model. 

Note that the requirement of a single vacuum with rank one deck group allows for plenty of interesting models.  For instance the discussion below applies to the $N$th ``Gaussian matrix model'' with superpotential 
\be 
W = \sum_{i=1}^N z_i^2 + m \sum_{i<j} \text{log}(z_i-z_j)
\ee 
As discussed in \cite{Cecotti:2014wea}, this model has a unique vacuum on a target space where certain identifications are made.

\subsection{Koszul Duality} 

We first begin by proving a theorem regarding Koszul duality. We first need to recall the notion of the cobar complex of an augmented $A_{\infty}$-algebra. Let $A$ be an $A_{\infty}$-algebra with an $A_{\infty}$ map $\varepsilon: A \rightarrow \mathbb{C} $ \footnote{We consider strict $A_{\infty}$ morphisms here so that the only non-vanishing map is $\varepsilon_1$.} so that $A$ admits $\mathbb{C}$ as an $A_{\infty}$ module. Let 
\be 
\ov{A} = \text{ker}(\varepsilon)
\ee 
be the augmentation ideal, and let $m_n: \ov{A}^{\otimes n} \rightarrow \ov{A}$ be the induced maps restricted to $\bar{A}.$ The \textit{cobar complex } of $A$, denoted as $\text{C}(A)$ consists of the following space. As a vector space it is the tensor algebra of the shifted dual of $\bar{A}$: 
\begin{align}
C(A) &= T^{\bullet} (\bar{A}^{\vee}[1]) \\ &= \mathbb{C} \oplus \bar{A}^{\vee}[1] \oplus (\bar{A}^{\vee}[1])^{\otimes 2} \oplus \dots.
\end{align} We equip $C(A)$ with the structure of a differential graded algebra. The product structure is given simply by taking the tensor product of words in the tensor algebra. To write the differential we choose a homogeneous basis $\{ e_a\}$ for $\bar{A}$ 
and define ``matrix elements'' of the multiplications by \be m_k(e_{a_1}, \dots, e_{a_k}) = \sum_a  (m_k)^a_{a_1 \dots a_k} e_a.\ee
Then we can write the differential as  
\be \label{cobardifferential}
d e^a = \sum_{k \geq 1} (m_k)^a_{a_1 \dots a_k} e^{a_1} \otimes \dots \otimes e^{a_k}.
\ee 
where $\{ e^a\} $ is the dual basis. It is extended to act on an arbitrary element of $C(A)$ by requiring that $d$ be a derivation of the tensor algebra product \be d(e^a \otimes e^b) = de^a \otimes e^b +(-1)^{|e_a|+1} e^a \otimes \text{d}b.\ee Note that (denoting the cohomological degree of an element $v$ as $\vert v \vert$), the degrees obey \be |e^a| = 1-|e_a|,\ee so that $d$ is a degree $+1$ map if and only if $m_n$ carries degree $2-n$ for each $n\geq1.$ The nilpotence
\be
d^2 =0
\ee 
then holds if and only if $A_{\infty}$-associativity relations hold.

One usually refers to the Koszul dual $A^{!}$ of $A$ as the cohomology of the cobar complex $C(A)$ with respect to the cobar differential $d$, equipped with the $A_{\infty}$-structure coming from transferring the tensor algebra structure to the cohomology.

For the homological origins of the cobar complex and some worked examples see the relevant section in \cite{Gaiotto:2023dvs}, which applies this notion to the problem of categorifying pentagon relations related to wall-crossing in four-dimensional $\mathcal{N}=2$ theories. See also \cite{brantner} for a more extensive discussion of this Koszul duality and its various applications in mathematics. We also refer to \cite{Paquette:2021cij} and references therein for other applications of this idea in quantum field theory and string theory.

Suppose then we're given a one-form $\alpha$ on $X$ with a single period and a single Morse vacuum. There is a cyclic cover $p:\widetilde{X} \rightarrow X$ such that the pullback of $\alpha$ is exact, $\pi^*(\alpha) = dW$, and the vacua on $\widetilde{X}$ form a  $\mathbb{Z}$-torsor.  We can decompose the superpotential $W$ (non-canonically) as
\be 
W = Y + F
\ee 
where, if we set the twisted mass $m=1$ (as we may), $Y$ undergoes a shift under deck transformations by $2\pi i$, and $F$ is a $\mathbb{Z}$-invariant function
\footnote{We thank E.~Witten for suggesting this decomposition.} 
on $\widetilde{X}$. Suppose we make a perturbation of $W$ to 
\be\label{eq:GenWPert}
W^{-\epsilon} = W-\epsilon f(Y)
\ee 
where $f(Y) = \frac{1}{2} Y^2$ with $\text{Re}(\epsilon)>0,$ so that (to first order in $\epsilon$) the critical values are shifted to look like a parabola $y = x^2$ that we rotate clockwise by 90 degrees. Again, no single value of $\epsilon$ will work uniformly for all 
``matrix elements'' involving solitons but for a fixed ``matrix element'' it will work 
provided we take $\epsilon$ sufficiently small with $\text{Re}(\epsilon)>0$. 

Before proceeding we remark that we expect that our result, Theorem \ref{thm:KoszulDualityTheorem}, will hold for a large family of perturbing functions $f(Y)$, so long as they deform the critical values to a convex set. 
For simplicity, in this section we just consider $f(Y) = \frac{1}{2} Y^2$ and $\text{Re}(\epsilon)>0$.

We now let $ R^-_{k,l}$
be the space of stable solitons for this perturbation, namely as before ones that exist for $\epsilon$ arbitrarily small (but non-zero) interpolating between the perturbation of the $k$th and $l$th critical point, equipped with the pairings $K_{k,l}:R^-_{k,l}\otimes R^-_{l,k} \rightarrow \mathbb{C}$. This will lead to a soliton spectrum with natural isomorphisms
\be 
R^-_{k,l} \cong R^-_{k+n, l+n}.
\ee 
This is again because we may identify the stable soliton spectrum with the intersection points of thimbles, and the latter set carries an action of $\mathbb{Z}.$ There is now an $n$-gon associated to integers $ k_1 > \dots > k_N$ where the corresponding cyclic fan of vacua is 
\be 
(k_1, k_N, k_{N-1}, \dots, k_{2}).
\ee 
Counting stable $\zeta$-instantons which fill in this $n$-gon leads to a $\mathbb{Z}$-invariant interior amplitude 
\be 
\beta = \sum_{k_1 > \dots > k_{N}} N_{a_1 \dots a_N}(k_1, \dots, k_N) \,\phi^{a_1}_{k_1, k_N} \otimes \phi^{a_2}_{k_N, k_{N-1}} \otimes \dots \otimes \phi^{a_N}_{k_2, k_1}.
\ee 
The stable $n$-gons and the spaces that we associate to them have a $\mathbb{Z}$-action and so it makes sense to claim that $\beta$ is $\mathbb{Z}$-invariant.

The category of boundary conditions for the right half-plane and $\zeta = -\text{i}$ is now easy to determine. 
The phase ordering of the central charges is determined by 
\be 
z_{a,b} \cong 2 \pi (a-b) e^{-i \pi \epsilon(a+b) - \epsilon Y_0} 
\ee
where $Y_0$ is a complex number so that $Y_k = Y_0 + 2\pi i k$, and recall we have set the twisted mass $m=1$. Comparing with 
\eqref{eq:z-freechir-minuspert} we see that, as far as phase orderings of central charges  and fans are concerned, the discussion will be identical to the case of the free chiral with the minus perturbation. Therefore, the ordering induced on the vacua is opposite the ordering of the integers. The only half-plane fans are given by single rays, so that we have 
\be 
\widehat{R}_{k,l} = R_{k,l}, \,\,\,\,\, k>l.
\ee 
The only taut half-plane webs with this perturbation are given by webs with arbitrarily many boundary vertices and a single bulk vertex. If there are $n$ boundary vertices the bulk vertex is associated to an $(n+1)$-gon, and leads to a map 
\be 
m_n: \widehat{R}_{k_0, k_1} \otimes \dots \otimes \widehat{R}_{k_{n-1},k_n} \rightarrow \widehat{R}_{k_0, k_n}
\ee 
for each sequence $k_0> k_1 > \dots > k_n$ given by 
\be m_n(\phi_{k_0 k_1}, \dots, \phi_{k_{n-1}, k_n}) = \sum_{a} N(\phi_{ k_1, k_0}, \dots, \phi_{k_{n} k_{n-1}}, \phi^a_{k_n, k_0} )   \phi^a_{k_0, k_n}
\ee
This defines an $A_{\infty}$-category, as the interior amplitude satisfies the Maurer-Cartan equation. 

\begin{figure}
    \centering
    \includegraphics[width=0.73\linewidth]{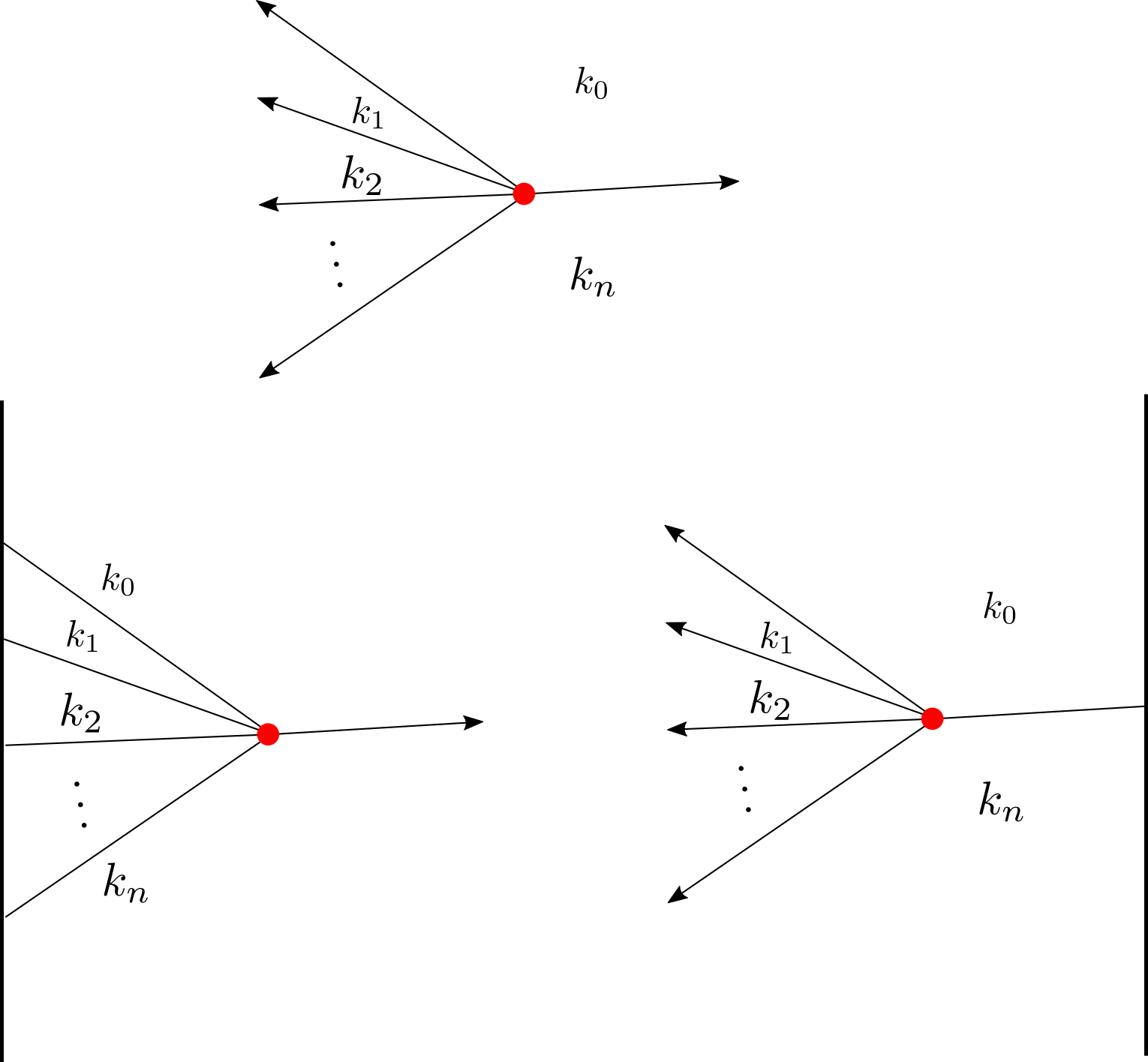}
    \caption{The figure describes the essential idea behind the Koszul duality between left and right half-plane web algbras. The top depicts a bulk vertex of the (family of) perturbed models. This bulk vertex leads to a non-trivial $A_{\infty}$-map $m_n$ for the right half-plane. The same web upon ``sliding the boundary through" gives us a contribution to a differential on the other side. An $A_{\infty}$-algebra and its cobar complex are related exactly by this kind of exchange. }
    \label{koszulidea}
\end{figure}

The orbit category consists of an object with endomorphism algebra $A^{\text{R}}$ where
\be
A^{\text{R}} = \bigoplus_{n \geq 0} R_{n,0}
\ee
is an augmented $A_{\infty}$-algebra, with the augmentation given by projecting to the $\widehat{R}_{0,0} \cong \mathbb{C},$ and the $A_{\infty}$-maps $m_n : A^{\otimes n} \rightarrow A$ all determined from the webs shown in the bottom left of Figure \ref{koszulidea}
with the interior amplitude plugged in. For $N$ free chirals as discussed in the previous section, $A^{\text{R}}$ coincided with the exterior algebra in $N$ variables. For a more general $\alpha$ it will be a more general augmented $A_{\infty}$-algebra.

We now come to the left half-plane. Here the ordering on the vacua coincides with the ordering of the integers that labels them, and the fans that contribute to $\widehat{R}_{k,l}$ for $k<l$ are in one-to-one correspondence with sequences of integers, $k< j_1 < \dots < j_n < l $ so that we have
\be \widehat{R}^{\text{L}}_{k,l} = \bigoplus_{\substack{k< j_1 < \dots < j_n < l}} R_{k,j_1} \otimes \dots \otimes R_{j_n, l}.
\ee
The taut left half-plane webs that give this an $A_{\infty}$-structure are simple to enumerate. It is a straightforward generalization of the webs we encountered in the mirror theory to the free chiral. They lead to two non-trivial maps, the differential $d$ and the product $m_2$. The differential $d$ is obtained by summing over all webs with a single boundary vertex and a single bulk vertex. More precisely a given web is enumerated as follows. The boundary 
is associated to a sequence of integers $k_0 < k_1 < \dots < k_n$ and the web then has a bulk vertex which for some $i$ splits the pair $k_i, k_{i+1}$ by inserting another tuple 
\be 
k_i < j_1 < \dots < j_m < k_{i+1}.
\ee 
This leads to a contribution to the differential $d: \widehat{R}^{\text{L}}_{k_0, k_n} \rightarrow \widehat{R}^{\text{L}}_{k_0, k_n}$ 
of the form 
\begin{align} \label{leftplanedifferential}
\begin{split}
d(e^{a_1}_{k_0, k_1}\otimes \dots \otimes e^{a_n}_{k_{n-1} k_n}) = \,\,\,\,\,\,\,\,\,\,\,\,\,\,\,\,\,\,\,\,\,\,\,\,\,\,\,\,\,\,\,\,\,\,\,\,\,\,\,\,\,\,\,\,\,\,\,\,\,\,\,\,\,\,\,\,\,\,\,\,\,\,\,\,\,\,\,\,\, \,\,\,\,\,\,\,\,\,\,\,\,\,\,\,\,\,\,\,\,\,\,\, \,\,\,\,\,\,\,\,\,\,\,\,\,\,\,\,\,\,\,\,\,\,\,  \\ \sum_{\substack{0 \leq i \leq n \\ k_i < j_1 < \dots < j_m < k_{i+1}}} \pm N^{a_i}_{b_1,\dots, b_m} e^{a_1}_{k_0, k_1} \otimes \dots \otimes e^{b_1}_{k_i, j_1} \otimes \dots \otimes e^{b_m}_{j_m, k_{i+1}} \otimes \dots \otimes e^{a_n}_{k_{n-1}, k_n}. \end{split}
\end{align}
See bottom right of Figure \ref{koszulidea} for the case $n=1$. 
The other taut web with two boundary vertices and no bulk vertex simply leads to an ordinary tensor product. 

We now pass to the orbit category to compute $A^{\text{L}}$ the boundary algebra for the left half-plane. It is given by 
\be 
A^{\text{L}} = \bigoplus_{n \geq 0} \widehat{R}_{0,n}^{\text{L}}.
\ee In fact what we have reproduced coincides with the cobar complex of the $A_{\infty}$-algebra $A^{\text{R}}$. To see this we recall that the underlying vector space of the cobar complex is given by the tensor algebra of the shifted dual of the augmentation ideal of $A^{\text{R}}$
\begin{align}
C(A^{\text{R}}) &= T^{\bullet}\big(\ov{A^{\text{R}}}^{\vee}[1] \big), \\ &= T^{\bullet}\big( \oplus_{ n \geq 1} R_{n,0}^{\vee}[1] \big).
\end{align} By using the collection of pairings $\{K_{n,0} \}_{n \geq 1}$ we can rewrite this as 
\be
C(A^{\text{R}} ) \cong T^{\bullet} \big(\oplus_{n \geq 1} R_{0,n} \big).
\ee
Expanding this out and using the shift symmetry isomorphisms $R_{a,b} \cong R_{a+k, b+k} $ we obtain 
\begin{align}
C(A^{\text{R}}) &= \mathbb{C} \oplus \left(\bigoplus_{n \geq 1} R_{0,n}\right) \oplus \left(\bigoplus_{n,m \geq 1} R_{0,n} \otimes R_{0,m}\right) \oplus \left(\bigoplus_{n,m,k \geq 1} R_{0,n} \otimes R_{0,m} \otimes R_{0,k}\right) \oplus \dots \label{eq:example_label} \\
&\cong \mathbb{C} \oplus \left(\bigoplus_{n} R_{0,n}\right) \oplus \left(\bigoplus_{n,m} R_{0,n} \otimes R_{n,n+m}\right) \oplus \left(\bigoplus_{n,m,k} R_{0,n} \otimes R_{n,n+m} \otimes R_{n+m, n+m+k}\right) \oplus \dots \notag \\
&= \bigoplus_{N \geq 0} \left(R_{0,N} \oplus \left(\bigoplus_{0<k<N} R_{0,k} \otimes R_{k,N}\right) \oplus \left(\bigoplus_{0<k_1<k_2<N} R_{0,k_1} \otimes R_{k_1, k_2} \otimes R_{k_2, N}\right) \oplus \dots\right) \notag \\
&= \bigoplus_{N \geq 0} \widehat{R}^{\text{L}}_{0,N} \notag \\
&= A^{\text{L}}. \notag
\end{align}
Moreover, by comparing \eqref{leftplanedifferential} with \eqref{cobardifferential}, we see that the differential produced by webs coincides precisely with the cobar differential. 

We have thus proven the following
\begin{theorem}\label{thm:KoszulDualityTheorem}
    Let $\alpha$ be one-form on $X$ with a single period and single Morse vacuum. Then the web based algebras associated to left and right half-planes $A^{\text{R}}(X,\alpha)$ and $A^{\text{L}}(X, \alpha)$ are Koszul dual, when they are defined by the perturbation \eqref{eq:GenWPert}
    with $\text{Re}(\epsilon)>0$. 
\end{theorem}

As noted above, a similar theorem will hold for the web algebras defined by the perturbation with $\text{Re}(\epsilon)<0$, although the details of the categories would be quite different, and moreover, we could perturb by a much larger family of functions $f(Y)$ and obtain analogous results. 

\subsubsection{On Bulk Observables} We now make some initial remarks on bulk observables, leaving a more detailed discussion for another occasion. Starting from the boundary observables of the right half-plane $A^{\text{R}}$ one expects on general grounds that the bulk local observables can be obtained by taking Hochschild cohomology of $A$. More specifically, for an augmented $A_{\infty}$-algebra $A$ with augmentation ideal $\ov{A}$ the relative\footnote{Here the term ``relative" refers to the augmentation $\varepsilon$ of $A$.} Hochschild complex is given by \be CC^{*}(A, \mathbb{C}, A) = \bigoplus_{n  = 0}^{\infty} \text{Hom}(\ov{A}^{\otimes n}, A)\ee with the differential given by a familiar sort of formula which we will not write out explicitly. We can see that this is precisely the vector space that is reproduced if we take the product of the left half-plane algebra $A^{\text{L}}$ with the right half-plane algebra $A^{\text{R}}$
\be
A^{\text{L}} \otimes A^{\text{R}} = T^\bullet\big( \ov{A^{\text{R}}}^{\vee}[1] \big) \otimes A^{\text{R}} = CC^*(A^{\text{R}}, \mathbb{C}, A^{\text{R}}). 
\ee
Moreover we expect that the bulk webs will reproduce the formula for the Hochschild differential and Lie bracket reproducing on the nose the differential graded Lie algebra given by the relative Hochschild complex of $A^{\text{R}}$. For $N$ free chirals we expect,  ultimately, to recover the differential graded Lie algebra of polyvector fields on $\mathbb{C}^N$.

\subsection{Emergence Of Fock Spaces} We demonstrated explicitly for the Landau-Ginzburg model of section \ref{subsec:MultipleChirals} that for a small perturbation, the (stable) soliton spectrum is expressed in terms of Fock spaces of a fixed finite-dimensional vector space. Here we conjecture this to be a general feature.

We again let $\alpha$ be a closed holomorphic one-form on $X$ with a unique Morse vacuum which defines a cyclic holomorphic cover $\widetilde{X}$, so that $\Gamma = \mathbb{Z} \langle \gamma \rangle.$ We let $R_{k\gamma}$ be the space of closed solitons of charge $k$ where $k \in \mathbb{Z}$ is a (non-zero) integer. 

Let $W$ be the superpotential defined by $\alpha$ on $\widetilde{X}$. As before we split 
\be W = Y + F\ee where $F$ is $\mathbb{Z}$-invariant and $Y$ undergoes shifts by $2\pi i$ and introduce the perturbed superpotential 
\be W^{-\epsilon} = W - \epsilon f(Y)\ee 
where we now let $f$ be any holomorphic function such that its restriction to the imaginary axis is a positive convex function. 
We claim that the spectrum of stable solitons for $W^{-\epsilon}$ is expressed in terms of Fock spaces of closed solitons. More precisely, introduce the space 
\be V = \bigoplus_{k \geq 1} R_{k\gamma}\ee 
considered as a vector space doubly graded by the (integral) fermion number and the deck group $\Gamma$. Letting $V_0$ and $V_1$ be the even and odd parts of $V$, let 
\be
 \mathscr{F}[V] =  S^*(V_0) \otimes \Lambda^*(V_1)
 \ee
 denote the $\mathbb{Z}_2$-graded Fock space of $V$. Decompose $\mathscr{F}[V]$ into components of definite deck degree so that
\be \mathscr{F}[V] = \bigoplus_{k \geq 1} \mathscr{F}^k[V].\ee 
Let $Y_k(\epsilon)$ denote the (perturbation of the) $k$th vacuum of $W$, and  $R^-_{k,l}$ denote the space of stable solitons for 
$\text{Re}(\epsilon)>0$. Similarly let $R^+_{k,l}$ denote the space of stable solitons for $\text{Re}(\epsilon)<0$. (As usual, ``stable'' refers to the space for
a sufficiently small value of $\vert \epsilon \vert $ for fixed values of $k,l$.) We then conjecture that 
\be R^-_{k,l} \cong \mathscr{F}^{k-l}[V] \ee
for $k<l$. CP would then also require
\be R^-_{k,l} \cong (\mathscr{F}^{l-k}[V])^{\vee}[1]\ee
for $k>l$.  Here $\cong$ denotes quasi-isomorphism of chain complexes.
Similarly, we conjecture that for the opposite convexity namely for $\text{Re}(\epsilon)<0$ the spectrum of stable solitons is
\be R^+_{k,l} \cong \mathscr{F}^{k-l}\big[V^{\vee}[1] \big],
\ee
for $k>l$. CP would then require that
\be R^+_{k,l} \cong \big(\mathscr{F}^{l-k}\big[V^{\vee}[1] \big] \big)^{\vee}[1]
\ee 
for $l>k$.

In other words we are claiming that Fock space combinatorics governs the ``gluings" of closed solitons to obtain stable solitons of the perturbed models. For a related discussion of gluings of closed orbits in the context of circle-valued Morse theory see \cite{hutchings4}. 

Let us mention a related conjecture concerning the geometry of Lefschetz thimbles. We saw how the spectrum of stable solitons can be read off from the intersection numbers of unperturbed Lefschetz thimbles on the covering space $\widetilde{X}.$ Since we are claiming that the spectrum of stable solitons are charged components of Fock spaces of closed solitons, our conjecture may be translated to a more geometric claim about how Lefschetz thimbles of a given one-form with a single Morse zero intersect. 

Suppose as before $\alpha$ is a one-form on $X$ with a single Morse zero and $\widetilde{X}$ is the minimal cover on which we can define $\widetilde{W}$. Suppose $\{L_k(\zeta) \}_{k\in \mathbb{Z}}$ are the collection of Lefschetz thimbles on $\widetilde{X}$. Define 
\be \label{intersectionnumber} A_k(\zeta) = L_0(\zeta e^{i \epsilon}) \circ L_{k}(\zeta e^{-i\epsilon}), \,\,\,\,\,\,\,\,\,\,\, k \in \mathbb{Z}\ee
for $\epsilon$ a sufficiently small positive number. (Note that there is nothing special about $L_0(\zeta)$. We could have defined 
\be A_{i,k}(\zeta) =L_i(\zeta e^{i\epsilon}) \circ L_{i+k}(\zeta e^{-i\epsilon}) \ee 
which is independent of $i$ by deck symmetry, so we simply set $i=0$.) We form the generating series of intersection numbers
\be S(q; \zeta) := \sum_{k \in \mathbb{Z}}  A_k(\zeta) q^k. \ee
Note that by definition of the intersection numbers \eqref{intersectionnumber}, $S(q; \zeta)$ is a formal 
series concentrated in either non-negative or non-positive powers of $q$ depending on whether $\text{Re}(\zeta)>0$ or $\text{Re}(\zeta)<0.$ Let \be \mu_{k\gamma} = \chi(R_{k\gamma}) \ee be the Euler character of the space of closed solitons of charge $k$. Our conjecture then states that 
\begin{align}
    S(q; \zeta) = 
    \begin{cases}  
    \displaystyle\prod\limits_{k \geq 0} (1-q^k)^{-\mu_{k\gamma}},  & \,\,\,\,\,\,\, \text{Re}(\zeta) > 0, \\ 
    \displaystyle\prod\limits_{k \geq 0}(1-q^{-k})^{\mu_{k\gamma}}, & \,\,\,\,\,\,\, \text{Re}(\zeta)<0. 
    \end{cases}  
\end{align}
These are formal power series in   $q$ and $q^{-1}$, respectively.
(The replacement of $q$ by $q^{-1}$ is consistent with the 
behavior of the equivariant grading under duality.)  We expect that they will in fact converge to analytic functions on the domains $\vert q \vert <1$ and $\vert q \vert >1$, respectively.

Thus the intersection pattern of thimbles reproduce Fock space characters. Moreover the generating series for thimble intersections in the opposite half $\zeta$-plane are related by characters of Koszul dual Fock spaces. To our knowledge a statement along these lines has not been discussed in the (rather vast) literature on Picard-Lefschetz theory. 

\section{Conclusions And Future Directions} \label{conc} 

In this paper we have initiated the program of generalizing the algebra of the infrared to two-dimensional $\mathcal{N}=(2,2)$ quantum field theories with twisted masses.
The generalization is not at all straightforward, as we explain in detail in section \ref{subsec:GeneralStrategy}. 
In that same section we outline a proposed solution to the difficulties. We then implement this outlined 
solution in some simple cases. In particular, we demonstrate how, by working on the cover, making a special kind of perturbation of the covering superpotential, and then considering fans, webs, solitons and instantons that are stable under the perturbation, we can recover boundary algebras for the mirror Landau-Ginzburg models to free chirals via a web-based formalism. The resulting construction   allows us to recover the Koszul duality of boundary algebras associated to the left and right half-planes. The emergence of Koszul duality appears very elegant and natural to us.

There are several points in our discussion which deserve further clarification. For examples,

\begin{enumerate} 
%
%

\item In the models we have considered we considered certain special perturbations of the superpotential $\tilde W$ on the covering space $\tilde X$. It is very desirable to have a precise characterization of the perturbations for which the crucial equation \eqref{eq:DeckInvPertRab} holds.  We also found interesting dependence on the sign of $Re(\epsilon)$ for perturbations of the form $\Delta \tilde W = \frac{1}{2} \epsilon Y^2$. Similarly, one could study how the category behaves under rotation of the half-plane $H$ (or equivalently, rotation of the phase $\zeta$). This strongly suggests there is an interesting wall-crossing phenomenon in the space of perturbations. 

\item One of the motivations for this project is the goal of categorifying the 2d4d wall-crossing formula of 
\cite{Gaiotto:2011tf,Gaiotto:2023dvs,Kontsevich:2008fj}. A crucial idea for the categorification of these wall-crossing formulas is the construction of wall-crossing interfaces. Since rotation of the half-plane is an example of one such interface the results of this paper allow us to conclude that Koszul duality is an essential new ingredient when discussing the wall-crossing formalism for $\mathcal{N}=(2,2)$ theories with twisted masses. We would like to combine this kind of ``$\mathcal{K}$ wall-crossing" with the formalism of $\mathcal{S}$-wall crossing as discussed in \cite{Gaiotto:2015aoa} into a single coherent framework.  

\item In the   examples we considered we focused on the $A_\infty$ category associated to half-planes. But, already in these simple examples, the $L_\infty$ algebra associated with local operators will be very interesting, and we expect a web construction of algebras of polyvector fields to emerge, possibly making contact with  \cite{Cattaneo:1999fm,Kontsevich:1997vb}. 

\item It would be highly desirable to understand more deeply the emergence of Fock space combinatorics in the gluing of closed solitons. We have observed both bosonic and fermionic Fock spaces emerging so the gluing theorems are going to be nontrivial. For some perturbations of the superpotential there will be infinitely many ways to concatenate closed solitons and perturb to a single soliton (resulting in a bosonic Fock space), while for others there will only be finitely many ways to concatenate and perturb. 

\end{enumerate} 

We plan to discuss \cite{WIP} more general models with twisted masses in a future paper. There are two ways one can generalize: We can increase the number of vacua beyond one, and we can increase the rank of the deck group $\Gamma$ beyond one. Both generalizations are illustrated in Appendix \ref{examples}. The example of the   $\mathbb{CP}^1$ sigma model gives a particularly interesting example with intricate wall-crossing
formally analogous to that of 4d SU(2) N=2 supersymmetric Yang-Mills theory
\cite{Dorey:1998yh}.
We expect to  show how our formalism allows one to categorify the rich wall-crossing formula that controls the jumping behavior in this model as we move from a weakly coupled region of parameter space to a strongly coupled one.
\footnote{An upgrade of the wall-crossing formula to an equivalence of $A_\infty$ categories was already presented in 
\cite{Khan:2021hve}. We hope to reproduce those results from the perturbed web formalism. } 
More generally we hope to show how the  web construction of the closed string sector and  interfaces generalizes in the presence of twisted masses.

\appendix

\section{An Analogy With $BF$ Theory}\label{app:BF-Analogy}

An analogous situation is to consider an interaction vertex in two-dimensional topological BF theory, where the fields consist of $B$, a $V^{\vee}$-valued scalar for some finite-dimensional vector space $V$, and $A$ a $V$-valued one-form on the worldsheet $\Sigma$. The free action of the $BF$ theory is \be S_{\text{free}} = \int_{\Sigma} B \text{d}A \ee with $d$ being the deRham differential on $\Sigma$. A trivalent vertex that couples one $B$ and two $A$'s consists of an element 
\be 
f \in V\otimes V^{\vee} \otimes V^{\vee} 
\ee 
and turns the free theory to an interacting one by adding to the action $S_{\text{free}}$ the interaction term \be S_{\text{int}} = \int_{\Sigma} f^a_{bc} B_a A^b A^c .\ee The element $f$ is analogous to the interior amplitude $\beta$. To define a consistent interacting gauge theory theory however, any old map $f$ does not work: instead it has to satisfy the Jacobi identity (thus giving $V$ the structure of a Lie algebra). This is analogous to the condition that $\beta$ satisfies the Maurer-Cartan equation. In fact in a proper discussion involving the Batalin-Vilkovisky formalism, one would see that this is more than just an analogy: in the BV formalism local observables carry a gradation by the ghost number and also possess an odd Lie bracket, and the element $f$ above indeed corresponds to an observable in ghost number $+2$ that satisfies the Maurer-Cartan equation.

For more details on BF theory and its relation to Koszul duality along the lines of the results in this paper see \cite{khantalk}.

\section{Examples Of Theories With Twisted Masses } \label{examples}

\subsection{The Free Chiral And Its Mirror} \label{freechiral}

\begin{figure}
\centering
\includegraphics[width=0.3\textwidth]{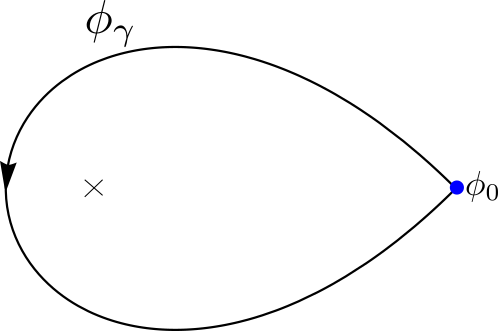}
\caption{Image in the $\phi$-plane of the closed soliton in the free chiral mirror at $m=1$. The cross denotes the singular point $\phi = 0$.}
\label{fc}
\end{figure}

Let $m$ be a non-zero complex number. The simplest Landau-Ginzburg model that exhibits a non-trivial closed soliton is the theory with target space $\mathbb{C}^*$ and  one-form 
\bea \text{d}W = \Big(\frac{m}{\phi}-1 \Big) \text{d}\phi.\eea 
The deck group $\Gamma = H_1(\mathbb{C}^*, \mathbb{Z})$ is rank one and generated by the cycle $\gamma$ that winds around the origin once in the counter-clockwise direction. $\text{d}W$ has a non-trivial period around it and the central charge is given by 
\bea Z(\gamma) = 2\pi \text{i} m.\eea 
The one-form $\text{d}W$ has a single zero located at $\phi_1 = m$, and there is a non-trivial BPS trajectory form $\phi_1$ to itself for a suitable phase $\zeta$, as
depicted in Figure \ref{fc}. We therefore have 
\bea R(\gamma) = \mathbb{Z}.\eea
Moreover $R(n\gamma) = 0 $ for $\vert n \vert > 1$ as one can see since the trajectory cannot self-intersect.

The Landau-Ginzburg $A$-model above has a mirror 
dual being the $B$-model of the \textit{free} theory of a chiral superfield $\Phi$ with a standard kinetic term $\int \text{d}^4 \theta \, \Phi\ov{\Phi}$  , and with a non-trivial twisted mass $\widetilde{m} = 2\pi \text{i}m$ turned on for the global symmetry $\Phi \rightarrow e^{\text{i} \xi } \Phi$. The explicit action in terms of the component fields $(\phi, \psi_{+}, \psi_-)$ of $\Phi$ is 
\begin{align} \begin{split} S=  \int \text{d}^2 x \big( |\del_t \phi|^2 - |\del_x \phi|^2 - |\widetilde{m}|^2 |\phi|^2 +  \text{i} \ov{\psi}_+ \del_- \psi_- + \text{i} \ov{\psi}_- \del_+ \psi_-  \\ - \ov{\widetilde{m}} \,\ov{\psi}_+ \psi_- - \widetilde{m} \ov{\psi}_- \psi_+ \big).\end{split} 
\end{align}
Being a free theory, it can be easily solved, and so the BPS spectrum is readily determined. 

The conserved Hamiltonian and momentum (our convention is $P_{\mu} = T^0_{\,\mu}$) are
\bea \begin{split} H = \int \text{d}x \big( |\pi|^2 + |\del_x\phi|^2 +|m|^2 |\phi^2| + \text{i} \ov{\psi}_+ \del_x \psi_+ - \text{i} \ov{\psi}_- \del_x \psi_-  \\ + \ov{m} \ov{\psi}_+ \psi_- + m \ov{\psi}_- \psi_+ \big), \end{split},\eea \bea P = \int \text{d}x \big(\pi \,\del_x \phi + \del_x \ov{\phi} \,\ov{\pi} + \text{i} \ov{\psi}_+ \del_x \psi_+ + \text{i} \ov{\psi}_- \del_x \psi_- \big), \eea 
The conserved $U(1)$ charge corresponding to $\phi \rightarrow e^{\text{i} \xi} \phi, \,\, \psi_{\pm} \rightarrow e^{\text{i} \xi} \psi_{\pm}$ is \bea J &=& \int \text{d}x \big( \text{i} \pi  \phi - \text{i} \ov{\phi} \ov{\pi} - \ov{\psi}_+ \psi_+ - \ov{\psi}_- \psi_-\big).\eea 
The vector Fermion number acting trivially on $\phi$ and sending $\psi_{\pm} \rightarrow e^{-\text{i} \alpha } \psi_{\pm}$ is 
\bea F_V = \int \text{d}x \big(\ov{\psi}_+ \psi_+ + \ov{\psi}_- \psi_- \big). \eea Finally the supercharges are \bea Q_+ &=& \int \text{d}x \big((\pi + \del_x \ov{\phi}) \psi_+ + \text{i} \ov{m} \ov{\phi} \psi_- \big), \\ Q_- &=& \int \text{d}x \big( (\pi - \del_x \ov{\phi}) \psi_- + \text{i} m \ov{\phi} \psi_+ \big), \\ \ov{Q}_+ &=& \int \text{d}x \big( (\ov{\pi} + \del_x \phi) \ov{\psi}_+ - \text{i} m \phi \ov{\psi}_- \big), \\ \ov{Q}_- &=& \int \text{d}x \big( (\ov{\pi} - \del_x \phi) \ov{\psi}_- - \text{i} \ov{m} \phi \ov{\psi}_+ \big). \eea One can diagonalize the Hamiltonian as usual by mode decomposition of the elementary fields $\phi, \psi_{\pm}$. Letting \bea E_p = \sqrt{p^2 + |m|^2} \, , \eea
the mode expansions for the bosons are (all integrals over $p$ are over the real line): 
\bea \phi(x) &=& \int \frac{\text{d}p}{\sqrt{2\pi}} \frac{1}{\sqrt{2E_p}}\big(a_p e^{-\text{i}px} + b_p^{\dagger} e^{\text{i}px} \big), \\ \pi(x) &=& -\text{i}\int \frac{\text{d}p}{\sqrt{2\pi}} \sqrt{\frac{E_p}{2}}\big(b_p e^{-\text{i}px} - a_p^{\dagger} e^{\text{i}px} \big), \eea 
with the modes satisfying the standard commutation relations 
\bea [a_p, a^{\dagger}_q] = [b_p, b^{\dagger}_q] = \delta(p-q)\eea
with other commutators vanishing. 
The mode expansions for the fermions read 
\bea \psi_+(x) &=& \frac{1}{\sqrt{2}}  \Bigg(\frac{\ov{m}}{|m|} \Bigg)^{\frac{1}{2}} \int \frac{\text{d}p} {\sqrt{2\pi}}  \sqrt{1+\frac{p}{E_p}}\big(\chi_p e^{-\text{i}px} + \lambda_p^{\dagger}e^{\text{i}px}\big), \\ \psi_-(x) &=& \frac{1}{\sqrt{2}} \Bigg(\frac{m}{|m|} \Bigg)^{\frac{1}{2}} \int \frac{\text{d}p }{\sqrt{2\pi}} \sqrt{1-\frac{p}{E_p}} \big(\chi_p e^{-\text{i}px} - \lambda_p^{\dagger} e^{\text{i}px} \big),\eea
with the modes satisfying 
\bea \{\chi_p, \chi_q^{\dagger} \} = \{\lambda_p, \lambda_q^{\dagger} \} = \delta(p-q)\eea
and other anti-commutators vanish. 
In terms of these modes then, the Hamiltonian, momentum, and global $U(1)$ charge are written as 
\bea H &=& \int \text{d}p \big(E_p a_p^{\dagger} a_p + E_p b_p^{\dagger}b_p + E_p \chi_p^{\dagger} \chi_p + E_p \lambda_p^{\dagger} \lambda_p \big), \\ 
P &=& \int \text{d}p \big(p \, a_p^{\dagger} a_p + p\, b_p^{\dagger}b_p + p\, \chi_p^{\dagger} \chi_p + p\, \lambda_p^{\dagger} \lambda_p \big), \\ 
q &=& \int \text{d}p \big(b_p^{\dagger} b_p - a_p^{\dagger} a_p + \lambda_p^{\dagger} \lambda_p - \chi_p^{\dagger} \chi_p \big).\eea 
The vector fermion number is \bea F_V = \int \text{d}p \big(\chi_p^{\dagger} \chi_p - \lambda_p^{\dagger} \lambda_p). \eea Finally the supercharges in terms of the modes are \bea Q_+ &=& -\text{i} \bigg(\frac{\ov{m}}{|m|} \bigg)^{\frac{1}{2}} \int \text{d}p \sqrt{E_p+p} \big(b_{p} \lambda_p^{\dagger} - a_p^{\dagger} \chi_p \big), \\ Q_- &=& \text{i} \bigg(\frac{m}{|m|} \bigg)^{\frac{1}{2}}\int \text{d}p \sqrt{E_p - p} \big(b_p \lambda_p^{\dagger} + a_p^{\dagger} \chi_p \big), \\ \ov{Q}_+ &=& \text{i} \bigg(\frac{m}{|m|} \bigg)^{\frac{1}{2}} \int \text{d}p \sqrt{E_p+p} \big(b_p^{\dagger} \lambda_p - a_p \chi_p^{\dagger} \big), \\ \ov{Q}_- &=& -\text{i} \bigg( \frac{\ov{m}}{|m|} \bigg)^{\frac{1}{2}} \int \text{d}p \sqrt{E_p - p} \big( b_p^{\dagger} \lambda_p + a_p \chi_p^{\dagger} \big).\eea 
The supersymmetric vacuum is then the unique state defined by 
\bea a_p|0 \rangle = b_p |0 \rangle = \chi_p |0\rangle = \lambda_p |0\rangle = 0\eea 
for all $p\in \IR$. The vacuum state $|0 \rangle$ as usual preserves all four supercharges. The BPS states on the other hand, are by definition states are annihilated by the supercharges \bea Q_{\text{B}}(\zeta) := \ov{Q}_+ + \zeta \ov{Q}_- , \,\,\,\, \ov{Q}_{\text{B}}(\zeta) = Q_+ + \zeta^{-1} Q_-.\eea Because the central charge is $\widetilde{Z} = -mJ,$ the values we consider for $\zeta$ are \bea \zeta_{\pm} = \pm \frac{m}{|m|} \eea For $\zeta_+$ we have 
\begin{align}
\begin{split} Q_B(\zeta_+) = -\text{i} \bigg(\frac{m}{|m|} \bigg)^{\frac{1}{2}} \int \text{d}p \big( (\sqrt{E_p+p} -\sqrt{E_p-p}) b_p^{\dagger} \lambda_p \,\,\,\,\,\,\,\,\,\,\,\,\,\,\,\,\,\,\,\, \\- (\sqrt{E_p+p} + \sqrt{E_p-p} ) a_p \chi_p^{\dagger} \big), \end{split} 
\end{align}
\begin{align}
\begin{split} \ov{Q}_B(\zeta_+) = \text{i} \bigg(\frac{\ov{m}}{|m|} \bigg)^{\frac{1}{2}} \int \text{d}p \big( (\sqrt{E_p+p} -\sqrt{E_p-p}) b_p \lambda_p^{\dagger} \,\,\,\,\,\,\,\,\,\,\,\,\,\,\,\,\,\,\,\, \\ - (\sqrt{E_p+p} + \sqrt{E_p-p} ) a^{\dagger}_p \chi_p \big) . \end{split} 
\end{align}
States that are killed by these two supercharges are then zero momentum states such that there are no particles of negative charge. In other words, the BPS states for $\zeta_+$ are 
\bea (b_0^{\dagger})^k | 0 \rangle, \,\,\,\,\, (b_0^{\dagger})^k \lambda_0^{\dagger} |0 \rangle, \eea 
for $k=0,1,2,\dots$.  In a similar way, one can determine the BPS states for $\zeta_-$. They are of the form 
\bea (a_0^{\dagger})^{k} |0 \rangle,\,\,\,\,\, (a_0^{\dagger})^k \chi_0^{\dagger} |0 \rangle.\eea  
The two classes of BPS states are exchanged under charge conjugation. 

\bigskip
\noindent
\textbf{Remark}: We note that these states are zero-momentum ``particles'' - hence spread out over space - and in fact are not normalizable, so they are not strictly states. (Taking the spatial manifold to be a circle the analogous states are normalizable.)  Moreover, one cannot form BPS wavepackets because the BPS condition constrains the momentum to be zero. We set aside the many conceptual problems raised by such observations and proceed formally. 
\bigskip 

It is clear then that there is a bose-fermi superpair of particles of mass $|\widetilde{m}|$ in the spectrum of this theory, one for each charge. Since the central charge   is 
\bea \widetilde{Z} = \widetilde{m}q \eea
these are BPS particles. We thus see that under the mirror duality, the closed soliton gets mapped to an elementary excitation $a_0^\dagger$ or $b_0^\dagger$ of the chiral superfield $\Phi$. 

\subsection{A Rank Two Model with a Unique Vacuum}

\begin{figure}
\centering
\includegraphics[width=0.35\textwidth]{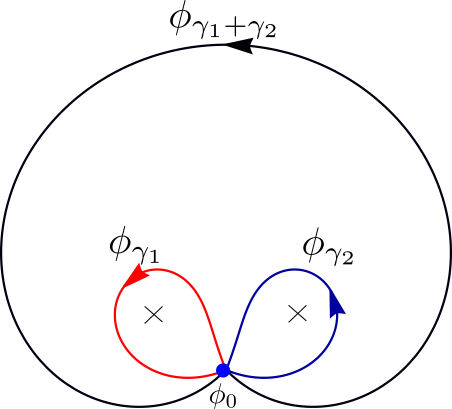}
\caption{Images in the $\phi$-plane of the closed solitons in the Double Penner model at $(m_1, m_2) = (1,i)$. The crosses denote the singular points $\phi=0,1$.}
\label{double}
\end{figure}

As our next example we illustrate a Landau-Ginzburg model where the deck group $\Gamma$ is a rank two lattice and once again there are only closed solitons in the BPS spectrum.  Let $m_1$ and $m_2$ be non-zero, and non-parallel complex numbers. The target space is now the twice punctured complex plane $X=\mathbb{C}\backslash \{0,1\}$ and Landau-Ginzburg one-form is \bea \text{d}W = \Big(\frac{m_1}{\phi} + \frac{m_2}{\phi-1} \Big) \text{d}\phi.\eea $\text{d}W$ again has a unique zero located at \bea \phi = \frac{m_1}{m_1+m_2}.\eea $\Gamma = H_1(X, \mathbb{Z})$ is spanned by the cycles $\gamma_1$ and $\gamma_2$ that go around each of the punctures in a counter-clockwise direction and so we have \bea Z(\gamma_1) = 2\pi \text{i} m_1, \,\,\,\,\,\, Z(\gamma_2) = 2\pi \text{i} m_2 .\eea The model has a soliton spectrum consisting of three closed solitons carrying charges $\gamma_1, \gamma_2$ and $\gamma_1 + \gamma_2$ (since there's only a single vacuum we can safely drop the subscript $ii$ in $\gamma_{ii}$, the charge of an $ii$-soliton). Therefore we have \bea R(\gamma_{1}) &=& \mathbb{Z} \langle \phi_{\gamma_1} \rangle, \\ R(\gamma_{2}) &=& \mathbb{Z} \langle \phi_{\gamma_2} \rangle, \\ R(\gamma_1 + \gamma_2) &=& \mathbb{Z} \langle \phi_{\gamma_1 + \gamma_2} \rangle,\eea
each being rank one. 
%
%
The spaces for higher multiples of the charges vanish, again because the soliton trajectory cannot 
self-intersect. 

Our next two examples will illustrate that even when the Landau-Ginzburg one-form has non-trivial periods, the spectrum of closed solitons can still be empty. In such a case the spectrum of traditional BPS solitons can still have some novel properties (as will be discussed at a later point in the paper).

\subsection{The $\mathbb{CP}^1$ Model At Strong Coupling}\label{subsec:CP1-Strong}

\begin{figure}
\centering
\includegraphics[width=0.3\textwidth]{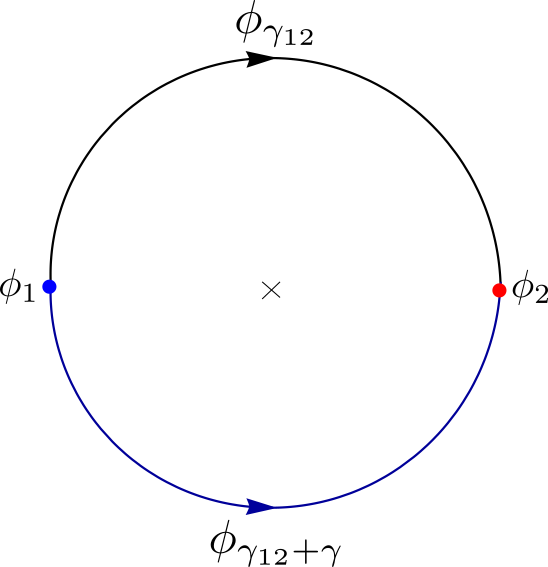}
\caption{Images in the $\phi$-plane of the solitons in the $\mathbb{CP}^1$ model at $(m,\Lambda) = (0.1,1)$, a point in the strong coupling regime.}
\label{p1sol}
\end{figure}

Let's consider the mirror \cite{Hori:2000kt} to the $\mathbb{CP}^1$ sigma model with twisted masses. This model has target the punctured plane and LG one-form 
\bea \alpha = \Big( 1 - \frac{m}{\phi} - \frac{\Lambda^2}{\phi^2} \Big) d\phi ,\eea
The model depends on two parameters $(m,\Lambda)\in \IC^2$. We work away from the ``discriminant locus'' where 
$m^2 + 4 \Lambda^2 = 0$. In general, all the structures we describe will be fibered over this space and are subjected to nontrivial monodromy. Some of our labelings of vacua and flavor charges therefore depend on making noncanonical choices. 

The form $\alpha$ has two zeroes which, for a fixed choice of $(m,\Lambda)$ can be (noncanonically) labeled: 
\bea \phi_1 &=& \frac{1}{2} \big(m-\sqrt{m^2 + 4\Lambda^2} \big), \\ \phi_2 &=& \frac{1}{2} \big(m+ \sqrt{m^2 + 4 \Lambda^2} \big).\eea
An important property of this model is that it has $\mathbb{Z}_2$-symmetry which acts on $\phi$ via
\bea \phi \rightarrow -\Lambda^2/\phi \eea 
taking $\alpha \to - \alpha$. The two vacua $\phi_1$ and $\phi_2$ are intercharged under the $\mathbb{Z}_2$-symmetry and under monodromy around the discriminant locus. 

In a subsequent section we will consider the wall-crossing properties of BPS states in this model. For now we consider the model in the so-called  ``strong coupling regime,'' defined to be the regime where  
$|\frac{m}{\Lambda} | << 1$. Let $\gamma_{12} \in \Gamma_{12}$ denote the simple cycle going from $\phi_1$ to $\phi_2$ 
shown in figure 3, and $\gamma$ denote the cycle going around the origin in a counter-clockwise direction with winding number 1. We have 
\bea \label{solitonchargep1} 
  Z(\gamma_{12} ) &=& 2\sqrt{m^2 + 4 \Lambda^2 } + m \, \text{log} \Bigg(\frac{m-\sqrt{m^2+4\Lambda^2}}{m+ \sqrt{m^2 + 4\Lambda^2}} \Bigg), \\    Z(\gamma) &=& - 2\pi \text{i} m.\eea 
The soliton spectrum at strong coupling is known to be \cite{Gaiotto:2015aoa, Hori:2000kt}
\bea R(\gamma_{12}) &=& \mathbb{Z} \langle \phi_{\gamma_{12}} \rangle, \\ R(\gamma_{12} + \gamma) &=& \mathbb{Z} \langle \phi_{\gamma_{12} + \gamma} \rangle,\eea
each space being concentrated in degree zero. All other $R(\gamma_{ij})$ vanish. The images of these solitons for instance at $(m, \Lambda) = (0.1,1)$ are depicted in Figure \ref{p1sol}.  There are no closed solitons at this point in parameter space.
The weak coupling regime will be described in section \ref{subsec:CP1-Weak} below.

\subsection{The $\mathbb{CP}^1$ Model At Weak Coupling}\label{subsec:CP1-Weak}

\begin{figure}
\centering
\includegraphics[width=0.55\textwidth]{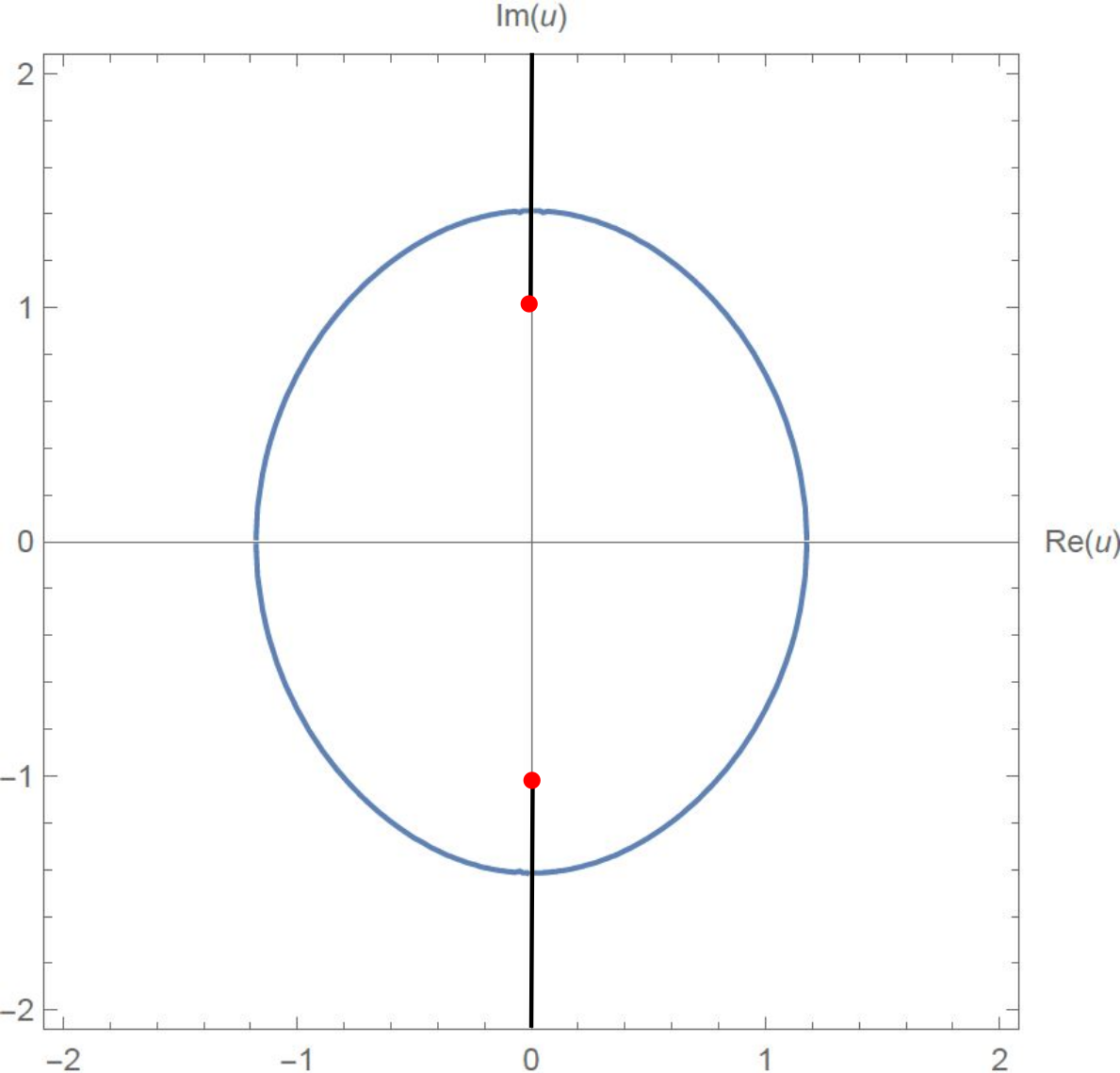}
\caption{The curve of marginal stability for the $\mathbb{CP}^1$ model with twisted masses in the $u= \frac{2\Lambda}{m}$-plane. By definition this is the region where $Z(\gamma_{12})$ is parallel or anti-parallel to $Z(\gamma)$. The closed curve plotted in blue separates the $u$-plane into two regions. The region outside of the curve is the strong coupling regime and the region inside is the weak coupling regime. }
\label{wall}
\end{figure}

\begin{figure}
\centering
\includegraphics[width=\textwidth]{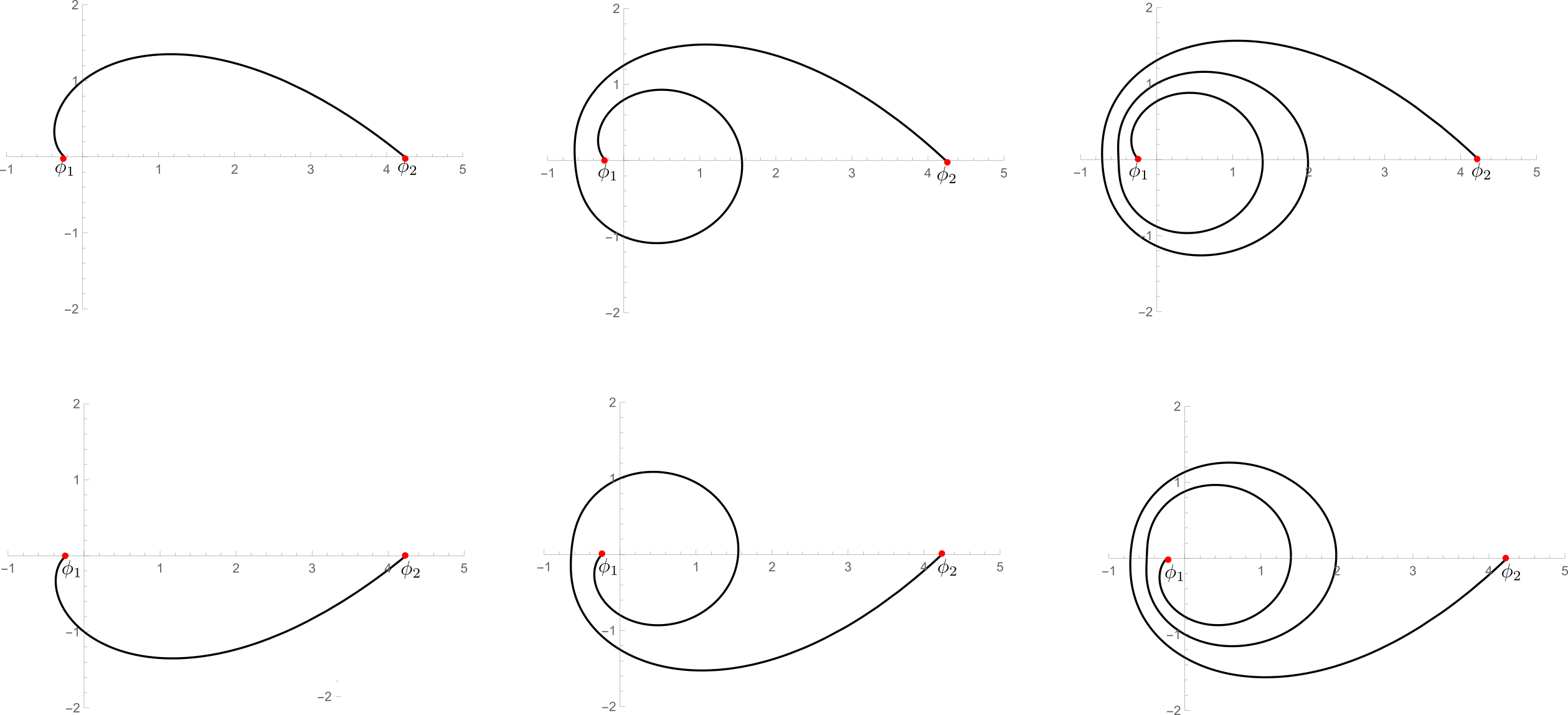}
\caption{Some non-closed solitons for the point in the weak coupling region $(m,\Lambda) =(4,1)$. We have taken $m>0$ and $\Lambda$ real and $\phi_1 \cong -\Lambda^2/m + \cdots $ close to zero and $\phi_2 \cong m + \Lambda^2/m + \cdots $ large and positive. The central charges of the solitons going from $\phi_1$ to $\phi_2$  are of the form 
\bea 2 \sqrt{m^2 + 4 \Lambda^2} + m \log(\Lambda^2/m^2) + 
i \pi (2n+1) m \eea
where we take the principal branch of the logarithm and $n$ is an integer. The upper row corresponds to $n=0,1,2$ respectively as we go from left to right, and the lower row corresponds to $n=-1,-2,-3$ respectively as we go from left to right. }
\label{nonperp1}
\end{figure}

The BPS spectrum of the $\mathbb{CP}^1$ model is subject to wall-crossing phenomenon. The wall is located at the region in $(m,\Lambda)$-space where the central charge of the soliton $Z(\gamma_{12})$ becomes parallel or anti-parallel to the central charge of the closed soliton $Z(\gamma)$. Letting 
\bea
u=\frac{2\Lambda}{m},
\eea 
this is the region where 
\bea \text{Re} \, f(u) = 0,
\eea
where 
\bea f(u) = \sqrt{1+u^2} + \text{log} \Bigg(\frac{1-\sqrt{1+u^2}}{1+\sqrt{1+u^2}} \Bigg).
\eea This locus, known as the wall of marginal stability is depicted in Figure \ref{wall}. The closed curve part of the region divides the $u$-plane into two regions. The region which contains the origin is where $m$ is large and the model is weakly coupled. We therefore call this the weak coupling regime. Here we will analyze the BPS spectrum of the model at a point in this regime.

\begin{figure}
\centering
\includegraphics[width=0.6\textwidth]{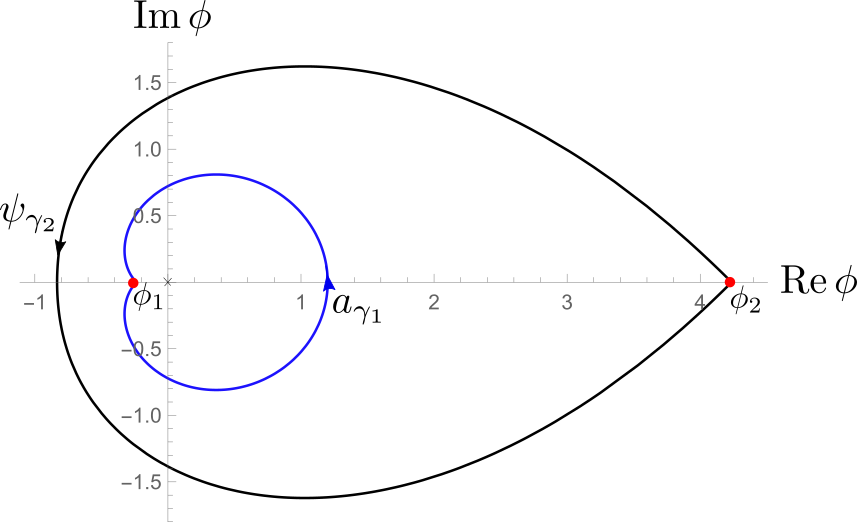}
\caption{Trajectories of $a_{\gamma_1}$ and $\psi_{\gamma_2}$, the closed soliton states for the $\mathbb{CP}^1$ mirror Landau-Ginzburg model at a point in the weak coupling regime. The origin of this notation is explained in \cite{Khan:2021hve}.}
\label{perp1}
\end{figure}

At a point in the weak coupling regime, which is when $| \frac{m}{\Lambda}|>> 1$, the spectrum of solitons is as follows. First we find that there is a non-trivial, and unique flow trajectory for \textit{every} charge in $\Gamma_{12}$ and $\Gamma_{21}$, 
so that 
\bea \label{weakspec} R(\gamma_{12}) &\cong& \mathbb{Z} \,\,\,\,\text{   for any } \gamma_{12} \in \Gamma_{12}, \\ R(\gamma_{21}) &\cong& \mathbb{Z}[1] \text{ for any } \gamma_{21} \in \Gamma_{21} . \eea
The trajectories that correspond to these states are depicted in Figure \ref{nonperp1}. Moreover, there are also closed solitons at such a point in parameter space: there is a non-trivial flow trajectory between both $\phi_1$ and itself, and $\phi_2$ and itself leading to the spaces 
\bea R(\gamma_1) &\cong& \mathbb{Z}, \,\,\,\,\,\,\,\,\,\,\,\,\,\,
R(\gamma_2) \cong \mathbb{Z}[1], \\ \label{weakspec2} 
R(-\gamma_1) &\cong& \mathbb{Z}[1],  \,\,\,\,\,\,\,\, 
R(-\gamma_2) \cong \mathbb{Z}, \eea 
where $\gamma$ is the counter-clockwise generator and $\gamma_1,\gamma_2$ are the corresponding generators of $\Gamma_{11}$ and $\Gamma_{22}$, respectively. The precise flow trajectories corresponding to these spaces are depicted in Figure \ref{perp1}. The complexes of closed solitons are trivial otherwise. 
\footnote{In fact, the charge lattice and all the complexes $R(\gamma_{ij})$ form local systems over the complement of the discriminant locus in the space of parameters $(m,\Lambda)$. Here we are making local choices and ignoring that very interesting class of phenomena. }

As mentioned before, the mirror dual
\footnote{In the topological theories we use an $A$-twist of the LG model and a $B$-twist of the sigma model.}
to the Landau-Ginzburg model above is the sigma model with target space $\mathbb{CP}^1$, where twisted masses are turned on with respect to rotations around the $z$-axis, so that the potential term is 
\bea S_{V} = \int \text{d}^2 x \, g_{i\bar{j}} V^i \ov{V}^{\bar{j}}\eea where $V$ is the vector field that generates rotations around the $z$-axis. The BPS spectrum that we wrote above can nicely be reproduced by semi-classical methods. The semi-classical vacua are given by the zeros of $V$, which coincide with the north and south pole. The spaces of closed solitons $R(\pm \gamma_{11})$ and $R(\pm \gamma_{22})$, as in the free theory discussed in Section, in this dual picture arise as the elementary field excitations around the north and south poles respectively. Moreover, as remarked briefly in section \ref{freechiral}, the classical BPS equation of the B-model is the gradient flow equation of the moment map $\mu$ with respect to the classical Fubini-Study metric. It is a familiar fact that there is an $S^1$-worth of classical trajectories that interpolate between the north ($\phi = 0$) and south poles $(\phi= \infty)$: \bea \phi(x, \alpha) = e^{x -x_0 + \text{i}\alpha},\eea where $\alpha$ is an angle that parametrizes the $S^1$-family.  When we quantize this family we get an integer worth of semi-classical BPS states between the vacuum $1$ and $2$. By CP, we also find an integer worth of states between the vacua $2$ and $1$. This is precisely what the pattern of BPS trajectories looks like in the mirror Landau-Ginzburg theory.

Yet another way to determine the BPS spectrum (at least the indices) that applies at both strong and weak coupling is the method of spectral networks, applied to the spectral curve. To see this carried out we refer the reader to sections 8.2 and 8.3 of \cite{Gaiotto:2011tf}.

\subsection{Weierstrass Model}

We now discuss a model with two vacua, and a rank two deck group. Consider the model\footnote{We thank S.~Cecotti for suggesting this Landau-Ginzburg model as an interesting example to investigate.} with target being a punctured elliptic curve $X = T^2_{\tau} \backslash \{0\}$ with modular parameter $\tau$, and Landau-Ginzburg  one-form 
\bea \text{d}W = \wp(\phi, \tau) \text{d}\phi .\eea 
The Weierstrass $\zeta$-function gives us a multivalued superpotential which on the universal cover is 
\bea 
\tilde W = -\zeta(\phi,\tau).\eea 
The model has two vacua, and in general it is difficult to write down explicit expressions for the zeros of $\wp(z,\tau)$. However one can work at the enhanced symmetry point $\tau = \tau_0 := e^{2\pi i /3}$ where the model has $\mathbb{Z}_6$ symmetry.
\footnote{Recall that $\wp(z,\tau)$ has a Laurent series about $z=0$ with coefficients in the ring generated by Eisenstein series $G_4$ and $G_6$, so writing it as $\wp(z,G_4, G_6)$ we have the scaling law 
\be 
\wp(\lambda z , \lambda^{-4} G_4, \lambda^{-6} G_6) = \lambda^{-2} \wp(z,G_4, G_6).
\ee Using the modular transformation law of the Eisenstein function one shows that $G_4(e^{2\pi i /3})=0$ and   $\partial_{\phi} W $ transforms as $\lambda^{-2}$ when $\lambda$ is a sixth root of unity. Thus by suitable rotation of $\theta, \bar\theta$ the action is invariant. }
Making use of this symmetry, it is possible to locate the exact zeroes. On the one hand the $\mathbb{Z}_6$ symmetry implies that if $\phi_0$ is a zero, then so is $\omega^k \phi_0$ where $\omega = e^{\frac{\pi \text{i} }{3}}$ and $k=0,1,\dots,5$. On the other hand there are only two zeroes in the fundamental domain by Riemann-Roch. The other zero must be $\omega\phi_0$, and since  the set of numbers $\omega^{2k}\phi_0$ are lattice translates of $\phi_0$ it must be that
$\omega^2 \phi_0 = \phi_0 + n + m \tau_0$ for some  integers $n,m$. Choose the standard fundamental domain, the parallelogram defined by $1$ and $\tau_0$. Rotation of $\phi_0, \omega\phi_0$ by 
$2\pi/3$ takes these points outside the fundamental domain. Let $\phi_1$ denote the zero with phase smaller than $\pi/3$. Then we must have:   
\bea \omega^2 \phi_1 = \phi_1 - 1 \eea 
which means that $\phi_1 = \frac{1}{\sqrt{3}} e^{ \frac{\text{i}\pi}{6}}$. The other zero in the fundamental domain is then given by  given by $\omega \phi_1$. Thus the two zeroes are located at 
\bea \phi_1 &=& \frac{1}{\sqrt{3}} \text{exp}\Big(\frac{\text{i} \pi}{6} \Big), \\ \phi_2 &=& \frac{1}{\sqrt{3}} \text{i}.\eea
The punctured torus has 
\bea H_1(T^2_{\tau_0 } \backslash \{0\}, \mathbb{Z}) \cong \mathbb{Z} \langle \gamma_1 \rangle \oplus \mathbb{Z} \langle \gamma_2 \rangle \eea
where $\gamma_1$ is the homology classes of an oriented path going from $\phi_*$ to $\phi_*+ 1 $ in a straight line that avoids the lattice $\mathbb{Z} + \tau_0 \mathbb{Z}$.  The choice of path does not matter because the form $\alpha$ does not have a residue around lattice points. 
Similarly,  $\gamma_2$ is the homology class of an  oriented path going from $\phi_*$ to   $\phi_* + \tau_0 $   in a straight line avoiding lattice points. 
In other words $\gamma_1$ and $\gamma_2$ are the standard $a$ and $b$-cycles respectively.  
In order to derive the periods we make use of the $\mathbb{Z}_6$ symmetry and the homogeneity property of the $\wp$ function. 
We can define a single-valued choice of $\zeta(\phi, G_4 , G_6)$ on the covering space of the torus by taking 
\be 
\tilde W(\phi,G_4,G_6) = - \zeta(\phi, G_4 , G_6) = -\frac{1}{\phi} + \sum_{k=1}^\infty P_{2k+2}(G_4, G_6) \phi^{2k+1} 
\ee
where $P_{2k+2}(G_4, G_6) = G_{2k+2}$. It then follows that 
\be 
\tilde W(\lambda \phi, \lambda^{-4} G_4 , \lambda^{-6} G_6) = \lambda^{-1}  \tilde W( \phi, G_4 , G_6) 
\ee
We apply this with $\tau= \tau_0$ to that $G_4 =0$ with $\lambda$ various powers of $\omega= e^{i \pi/3}$. 
Note that any point in the $\mathbb{Z}_6$-orbit $\{\omega^k \phi_1\}_{k=0, \dots, 5}$ may be expressed as a lattice translate of either $\phi_1$ or $\phi_2$:
\be 
\begin{split}
\omega^0 \phi_1 &= \phi_1,\\ 
\omega \,\phi_1 &= \phi_2, \\
\omega^2 \phi_1  & = \phi_1 -1, \\ 
\omega^3 \phi_1 & = \phi_2 - (1 + \tau_0), \\
\omega^4 \phi_1 & = \phi_1 - (1 + \tau_0), \\
\omega^5 \phi_1 & = \phi_2 - \tau_0 .\\
\end{split}
\ee
Therefore we see that (writing simply $\tilde W(\phi_1)$ for $\tilde W(\phi_1,0,G_6)$) 
\be 
\begin{split} 
Z(\gamma_1) & = \tilde W(\phi_1) - \tilde W(\phi_1-1) \\ 
& = \tilde W(\phi_1 ) - \tilde W(\omega^2\phi_1) \\
& = (1- \omega^{-2}) \tilde W(\phi_1) \\
& = (2+\tau_0) \tilde W(\phi_1) \\ 
\end{split}
\ee
Similarly we find 
\be 
Z(\gamma_2) = (1+2\tau_0) \tilde W(\phi_1) 
\ee
and for the open chain
\bea 
Z_{\gamma_{12}} = \tilde W(\phi_2) - \tilde W(\phi_1) = \tilde W(\omega \phi_1) - \tilde W(\phi_1) =  -(1+\tau_0) \tilde W(\phi_1) \eea 
so up to an overall normalization by $\tilde W(\phi_1)$ they lie in the lattice. 

\begin{figure}
\centering
\includegraphics[width=0.65\textwidth]{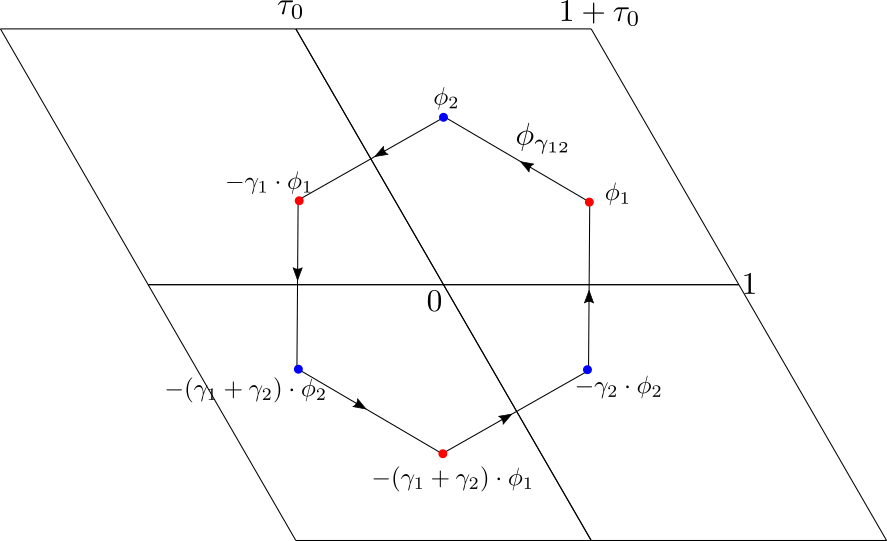}
\caption{The critical points and soliton paths of the $\alpha = \wp(\phi,\tau) d\phi$ model as depicted on the cover.}
\label{wpsol}
\end{figure}

The BPS spectrum consists of three   solitons of type $12$, all with distinct charges in $\Gamma_{12}$: \bea R(\gamma_{12}) &=& \mathbb{Z} \langle \phi_{\gamma_{12}} \rangle, \\ R(\gamma_{12}-\gamma_1) &=& \mathbb{Z} \langle \phi_{\gamma_{12}-\gamma_1} \rangle, \\ R(\gamma_{12} + \gamma_2) &=& \mathbb{Z} \langle \phi_{\gamma_{12} + \gamma_2} \rangle  \eea along with the three anti-solitons. The soliton trajectories are depicted in Figure \ref{wpsol} The spaces have degree $(1,0,1)$ as listed from top to bottom. There model has no closed solitons at this point in parameter space.


\subsection{Another Rank One Example with Two Vacua}

If we constrain $\alpha$ to be a holomorphic one-form on $\mathbb{C}^*$ that 

\begin{enumerate}
\item is cohomologically non-trivial $ H^{(1,0)}(X, \mathbb{C})\ni [\alpha] \neq 0.$
\item has exactly two Morse zeroes
\end{enumerate}

then we'll find that it must either be the  one-form for the $\mathbb{CP}^1$ mirror model or of the form 
\bea \alpha  = \Big( \phi - 2 \Lambda + \frac{\mu^2}{\phi}\Big) \text{d}\phi,\eea
where $\Lambda$ and $\mu$ are complex valued parameters. Once again, we work away from the discriminant locus where 
$\Lambda^2 - \mu^2 = 0$. 
\footnote{We can call this the ``$AD_2$ LG model'' based on the relation of $\alpha$ to the SW curve. 
(In this terminology the $\IC\IP^1$ model would be called the ``pure $SU(2)$ model.'') }

The zeroes of $\alpha$  are given by 
\bea \phi_1 &=& \Lambda - \sqrt{\Lambda^2 - \mu^2}, \\ \phi_2 &=& \Lambda + \sqrt{\Lambda^2 - \mu^2}\eea and the central charges are 
\bea Z_{\gamma_{12}} &=& 2\Lambda \sqrt{\Lambda^2 -\mu^2} + \mu^2 \text{log} \Bigg( \frac{\Lambda - \sqrt{\Lambda^2 - \mu^2}}{\Lambda + \sqrt{\Lambda^2 - \mu^2}} \Bigg), \\ Z_{\gamma} &=& 2 \pi \text{i} \mu^2.\eea The expressions above again lead to interesting marginal stability walls, but for this section we again restrict ourselves and work at a particular point in $(\Lambda, \mu)$ space away from the walls. Let's analyze the soliton spectrum at $(\Lambda, \mu) = (\frac{1}{2}, i)$. We have that the non-vanishing BPS complexes consist of $12$ BPS states 
\bea R(\gamma_{12}) &=& \mathbb{Z} \langle \phi_{\gamma_{12}} \rangle \\ 
R(\gamma_{12} +\gamma) &=& \mathbb{Z} \langle \phi_{\gamma_{12} + \gamma} \rangle \eea
along with a single closed soliton from $1 \rightarrow 1$ 
\bea R(\gamma_{11}) = \mathbb{Z} \langle \phi_{\gamma_1} \rangle.\eea

\begin{figure}
\centering
\includegraphics[width=0.6\textwidth]{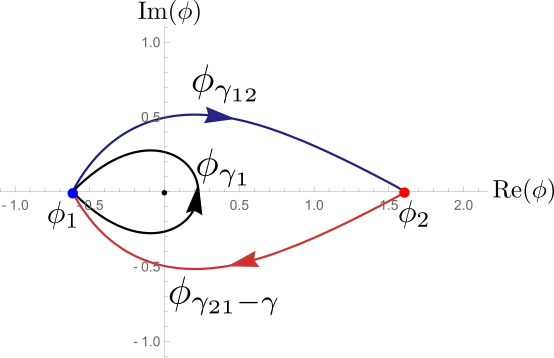}
\caption{The paths the solitons in the $AD_2$ LG model trace out at $(\Lambda, \mu) = (\frac{1}{2},i)$.}
\label{ad2sol}
\end{figure}

The soliton paths are depicted in Figure \ref{ad2sol}. The geometry in that image implies that 
\bea (\phi_{\gamma_1}, \phi_{\gamma_{12}}, \phi_{\gamma_{21}-\gamma})\eea
forms a ``cyclic fan of solitons.''  
Using the method of 
\cite{Khan:2020hir} one computes the total degree is two and therefore  
 the fermion degrees of these spaces satisfy 
 \bea f_{\gamma_1} + f_{\gamma_{12}} + f_{\gamma_{21}-\gamma} = 2.\eea
By the orientation of the thimbles we know that 
$f_{\gamma_{12}} = f_{\gamma_{21}-\gamma}\mod 2 $. 
As an educated guess we take the minimal choice 
\bea f_{\gamma_1} &=& 0,\\ f_{\gamma_{12}} &=& f_{\gamma_{21}-\gamma} = 1. \eea

\subsection{Infinite-dimensional Examples} This section is a slight aside from the main points of the paper. Here we simply remark that supersymmetric theories in higher dimensions can often be viewed as Landau-Ginzburg models with infinite-dimensional target spaces, and non-trivial twisted masses. 

In three dimensions, the $\mathcal{N}=4$ sigma model with target space the hyperK\"{a}hler manifold $Y$ is equivalent to the following Landau-Ginzburg model. Let $\Omega$ be a holomorphic symplectic form with respect to a fixed complex structure $I$ on $Y$. We consider the space \bea X = \{\varphi: \mathbb{R} \rightarrow Y\},\eea equipped with the complex structure induced by $I$,  and let Landau-Ginzburg one-form be \bea \alpha = \int_{\mathbb{R}} \varphi^*(\Omega). \eea The deck group $\Gamma$ is a (subgroup) of the second homology group \bea \Gamma \subset H_2(X, \mathbb{Z})\eea (mod torsion), and the $\zeta$-solitons are pseudoholomorphic maps with respect to some $\zeta$-dependent complex structure on $Y$. The point is that if $\Omega$ is a cohomologically non-trivial two-form on $Y$ then the deck group is indeed non-trivial. The cohomology class of the two-form $\Omega$ is indeed non-trivial for many examples of interest, such as hyperK\"{a}hler manifolds that arise as Higgs branches of theories of interest to geometric Langlands (for instance the theory $T(G)$ introduced in \cite{Gaiotto:2008ak} has this property).

One can also easily generalize to gauged Landau-Ginzburg models, where we suppose that there's a gauge group acting on the LG target $X$, such that the action of the gauge group preserves the one-form $\alpha$. A particularly interesting example that fits into this framework is five-dimensional $\mathcal{N}=2$ Yang-Mills. Letting $G$ be a compact gauge group, here the target space consists of the space of all $G_{\mathbb{C}}$-connections on a three-manifold $M_3$ 
\bea X = \{G_{\mathbb{C}}\text{-connections on } M_3 \}\eea
and the   one-form is \bea \alpha = \int_{M_3} \text{Tr} \, F^2.  \eea
The gauge group $\mathcal{G}$ acts by gauge transformations on $X$ in the usual way and we must treat this as a gauged LG model.  The deck group in this case is rank one, isomorphic to $\mathbb{Z}$. Indeed as discussed in \cite{Witten:2011zz}, the presence of this deck group is what leads to the second gradation on the Khovanov homology groups in their gauge theoretic formulation. More details on the Chern-Simons Landau-Ginzburg model and its relation to knot homology may be found in Section 18.4 of \cite{Gaiotto:2015aoa}. 
\footnote{In discussions  with Tudor Dimofte, Davide Gaiotto, Andy Neitzke, and Fei Yan we have explored relations to indices of 3d supersymmetric theories and even used this model to construct interesting invariants of knots and links. We hope to return to these questions in a future paper.   Some of the main ideas and results were mentioned in \cite{MooreTalks}. }

\section{Group actions and Orbit Categories} \label{App:groupactions}

Landau-Ginzburg models often enjoy interesting discrete symmetries. It is therefore natural 
to extend the web formalism to an ``equivariant web-formalism.'' Physically, the equivariant web-formalism would capture properties of the Landau-Ginzburg model when the discrete symmetries in question are gauged. Unfortunately this has not been developed in the literature. Here we initiate some basics of what would be involved, only developing the theory to the extent that it will be needed in sections \ref{subsec:GeneralStrategy}, \ref{toyexample} and \ref{koszul}. 


Let $\mathcal{C}$ be a (small) $\mathbb{C}$-linear category, and let $G$ be a finite (or discretely infinite) group that acts freely on $\mathcal{C}$ by autofunctors. By definition, this means that for each $g \in G$ we have a functor $F_g: \mathcal{C} \rightarrow \mathcal{C}$ such that 
\begin{equation}\label{eq:Fgh=FgFh}
F_{g} \circ F_{h} = F_{g h}.
\end{equation}
and $F_{1}$ is the identity functor. Thus,   $F_{g}$ is an autoequivalence. 
The functors $F_{g}$ are assumed to define a free $G$-action on the set of objects $\text{Ob}(\mathcal{C})$.  
As usual since group elements have inverses, the linear map 
\begin{equation}
F_g: \text{Hom}(X,Y) \rightarrow \text{Hom}(F_g(X), F_g(Y))
\end{equation}
is an isomorphism. 

When a group $G$ acts on $\mathcal{C}$ we can pass to the $\textit{orbit category}$ denoted as $\mathcal{C}/G$. The objects of $\mathcal{C}/G$ consist of $G$-orbits in $\text{Ob}(\mathcal{C})$. In order to discuss morphism spaces we need to recall the notion of coinvariants.

Let $B$ be an associative algebra over $\mathbb{C}$ equipped with an ``augmentation,'' i.e. an 
algebra homomorphism  $a: B \rightarrow \mathbb{C}$ and suppose $V$ is a left $B$-module. The linear space of \textit{coinvariants} of $V$, denoted as $V/B$ or $V_B$ is defined to be the space $V$ modulo the subspace of relations $R$ which is formed by taking the linear combinations of elements of the form 
\begin{equation}
\epsilon(b)v - b v
\end{equation}
for $v \in V$ and $b \in B$. This can be identified with $\IC\otimes_{B} V$ where $\IC$ is considered a right $B$-module via the augmentation.

In order to define the morphism space of two orbits $\alpha,\beta$ we consider the space 
\begin{equation}
\bigoplus_{x \in \alpha, y \in \beta} \text{Hom}(x,y)
\end{equation}
which is a $\mathbb{C}G$-module. Equipping $\mathbb{C}G$ with its canonical augmentation, i.e. $\epsilon(g)=1$ for all $g$, the morphism space in the orbit category is then defined as the coinvariants 
\begin{equation}
\text{Hom}(\alpha, \beta) = \left( \bigoplus_{x \in \alpha, y \in \beta} \text{Hom}(x,y)\right)_{\mathbb{C}G}.
\end{equation}

We can give a more concrete construction of the morphisms in the orbit category by picking two elements in each orbit, say $x_0 \in \alpha$ and  $y_0 \in \beta$. Since the action of $G$ on the objects is free one can 
write, noncanonically,  
\begin{equation}
\text{Hom}(\alpha, \beta) \cong \bigoplus_{g \in G} \text{Hom}(x_0, g y_0) 
\cong \bigoplus_{g \in G} \text{Hom}(gx_0,  y_0) 
\end{equation}
Morphisms can be defined by shifting using the $G$-action so that the source and target objects agree.

Put differently, the product 
\be 
\text{Hom}(\alpha,\beta) \times \text{Hom}(\beta,\gamma) \to \text{Hom}(\alpha,\gamma) 
\ee
can be defined by choosing points $x_0,y_0, z_0$ in the three orbits and taking 
\be 
\oplus_g \text{Hom}(g x_0, y_0) \times \oplus_{g'} \text{Hom}(y_0, g' z_0) \to \oplus_{g,g'} \text{Hom}(g x_0, g' z_0) 
\ee
and then projecting to the space of coinvariants 
\be 
\oplus_{g,g'} \text{Hom}(g x_0, g' z_0)  \rightarrow  \left( \oplus_{g,g'} \text{Hom}(g x_0, g' z_0) \right)_{\mathbb{C}[G]} = \text{Hom}(\alpha, \gamma) 
\ee
because $F_g$ define autoequivalences the product will be independent of the choices of  $x_0, y_0, z_0$.

\begin{remark}
    A more general construction would allow for natural transformations between  $F_g \circ F_h$ and $F_{gh}$. We will not need to work in this level of generality.
\end{remark}

Not much changes when discussing the generalization to $A_{\infty}$-categories. We will say that a group $G$  acts (strictly) on a small $A_{\infty}$-category if for each $g \in G$ we have   a strict $A_{\infty}$ functor $F_g$ \footnote{Recall that a general $A_{\infty}$-functor $\mathcal{F}$ has a collection of multilinear maps $\mathcal{F}_n$ for $n \geq 1$. ``Strictness" means that $\mathcal{F}_n = 0$ for $n \geq 2.$},   meaning that $F_g$  acts on objects freely and that there is a linear map 
\begin{equation}
F_g : \text{Hom}(X,Y) \rightarrow \text{Hom}(F_g X,F_g Y)
\end{equation}
of degree zero such that 
\footnote{The case $n=2$ was implicitly assumed above because in that case $F_g$ are functors.}
\begin{equation}
m_n(F_g(a_1), \dots, F_g(a_n)) = F_g(m_n(a_1, \dots, a_n))
\end{equation}
for all $n \geq 1,$ and 
\begin{equation} 
F_g \circ F_h = F_{gh}.
\end{equation}

We will now discuss the notion of a group $G$ acting on infrared data. We say that a group $G$ acts on the infrared data $(\mathbb{V}, z, R_{ij}, \beta)$ if the following conditions hold:

\begin{enumerate}
    \item $G$ acts freely on $\mathbb{V}$ and also acts freely on $z(\mathbb{V}) \subset \mathbb{C}$ such that $z: \mathbb{V} \rightarrow \mathbb{C}$ is a $G$-equivariant map \begin{equation}
        z_{g \cdot i} = g \cdot z_i.
    \end{equation} 
    \item For each $g \in G$ there is an isomorphism 
    \begin{equation}
    T_{ij}^g: R_{ij} \rightarrow R_{g \cdot i, g \cdot j}
    \end{equation}
    which is compatible with the group multiplication 
    \begin{equation}
    T^h_{gi, gj} \circ T_{ij}^g = T_{ij}^{hg}, 
    \end{equation}
    such that the pairings are $G$-equivariant in the sense that 
    \begin{equation}
    K_{gi, gj} \circ T^g_{ij} \otimes T^g_{ji} = K_{ij}.
    \end{equation}
    \item $G$ acts on fans, so that if 
    \begin{equation}
    \{i_0, i_1 \dots, i_n\} 
    \end{equation} is a cyclic fan of vacua, then so is 
    \begin{equation}
    \{g \cdot i_0, g \cdot i_1 \dots, g_ \cdot i_n \},
    \end{equation}
    for each $g \in G$.
    \item Items (2,3) above imply that the vector space $R_c$ is a representation of $G$. We require that the interior amplitude $\beta$ to be in the $G$-invariant part of $R_c$ \begin{equation}
        \beta \in (R_c)^G.
    \end{equation}
    
\end{enumerate}

We note that $R_c$ is not only a $G$-representation as a vector space, but the $L_{\infty}$-operations are also compatible with the $G$-action in the sense that 
\begin{equation}
g \cdot \lambda_n(a_1, \dots, a_n) = \lambda_n(g \cdot a_1, \dots g \cdot a_n). 
\end{equation}
This is because if the group $G$ acts on fans it also acts on webs, and the pairings satisfy the equivariance property. Moreover the $L_{\infty}$-structure deformed by the interior amplitude is also compatible with $G$ since $\beta$ is $G$-invariant.

Finally the group $G$ will also act on the $A_{\infty}$-category of boundary conditions if it acts on the collection of half-fans for the chosen half-plane: It acts on the objects in the same way it acts on the vacuum set $\mathbb{V}$ and the maps $T^g_{ij}$ will induce maps (linear isomorphisms) 
\begin{equation}
\widehat{T}^g_{ij}: \widehat{R}_{ij} \rightarrow \widehat{R}_{g \cdot i, g \cdot j}
\end{equation}
on the morphism spaces. Since by assumption $G$ acts on half-fans and fans, it acts on half-plane webs, and since $\beta$ is $G$-invariant the $A_{\infty}$ maps $\{m_n \}_{n=1,2, \dots}$ satisfy 
\begin{equation}
m_n(g \cdot a_1, \dots, g \cdot a_n) = g \cdot m_n(a_1, \dots, a_n).
\end{equation}
We may then pass to the orbit category. 

Our model situation for when this is the case is the following. We say that a Landau-Ginzburg model with target space and superpotential $(\widetilde{X}, \widetilde{W})$ is a cover of an LG model with target $(X,W)$ if there is a covering map $p: \widetilde{X} \rightarrow X$ such that $p^*(W) = \widetilde{W}$. The deck group $G$ of the covering will then act on the covering Landau-Ginzburg model by symmetries. 

Such ``Galois coverings" are defined and prove to be useful in the study of BPS states of four-dimensional $\mathcal{N}=2$ theories \cite{Cecotti:2015qha}.

\begin{remark}
    It is an interesting direction to show that the web-based formalism is functorial with respect to such coverings, namely the orbit category of the covering theory and the category associated to the downstairs theory are equivalent. We will not pursue this in the present paper. 
\end{remark}

\subsubsection{Example} Consider the Landau-Ginzburg model with target space $\mathbb{C}^*$ and superpotential 
\be 
\widetilde{W} = \phi^2 + \phi^{-2}.
\ee Theory has four vacua located at the fourth roots of unity 
\be \phi_k = e^{\pi i k/2}, \,\,\, k= 0,1,2,3\ee
and a single soliton between each adjacent vacua
\be R_{k,k+1} \cong \mathbb{C},\ee
with the non-adjacent spaces being zero-dimensional. 
We can show that $\mathbb{Z}_2$ acts on the infrared data as well as on the category of boundary conditions. The orbit category is identified as the thimble category of the LG model obtained by the covering map $p: \mathbb{C}^* \rightarrow \mathbb{C}^*$ where 
\be
z = p(\phi) = \phi^2 
\ee is the $\mathbb{CP}^1$ mirror model with the same target space but superpotential 
\be 
W=z+ z^{-1}.
\ee Theories related by this kind of covering were considered in \cite{Cecotti:1991me}, which also computes the soliton spectrum of the model at hand. 

\section{Koszul Duality with Multiple Vacua and Cobar Categories} \label{multivackoszul} We now introduce the notion of the \textit{cobar category}, which we expect is relevant to describe the Koszul duality statement applicable to theories with multiple vacua. 

Suppose $\mathcal{C}$ is an $A_{\infty}$-category with finitely many objects \be (O_1, \dots, O_n)\ee that form an exceptional collection. Introducing the notation 
\be
V_{ij} := \text{Hom}_{\mathcal{C}}(O_j, O_i),
\ee we recall that $(O_1, \dots, O_n)$ is said to form an exceptional collection if 
\begin{align}
V_{ii} &\cong \mathbb{C}, \\
V_{ij} &= \{0\}  \text{ for } i>j.
\end{align} The category $\mathcal{C}$ is defined by a collection of maps \be m_k^{i_0, \dots, i_k}: V_{i_0 i_1} \otimes \dots \otimes V_{i_{k-1} i_k} \rightarrow V_{i_0 i_k}\ee for each $k \geq 1$ and each $i_0 \leq \dots \leq i_k$ that satisfy the $A_{\infty}$-associativity axioms. 

We introduce the \textit{cobar category}, denoted by $\mathcal{C}^{!}$ as follows. It will have objects \be (P_1, \dots, P_n)\ee with morphism spaces given as follows. Denoting 
\be 
W_{ij} := \text{Hom}_{\mathcal{C}^{!}}(P_j, P_i)
\ee
we set 
\begin{align}
W_{ij} &= \{0\} \text{ for } i<j, \\ W_{jj} &= V_{jj} \cong \mathbb{C}.
\end{align} Finally if $i<j$ we'll set
\begin{align}
    W_{ji} &= \bigoplus_{\{(k_0, \dots, k_n) | i < k_0 < \dots < k_n < j \} } V_{i k_0}^{\vee}[1] \otimes \dots \otimes V_{k_n j}^{\vee}[1] .\\  &= V_{ij}^{\vee}[1] \oplus \big(\oplus_{i<k<j} V_{ik}^{\vee}[1] \otimes V_{kj}^{\vee}[1] \big)\oplus \dots. \\  
\end{align}
The cobar category $\mathcal{C}^{!}$ will by definition be a differential graded category with the dg structure defined as follows. The composition in $\mathcal{C}^!$ is given by taking the tensor product of elements in $W_{ab}$ and $W_{bc}$ to produce an element in $W_{ac}$ for any triple $(a,b,c)$ with $a>b>c$. The differential in the cobar category is given as follows. We define it on the subset of ``words" of unit length and extend it to an arbitrary word by imposing the Leibniz rule. On unit length words, it is given as follows. Choose a basis $\{e_a(i,j) \}$ for $V_{ij}$ and let $\{e^a(i,j) \}$ be the dual basis for $V_{ij}^{\vee}[1]$. Consider the matrix elements \be m_n\big(e_{a_1}(i_0, i_1), , \dots, e_{a_n}(i_{n-1}, i_n)  \big) = \sum_{a} (m^{i_0, \dots, i_n}_n)^{a}_{a_1 \dots a_n} e_a(i_0, i_n) .\ee We then set 
\begin{align}
d_{ij}(e^a_{ij}) = \sum_{\substack{n \geq 0 \\ i< k_1 < \dots < k_{n-1} < j} } \big(m_n^{i,k_1, \dots, k_{n-1}, j}\big)^a_{a_1 \dots a_n} e^{a_1}(i,k_1) \otimes \dots \otimes e^{a_n}(k_{n-1}, j). 
\end{align} This is nilpotent due to the $A_{\infty}$-relations in $\mathcal{C}$.
The cobar category $\mathcal{C}^{!}$ thus is an $A_{\infty}$-category with an exceptional collection \be (P_n, P_{n-1}, \dots, P_1)\ee that is \textit{dual} to the exceptional collection $(O_1, \dots, O_n)$ in $\mathcal{C}.$

As we have reviewed in the bulk text, the infrared data 
\be
(\mathbb{V}, z, \{R_{ij} \}, \{ K_{ij} \}, \beta) 
\ee
of a theory (without twisted masses), with a chosen half-plane $H$ defines an $A_{\infty}$ ``thimble" category $\mathcal{C}_H$ which is a category with an exceptional collection. We expect that the thimble category for the opposite half-plane $\mathcal{C}_{H_{\text{opp}}}$ is $A_{\infty}$ equivalent to the cobar category $(\mathcal{C}_{H})^{!}$: 
\be \mathcal{C}_{H_{\text{opp}}} \simeq (\mathcal{C}_H)^{!}. \ee
We believe that the proof of this statement will be closely related to the ``universality theorem" proved in \cite{Kapranov:2014uwa}. We expect the generalization to the case of multiple vacua and non-trivial twisted masses, when working with the covering theory, will involve exceptional collections with infinitely many objects.


\begin{thebibliography}{99}

\bibitem[Aga21]{Aganagic:2021ubp}
M.~Aganagic,
``Knot Categorification from Mirror Symmetry, Part II: Lagrangians,''
[arXiv:2105.06039 [hep-th]].

\bibitem[AF83]{AlvarezGaume:1983ab}
L.~Alvarez-Gaume and D.~Z.~Freedman,
``Potentials for the Supersymmetric Nonlinear Sigma Model,''
Commun. Math. Phys. \textbf{91}, 87 (1983)
doi:10.1007/BF01206053

\bibitem[AKO08]{Auroux:2008xno}
D.~Auroux, L.~Katzarkov and D.~Orlov,
``Mirror symmetry for weighted projective planes and their noncommutative deformations,''
Annals Math. \textbf{167}, no.3, 867-943 (2008)
doi:10.4007/annals.2008.167.867

\bibitem[BW82]{Bagger:1982fn}
J.~Bagger and E.~Witten,
``The Gauge Invariant Supersymmetric Nonlinear Sigma Model,''
Phys. Lett. B \textbf{118}, 103-106 (1982)
doi:10.1016/0370-2693(82)90609-8

\bibitem[PhysMath22]{Bah:2022wot}
I.~Bah, D.~S.~Freed, G.~W.~Moore, N.~Nekrasov, S.~S.~Razamat and S.~Schafer-Nameki,
``A Panorama Of Physical Mathematics c. 2022,''
[arXiv:2211.04467 [hep-th]].

\bibitem[BDGH16]{Bullimore:2016nji}
M.~Bullimore, T.~Dimofte, D.~Gaiotto and J.~Hilburn,
``Boundaries, Mirror Symmetry, and Symplectic Duality in 3d $\mathcal{N}=4$ Gauge Theory,''
JHEP \textbf{10} (2016), 108
doi:10.1007/JHEP10(2016)108
[arXiv:1603.08382 [hep-th]].

\bibitem[Br]{brantner}
L.~Brantner, 
``Topics in Koszul Duality," Lecture Notes available
\href{https://people.maths.ox.ac.uk/brantner/Koszul_lectures.html}{here}

\bibitem[CF]{Cattaneo:1999fm}
A.~S.~Cattaneo and G.~Felder,
``A Path integral approach to the Kontsevich quantization formula,''
Commun. Math. Phys. \textbf{212}, 591-611 (2000)
doi:10.1007/s002200000229
[arXiv:math/9902090 [math]].

\bibitem[CDZ15]{Cecotti:2015qha}
S.~Cecotti and M.~Del Zotto,
``Galois covers of $\mathcal{N}=2$ BPS spectra and quantum monodromy,''
Adv. Theor. Math. Phys. \textbf{20} (2016), 1227-1336
doi:10.4310/ATMP.2016.v20.n6.a1
[arXiv:1503.07485 [hep-th]].

\bibitem[CFIV92]{Cecotti:1992qh}
S.~Cecotti, P.~Fendley, K.~A.~Intriligator and C.~Vafa,
``A New supersymmetric index,''
Nucl. Phys. B \textbf{386}, 405-452 (1992)
doi:10.1016/0550-3213(92)90572-S
[arXiv:hep-th/9204102 [hep-th]].

\bibitem[CNV14]{Cecotti:2014wea}
S.~Cecotti, A.~Neitzke and C.~Vafa,
``Twistorial topological strings and a $\mathrm{tt}^*$ geometry for $\mathcal{N} = 2$ theories in $4d$,''
Adv. Theor. Math. Phys. \textbf{20} (2016), 193-312
doi:10.4310/ATMP.2016.v20.n2.a1
[arXiv:1412.4793 [hep-th]].

\bibitem[CV91]{Cecotti:1991me}
S.~Cecotti and C.~Vafa,
``Topological antitopological fusion,''
Nucl. Phys. B \textbf{367}, 359-461 (1991)
doi:10.1016/0550-3213(91)90021-O

\bibitem[CV92]{Cecotti:1992rm}
S.~Cecotti and C.~Vafa,
``On classification of N=2 supersymmetric theories,''
Commun. Math. Phys. \textbf{158}, 569-644 (1993)
doi:10.1007/BF02096804
[arXiv:hep-th/9211097 [hep-th]].

\bibitem[CV10]{Cecotti:2010qn}
S.~Cecotti and C.~Vafa,
``2d Wall-Crossing, R-Twisting, and a Supersymmetric Index,''
[arXiv:1002.3638 [hep-th]].

\bibitem[CS15]{Cordova:2015nma}
C.~Cordova and S.~H.~Shao,
``Schur Indices, BPS Particles, and Argyres-Douglas Theories,''
JHEP \textbf{01}, 040 (2016)
doi:10.1007/JHEP01(2016)040
[arXiv:1506.00265 [hep-th]].

\bibitem[CG21]{Costello:2021jvx}
K.~Costello and O.~Gwilliam,
``Factorization Algebras in Quantum Field Theory,'' Volume 1,
Cambridge University Press, 2021,
ISBN 978-1-316-67866-4, 978-1-107-16315-7
doi:10.1017/9781316678664

\bibitem[CG23]{Costello:2023knl}
K.~Costello and O.~Gwilliam,
``Factorization algebra,''
[arXiv:2310.06137 [math-ph]].

\bibitem[Dor98]{Dorey:1998yh}
N.~Dorey,
``The BPS spectra of two-dimensional supersymmetric gauge theories with twisted mass terms,''
JHEP \textbf{11}, 005 (1998)
doi:10.1088/1126-6708/1998/11/005
[arXiv:hep-th/9806056 [hep-th]].

\bibitem[FB96]{Ferrari:1996sv}
F.~Ferrari and A.~Bilal,
``The Strong coupling spectrum of the Seiberg-Witten theory,''
Nucl. Phys. B \textbf{469}, 387-402 (1996)
doi:10.1016/0550-3213(96)00150-2
[arXiv:hep-th/9602082 [hep-th]].

\bibitem[GK23]{Gaiotto:2023dvs}
D.~Gaiotto and A.~Khan,
``Categorical Pentagon Relations and Koszul Duality,''
[arXiv:2309.12103 [hep-th]].

\bibitem[GKW24]{Gaiotto:2024gii}
D.~Gaiotto, J.~Kulp and J.~Wu,
``Higher Operations in Perturbation Theory,''
[arXiv:2403.13049 [hep-th]].

\bibitem[GMN]{Gaiotto:2011tf}
D.~Gaiotto, G.~W.~Moore and A.~Neitzke,
``Wall-Crossing in Coupled 2d-4d Systems,''
JHEP \textbf{12}, 082 (2012)
doi:10.1007/JHEP12(2012)082
[arXiv:1103.2598 [hep-th]].

\bibitem[GMN12]{Gaiotto:2012rg}
D.~Gaiotto, G.~W.~Moore and A.~Neitzke,
``Spectral networks,''
Annales Henri Poincare \textbf{14}, 1643-1731 (2013)
doi:10.1007/s00023-013-0239-7
[arXiv:1204.4824 [hep-th]].

\bibitem[GMW15]{Gaiotto:2015aoa}
D.~Gaiotto, G.~W.~Moore and E.~Witten,
``Algebra of the Infrared: String Field Theoretic Structures in Massive $\mathcal{N}=(2,2)$ Field Theory In Two Dimensions,''
[arXiv:1506.04087 [hep-th]].

\bibitem[GMW15Sh]{Gaiotto:2015zna}
D.~Gaiotto, G.~W.~Moore and E.~Witten,
``An Introduction To The Web-Based Formalism,''
[arXiv:1506.04086 [hep-th]].

\bibitem[GW08]{Gaiotto:2008ak}
D.~Gaiotto and E.~Witten,
``S-Duality of Boundary Conditions In N=4 Super Yang-Mills Theory,''
Adv. Theor. Math. Phys. \textbf{13} (2009) no.3, 721-896
doi:10.4310/ATMP.2009.v13.n3.

\bibitem[GW11]{Gaiotto:2011nm}
D.~Gaiotto and E.~Witten,
``Knot Invariants from Four-Dimensional Gauge Theory,''
Adv. Theor. Math. Phys. \textbf{16}, no.3, 935-1086 (2012)
doi:10.4310/ATMP.2012.v16.n3.a5
[arXiv:1106.4789 [hep-th]].

\bibitem[GM16]{Galakhov:2016cji}
D.~Galakhov and G.~W.~Moore,
``Comments On The Two-Dimensional Landau-Ginzburg Approach To Link Homology,''
[arXiv:1607.04222 [hep-th]].

\bibitem[GGM]{Garoufalidis:2020xec}
S.~Garoufalidis, J.~Gu and M.~Marino,
``Peacock patterns and resurgence in complex Chern-Simons theory,''
[arXiv:2012.00062 [math.GT]].

\bibitem[HM96]{Harvey:1996gc}
J.~A.~Harvey and G.~W.~Moore,
``On the algebras of BPS states,''
Commun. Math. Phys. \textbf{197}, 489-519 (1998)
doi:10.1007/s002200050461
[arXiv:hep-th/9609017 [hep-th]].

\bibitem[IqVa12]{Iqbal:2012xm}
A.~Iqbal and C.~Vafa,
``BPS Degeneracies and Superconformal Index in Diverse Dimensions,''
Phys. Rev. D \textbf{90}, no.10, 105031 (2014)
doi:10.1103/PhysRevD.90.105031
[arXiv:1210.3605 [hep-th]].

\bibitem[LS03]{Losev:2003gs}
A.~Losev and M.~Shifman,
``N=2 sigma model with twisted mass and superpotential: Central charges and solitons,''
Phys. Rev. D \textbf{68}, 045006 (2003)
doi:10.1103/PhysRevD.68.045006
[arXiv:hep-th/0304003 [hep-th]].

\bibitem[HH97]{Hanany:1997vm}
A.~Hanany and K.~Hori,
``Branes and N=2 theories in two-dimensions,''
Nucl. Phys. B \textbf{513}, 119-174 (1998)
doi:10.1016/S0550-3213(97)00754-2
[arXiv:hep-th/9707192 [hep-th]].

\bibitem[HMW11]{Harlow:2011ny}
D.~Harlow, J.~Maltz and E.~Witten,
``Analytic Continuation of Liouville Theory,''
JHEP \textbf{12}, 071 (2011)
doi:10.1007/JHEP12(2011)071
[arXiv:1108.4417 [hep-th]].

\bibitem[Hay10]{Haydys:2010dv}
A.~Haydys,
``Fukaya-Seidel category and gauge theory,''
J. Sympl. Geom. \textbf{13}, 151-207 (2015)
doi:10.4310/JSG.2015.v13.n1.a5
[arXiv:1010.2353 [math.SG]].

\bibitem[HIV00]{Hori:2000ck}
K.~Hori, A.~Iqbal and C.~Vafa,
``D-branes and mirror symmetry,''
[arXiv:hep-th/0005247 [hep-th]].

\bibitem[MS1]{Hori:2003ic}
K.~Hori, S.~Katz, A.~Klemm, R.~Pandharipande, R.~Thomas, C.~Vafa, R.~Vakil and E.~Zaslow,
``Mirror symmetry.''

\bibitem[HV00]{Hori:2000kt}
K.~Hori and C.~Vafa,
``Mirror symmetry,''
[arXiv:hep-th/0002222 [hep-th]].

\bibitem[HL96]{hutchings1}
M.~Hutchings and Y.~Lee,
``Circle-valued Morse theory, Reidemeister torsion, and Seiberg-Witten invariants of 3-manifolds,"
Topology 38 (1999), 861-888
[arXiv:dg-ga/9612004 [math.DG]].

\bibitem[HL97]{hutchings2}
M.~Hutchings and Y.~Lee,
``Circle-valued Morse theory and Reidemeister torsion,"
Geom. Topol. 3 (1999) 369-396
[arXiv:dg-ga/9706012 [math.DG]].

\bibitem[H99]{hutchings3}
M.~Hutchings,
``Reidemeister torsion in generalized Morse theory,"
Forum Mathematicum 14 (2002), 209-244
[arXiv:math/9907066 [math.DG]].

\bibitem[HBlog]{hutchings4}
M.~Hutchings,
``Gluing a flow line to itself,"
Blog post, 
\href{https://floerhomology.wordpress.com/2014/07/14/gluing-a-flow-line-to-itself/}{https://floerhomology.wordpress.com/2014/07/14/gluing-a-flow-line-to-itself/}

\bibitem[KS]{Kajiura:2004xu}
H.~Kajiura and J.~Stasheff,
``Homotopy algebras inspired by classical open-closed string field theory,''
Commun. Math. Phys. \textbf{263}, 553-581 (2006)
doi:10.1007/s00220-006-1539-2
[arXiv:math/0410291 [math.QA]].

\bibitem[KKS14]{Kapranov:2014uwa}
M.~Kapranov, M.~Kontsevich and Y.~Soibelman,
``Algebra of the infrared and secondary polytopes,''
Adv. Math. \textbf{300}, 616-671 (2016)
doi:10.1016/j.aim.2016.03.028
[arXiv:1408.2673 [math.SG]].

\bibitem[KSS20]{Kapranov:2020zoa}
M.~Kapranov, Y.~Soibelman and L.~Soukhanov,
``Perverse schobers and the Algebra of the Infrared,''
[arXiv:2011.00845 [math.AG]].

\bibitem[Kel99]{Keller}
B.~Keller,
``Introduction to A-infinity algebras and modules,"
[arXiv:math/9910179 [math.RA]].



\bibitem[Kh21]{Khan:2021hve}
A.~Z.~Khan,
``Categorical aspects of BPS states,'' Rutgers PhD Thesis, 
doi:10.7282/t3-d4e7-n717

\bibitem[KhTalk]{khantalk}
A.~Z.~Khan, ``Algebra of the Infrared with Twisted Masses," IAS High Energy Theory Seminar, Available at \href{https://www.ias.edu/video/algebra-infrared-twisted-masses}{https://www.ias.edu/video/algebra-infrared-twisted-masses} 

\bibitem[KM20]{Khan:2020hir}
A.~Z.~Khan and G.~W.~Moore,
``Categorical Wall-Crossing in Landau-Ginzburg Models,''
[arXiv:2010.11837 [hep-th]].


\bibitem[KMWP]{WIP}
A.~Z.~Khan and G.~W.~Moore, Work in progress. 

\bibitem[Ko97]{Kontsevich:1997vb}
M.~Kontsevich,
``Deformation quantization of Poisson manifolds. 1.,''
Lett. Math. Phys. \textbf{66}, 157-216 (2003)
doi:10.1023/B:MATH.0000027508.00421.bf
[arXiv:q-alg/9709040 [math.QA]].

\bibitem[KoSo08]{Kontsevich:2008fj}
M.~Kontsevich and Y.~Soibelman,
``Stability structures, motivic Donaldson-Thomas invariants and cluster transformations,''
[arXiv:0811.2435 [math.AG]].

\bibitem[KoSo24]{Kontsevich:2024esg}
M.~Kontsevich and Y.~Soibelman,
``Holomorphic Floer theory I: exponential integrals in finite and infinite dimensions,''
[arXiv:2402.07343 [math.SG]].

\bibitem[Lur12]{lurie}
J.~Lurie,
``Higher Algebra,''
Available at \href{https://people.math.harvard.edu/~lurie/papers/HA.pdf}{https://people.math.harvard.edu/~lurie/papers/HA.pdf}.
2012.

\bibitem[M97]{Moore:1997ar}
G.~W.~Moore,
``String duality, automorphic forms, and generalized Kac-Moody algebras,''
Nucl. Phys. B Proc. Suppl. \textbf{67}, 56-67 (1998)
doi:10.1016/S0920-5632(98)00120-0
[arXiv:hep-th/9710198 [hep-th]].


%
\bibitem[MRTLK]{MooreTalks}
See talk \#  102 at  \href{https://www.physics.rutgers.edu/~gmoore/}{https://www.physics.rutgers.edu/~gmoore/} for video, powerpoint, and pdf versions of three talks summarizing the results. 
  

\bibitem[OINS99]{Oda:1999az}
H.~Oda, K.~Ito, M.~Naganuma and N.~Sakai,
``An Exact solution of BPS domain wall junction,''
Phys. Lett. B \textbf{471}, 140-148 (1999)
doi:10.1016/S0370-2693(99)01355-6
[arXiv:hep-th/9910095 [hep-th]].




\bibitem[PW21]{Paquette:2021cij}
N.~M.~Paquette and B.~R.~Williams,
``Koszul duality in quantum field theory,''
[arXiv:2110.10257 [hep-th]].

\bibitem[Park]{Park:2015kra}
D.~S.~Park,
``Topological charges in 2d N=(2,2) theories and massive BPS states,''
Phys. Rev. D \textbf{92}, no.2, 025044 (2015)
doi:10.1103/PhysRevD.92.025044
[arXiv:1505.01508 [hep-th]].

\bibitem[Pri70]{priddykoszul}
S.~B.~Priddy,
``Koszul Resolutions.''
Transactions of the American Mathematical Society , Nov., 1970, Vol. 152, No. 1
(Nov., 1970), pp. 39-60

\bibitem[S99]{Saffin:1999au}
P.~M.~Saffin,
 ``Tiling with almost BPS junctions,''
Phys. Rev. Lett. \textbf{83}, 4249-4252 (1999)
doi:10.1103/PhysRevLett.83.4249
[arXiv:hep-th/9907066 [hep-th]].


\bibitem[Sei94]{Seiberg:1994bp}
N.~Seiberg,
``The Power of holomorphy: Exact results in 4-D SUSY field theories,''
[arXiv:hep-th/9408013 [hep-th]].

\bibitem[SW94]{Seiberg:1994rs}
N.~Seiberg and E.~Witten,
``Electric - magnetic duality, monopole condensation, and confinement in N=2 supersymmetric Yang-Mills theory,''
Nucl. Phys. B \textbf{426}, 19-52 (1994)
[erratum: Nucl. Phys. B \textbf{430}, 485-486 (1994)]
doi:10.1016/0550-3213(94)90124-4
[arXiv:hep-th/9407087 [hep-th]].

\bibitem[SY09]{Shifman:2009zz}
M.~Shifman and A.~Yung,
``Supersymmetric Solitons,''
Cambridge University Press, 2023,
ISBN 978-1-00-940220-0, 978-1-00-940217-0, 978-1-00-940222-4, 978-0-511-51250-6, 978-0-521-51638-9
doi:10.1017/9781009402200

\bibitem[Sou18]{soukhanov}
L.~Soukhanov,
``2-Morse Theory and the Algebra of the Infrared,"
[arXiv:1810.08776 [math.AG]].

\bibitem[Wan22]{Wang:2022xdn}
D.~Wang,
``The Complex Gradient Flow Equation and Seidel's Spectral Sequence,''
[arXiv:2209.02810 [math.SG]].

\bibitem[WZ73]{Wess:1973kz}
J.~Wess and B.~Zumino,
``A Lagrangian Model Invariant Under Supergauge Transformations,''
Phys. Lett. B \textbf{49}, 52 (1974)
doi:10.1016/0370-2693(74)90578-4


\bibitem[Wit82]{Witten:1982im}
E.~Witten,
 ``Supersymmetry and Morse theory,''
J. Diff. Geom. \textbf{17}, no.4, 661-692 (1982)


\bibitem[Wit10]{Witten:2010cx}
E.~Witten,
``Analytic Continuation Of Chern-Simons Theory,''
AMS/IP Stud. Adv. Math. \textbf{50}, 347-446 (2011)
[arXiv:1001.2933 [hep-th]].

\bibitem[Wit11]{Witten:2011zz}
E.~Witten,
``Fivebranes and Knots,''
[arXiv:1101.3216 [hep-th]].

\bibitem[Zum79]{Zumino:1979et}
B.~Zumino,
``Supersymmetry and Kahler Manifolds,''
Phys. Lett. B \textbf{87}, 203 (1979)
doi:10.1016/0370-2693(79)90964-X
\end{thebibliography}
\end{document}